\documentclass[a4paper,11pt]{article}
\pdfoutput=1 % if your are submitting a pdflatex (i.e. if you have
\usepackage{jcappub} % for details on the use of the package, please
\usepackage[T1]{fontenc} % if needed

\usepackage{tikz}
\usepackage{hyperref}
\usepackage{subcaption}
\usepackage{graphicx}
\usepackage{xcolor}
\usepackage{multirow}
\usepackage{comment}
\usepackage{amsmath}

\newcommand{\cosmogrid}{\texttt{CosmoGridV1}}
\newcommand{\pkdgrav}{\texttt{PkDGraV3}}
\newcommand{\ufalcon}{\texttt{UFalconv2}}
\newcommand{\class}{\texttt{CLASS}}
\newcommand{\highmnu}{high $M_\nu$ scenario}

\title{$\mathbf{12\times2}$pt combined probes: pipeline, neutrino mass, and data compression}
\author[a]{Alexander Reeves,}
\author[b]{Andrina Nicola,}
\author[a]{Alexandre Refregier,}
\author[a]{Tomasz Kacprzak,}
\author[a]{and Luis Fernando Machado Poletti Valle}

\affiliation[a]{ETH Zürich \\ Institute for Particle Physics and Astrophysics, Wolfgang-Pauli-Strasse 27, CH-8093 Zürich, Switzerland}

\affiliation[b]{Argelander Institut f\"ur Astronomie, Universit\"at Bonn, Auf dem H\"ugel 71, 53121 Bonn, Germany}

\emailAdd{areeves@phys.ethz.ch}

\abstract{With the rapid advance of wide-field surveys it is increasingly important to perform combined cosmological probe analyses. We present a new pipeline for simulation-based multi-probe analyses, which combines tomographic large-scale structure (LSS) probes (weak lensing and galaxy clustering) with cosmic microwave background (CMB) primary and lensing data. These are combined at the $C_\ell$-level, yielding 12 distinct auto- and cross-correlations. The pipeline is based on \ufalcon{}, a framework to generate fast, self-consistent map-level realizations of cosmological probes from input lightcones, which is applied to the \cosmogrid{} N-body simulation suite. It includes a non-Gaussian simulation-based covariance for the LSS tracers, several data compression schemes, and a neural network emulator for accelerated theoretical predictions. We validate the pipeline by comparing the simulations to these predictions, and our derived constraints to earlier analyses. We apply our framework to a simulated $12\times2$pt tomographic analysis of KiDS, BOSS, and \textit{Planck}, and forecast constraints for a $\Lambda$CDM model with a variable neutrino mass. We find that, while the neutrino mass constraints are driven by the CMB data, the addition of LSS data helps to break degeneracies and improves the constraint by up to 35\%. For a fiducial $M_\nu=0.15\mathrm{eV}$, a full combination of the above CMB+LSS data would enable a $3\sigma$ constraint on the neutrino mass. We explore data compression schemes and find that MOPED outperforms PCA and is made robust using the derivatives afforded by our automatically differentiable emulator. We also study the impact of an internal lensing tension in the CMB data, parametrized by $A_L$, on the neutrino mass constraint, finding that the addition of LSS to CMB data including all cross-correlations is able to mitigate the impact of this systematic. \ufalcon{} and a MOPED compressed \textit{Planck} CMB primary + CMB lensing likelihood are made publicly available.\footnote{\ufalcon{}: \url{https://cosmology.ethz.ch/research/software-lab/UFalcon.html}, compressed \textit{Planck} CMB primary + CMB lensing likelihood: \url{https://github.com/alexreevesy/planck_compressed}}}

\begin{document}
\maketitle
\flushbottom

\section{Introduction} \label{sec:introduction}

With the onset of next-generation surveys such as Euclid~\cite{EUCLID:2011zbd}, the Large Synoptic Survey Telescope (LSST)~\cite{LSSTScience:2009jmu}, the Simons Observatory~\cite{SimonsObservatory:2018koc}, Roman Space Telescope ~\cite{Akeson:2019biv} and  CMB-S4~\cite{CMB-S4:2016ple}, we will be uniquely positioned to rigorously test our cosmological model. In this context of high-precision data, increasingly dominated by systematics, combined probes analyses offer several advantages. This method uses a common framework to analyze different survey datasets, allowing for a consistent exploration of discrepancies between them~\cite{Ruiz-Zapatero:2021rzl, Nicola:2017ryw}. In addition, combining cosmological probes can break parameter degeneracies inherent in single probe analyses, thus leading to tighter constraints~\cite{Kacprzak:2022oit,Rhodes:2013fyq}. Finally, by leveraging cross-correlations, we can calibrate systematic effects that do not correlate between the datasets~\cite{Weinberg:2013agg, Baxter:2022enq}.

It is no surprise, then, that combined probe analyses have been a central consideration in the planning of recent and future surveys. In the context of large-scale structure (LSS) data, $3\times2$pt analyses, which combine weak lensing (WL) and galaxy clustering data at the 2-point level, are now standard for current and recent optical surveys such as the Dark Energy Survey (DES)~\cite{DES:2017myr}, Hyper Suprime Cam (HSC)~\cite{HSC:2018mrq} and the Kilo-Degree Survey (KiDS)~\cite{Heymans:2020gsg, Hildebrandt:2018yau}. Cosmic microwave background (CMB) measurements have also been jointly analyzed with LSS data to derive strong constraints: There have been a number of recent studies considering the CMB lensing-galaxy clustering cross-correlation~\cite{Piccirilli:2022myi,Singh:2016xey,White:2021yvw,Doux:2017tsv,Krolewski:2019yrv, HerschelATLAS:2014txv}, the Integrated Sachs-Wolfe (ISW) cross-correlation~\cite{Planck:2015fcm,Krolewski:2021znk} and the CMB lensing-galaxy weak lensing cross-correlation~\cite{Robertson:2020xom}. 

The aspiration to create a comprehensive framework for analyzing all datasets, from both the CMB and LSS realms, remains a fundamental goal within combined probes analysis. In recent years a number of studies have emerged that move towards this goal~\cite{Fang:2023efj,Sgier:2021bzf, Garcia-Garcia:2021unp, Nicola:2016qrc, Nicola:2016eua, DES:2022xxr, Wenzl:2021rrq}. In particular, Ref.~\cite{Sgier:2021bzf} builds upon the map-based approach introduced in Refs.~\cite{Nicola:2016eua, Nicola:2016qrc} to combine CMB and LSS data. Recently, Ref.~\cite{Fang:2023efj} (see also~\cite{Krause:2016jvl, Eifler:2013fit}) provides a halo model-based scheme to combine galaxy clustering, weak lensing, CMB lensing, and thermal Sunyaev Zel'dovich (tSZ) data, resulting in 10 different 2-point functions. 

In this work, we introduce a novel combined probes pipeline, designed to incorporate several cosmological datasets at the $C_\ell$ level, inclusive of their respective cross-correlations. Specifically, this pipeline is designed to incorporate CMB primary, galaxy clustering, weak lensing (WL), and CMB lensing data, resulting in a suite of 12 distinct 2-point (2pt) functions once cross-correlations are included. The pipeline includes several advantages compared to previous approaches including a non-Gaussian simulation-based covariance for the LSS probes, methods for data compression, and a neural network emulator for accelerated parameter exploration. The pipeline uses \ufalcon{}, a new publicly available software that produces consistent map-level realizations of LSS probes (WL, galaxy clustering CMB lensing, and ISW) from an input lightcone. This is applied to the \cosmogrid{} N-body simulation suite to produce a set of correlated map-level realizations from which the LSS mock data vector and covariance matrix are derived.

We apply this framework to perform a forecast tomographic analysis of KiDS~\cite{Heymans:2020gsg, Hildebrandt:2018yau}, BOSS~\cite{BOSS:2016wmc}, and \textit{Planck}~\cite{Planck:2018vyg}. We validate our pipeline by comparing our constraints with the official results from each of the individual surveys and ensuring that we recover unbiased constraints. We then examine two applications of the framework. In the first, we use the full $12\times$2pt data vector and forecast its constraining power on the sum of the neutrino mass eigenstates, $M_\nu = \sum_i m_{\nu_i}$. We also analyze how the  `A-lens' parameter ($A_L$), first introduced in ~\cite{Calabrese:2008rt}, impacts this measurement\footnote{There is mild evidence of $A_L>1$ in \textit{Planck} PR3 data~\cite{Planck:2018vyg} which may bias the neutrino mass constraint towards low values ~\cite{DiValentino:2021imh}. See also ~\cite{Rosenberg:2022sdy, Tristram:2023haj} for recent \textit{Planck} PR4 analyses which show less evidence for $A_L>1$.}.

In a second application, we explore methods of data compression. This is especially important in the case of multiprobe analyses where uncompressed data vectors are typically large. We specifically compare the performance of two data compression techniques: MOPED ~\cite{Heavens:1999am} and Principal Component Analysis (PCA)~\cite{pearson:1901op}. We also apply MOPED to the full \textit{Planck} CMB primary and lensing likelihood, extending the work presented in Ref.~\cite{Prince:2019hse}.

\subsection{Overview of pipeline}
\begin{figure*}[htbp!]
\centering
\includegraphics[width=0.7\paperwidth]{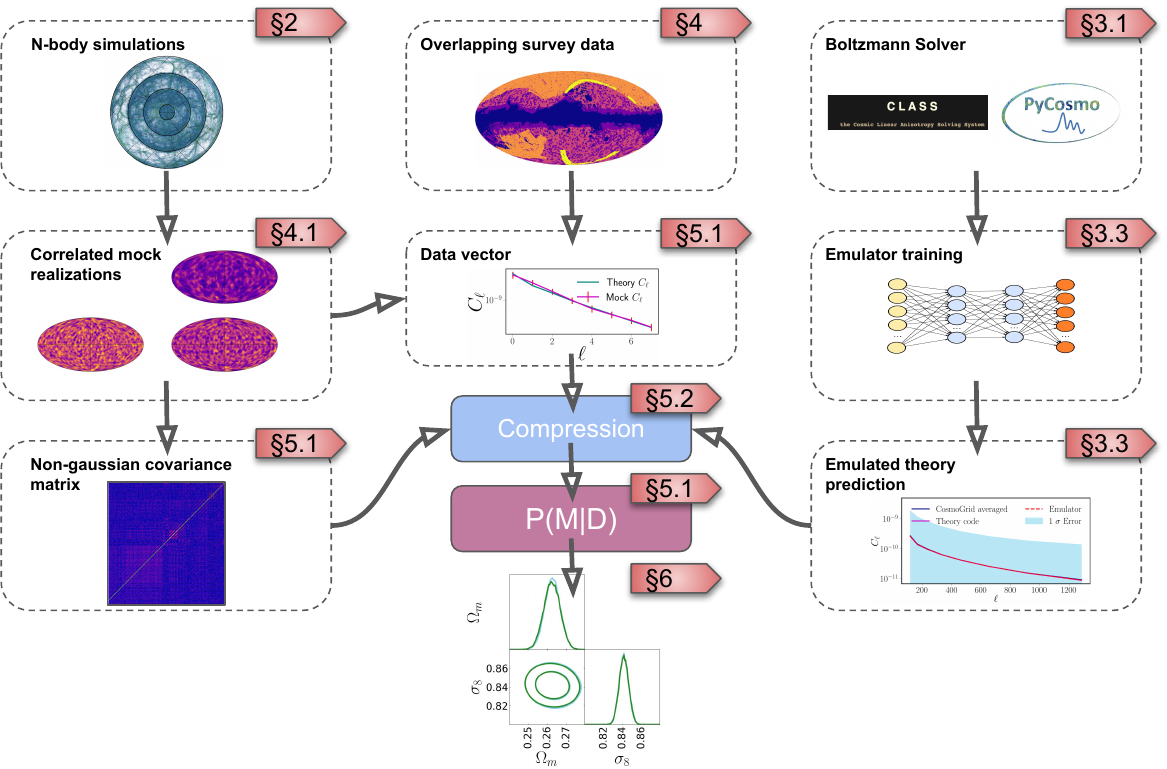}
\caption{An overview of the combined-probes pipeline. The left-hand channel represents the processing of simulations in this pipeline to produce a Non-Gaussian covariance, whilst the right-hand channel represents the process of training a neural network emulator for rapid theoretical predictions. Finally, the middle section shows how survey data is used to derive a data vector (in this case a mock data vector) before all three elements are combined and optionally compressed (pink panel) to compute the likelihood and eventually derive parameter constraints.\label{fig:pipeline}}
\end{figure*}

A schematic of our combined probes pipeline is shown in Fig.~\ref{fig:pipeline}. On the left-hand side, the framework is built upon \ufalcon{} which processes N-body simulations into correlated map-level realizations of each probe. These are used to produce both the fully non-Gaussian covariance matrix for the LSS tracers (bottom left panel in Fig.~\ref{fig:pipeline}) and in producing realistic mock data vectors (middle panel in Fig.~\ref{fig:pipeline}). We then optionally compress these data vectors, combine them with theoretical predictions obtained from a differentiable neural-network emulator (bottom right panel in Fig.~\ref{fig:pipeline}), and eventually derive the likelihood and cosmological parameter constraints (respectively bottom two panels in Fig.~\ref{fig:pipeline}). In this way, the pipeline is built in a modular fashion and the interested reader can skip to one of the descriptions in the sections shown in Fig.~\ref{fig:pipeline} for details.

The paper is structured as follows: We begin with a description of the suite of simulations in $\S$\ref{sec:simulations}, including links to the publicly available \ufalcon{} software. Then, in $\S$\ref{sec:theory_model_emu} we describe the theoretical modeling of probes in the pipeline and the training of our multiprobe neural network emulator.  In $\S$\ref{sec:survey_data}, we outline the data products from real surveys that are used to build our realistic mock observations, and in $\S$\ref{sec:combined_probes_like} how the multiprobe likelihood is computed and interfaced with the inference part of the pipeline. Finally, in $\S$\ref{sec:results} we describe and discuss the results of our analysis before concluding in $\S$\ref{sec:conclusion}.

\section{Simulations} \label{sec:simulations}
\subsection{\cosmogrid}  

In this section, we describe the simulations that we use for both estimating the covariance matrix and producing realistic mock data vectors. We use the simulations provided in the \cosmogrid \footnote{\url{http://www.cosmogrid.ai/}} ~\cite{Kacprzak:2022pww}. This is a large suite of simulations produced with the GPU-accelerated N-Body code \pkdgrav\ \cite{Potter:2016ttn}. The simulations have a box size of $L = 900$ Mpc/h, contain $832^3$ particles to sample the dark matter distribution, and are stored as lightcone output. The lightcone consists of 69 shells stored as \texttt{HEALPix}\footnote{\url{https://healpix.sourceforge.io/}} maps with \texttt{nside}=2048, in the redshift range $z=0-3.5$, which allows for accurate map projections for various cosmological probes. Whilst the full \cosmogrid{} consists of a suite of 2500 cosmologies with varied [$\Omega_m$, $\sigma_8$, $h$, $w_0$, $n_s$, $\Omega_b$], in this pipeline we make use only of the 200 fiducial simulations, run at a fixed set of cosmological parameters shown in  Table~\ref{tab:fidcosmo}. The effect of massive neutrinos is included in the simulations following the implementation from Ref.~\cite{Tram:2018znz} available in \pkdgrav. This method accurately models the effect of massive, but light neutrinos, on the matter clustering, using linear perturbation theory (see Ref.~\cite{Kacprzak:2022pww} for further details). 

\begin{table}[h]
\centering
\begin{tabular}{cc}
    \hline
    Parameter & Value \\
    \hline
    $\Omega_{m}$ & 0.26 \\
    $\sigma_{8}$ & 0.84 \\
    $n_{s}$ & 0.9649 \\
    $\Omega_{b}$ & 0.0493 \\
    $H_{0}$(Km/s/Mpc) & 67.3 \\
    $M_\nu$(eV) & 0.06 \\
    \hline
\end{tabular}
\caption{Fiducial Cosmological parameters used in the \cosmogrid.}
\label{tab:fidcosmo}
\end{table}

\subsection{Map making with \ufalcon{}}\label{subsec:UFalcon2}
\begin{figure}
    \centering
    \begin{subfigure}[b]{0.45\textwidth}
        \centering
        \includegraphics[width=\textwidth]{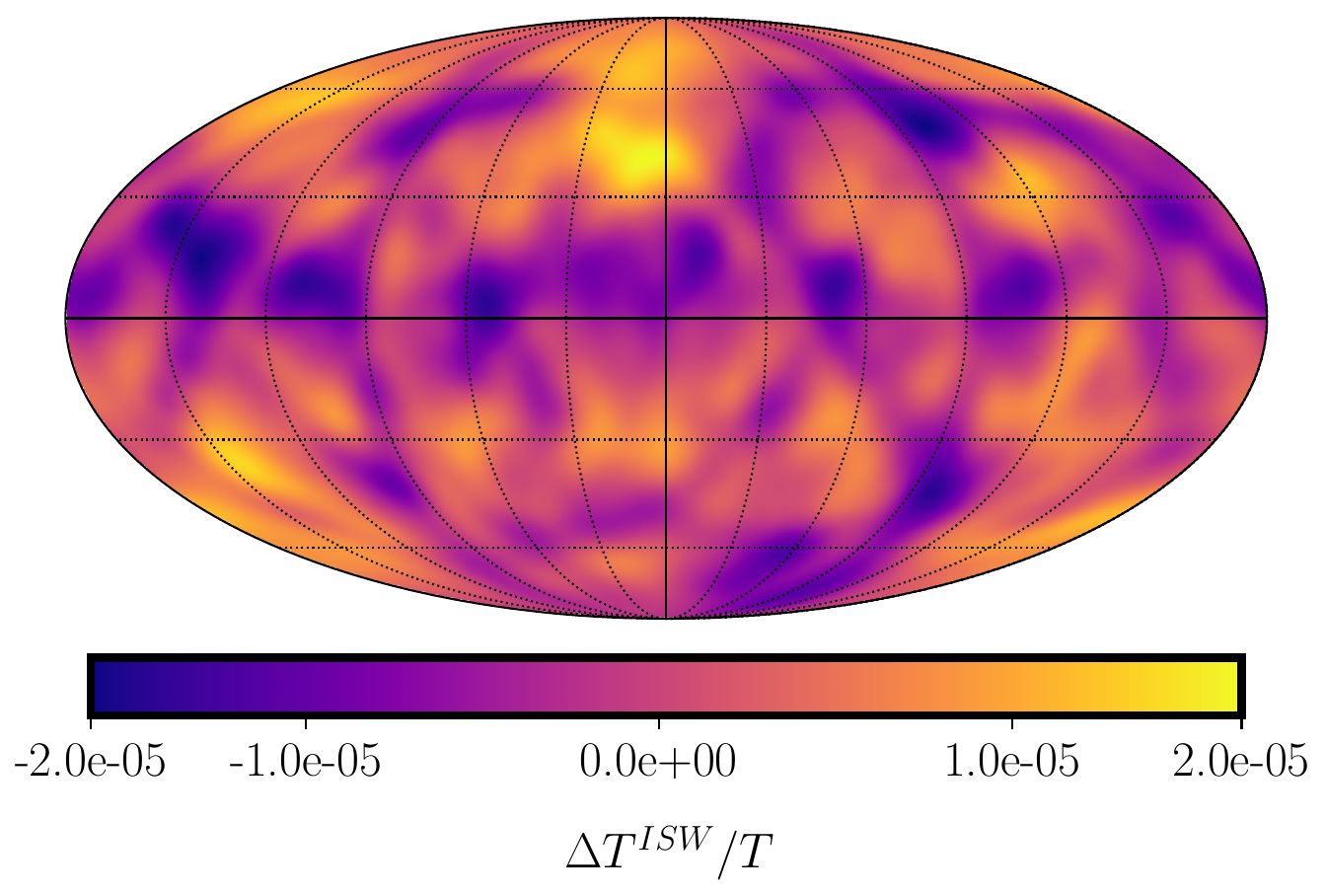}
        \caption{ISW map.}
    \end{subfigure}
    \hfill
    \begin{subfigure}[b]{0.45\textwidth}
        \centering
        \includegraphics[width=\textwidth]{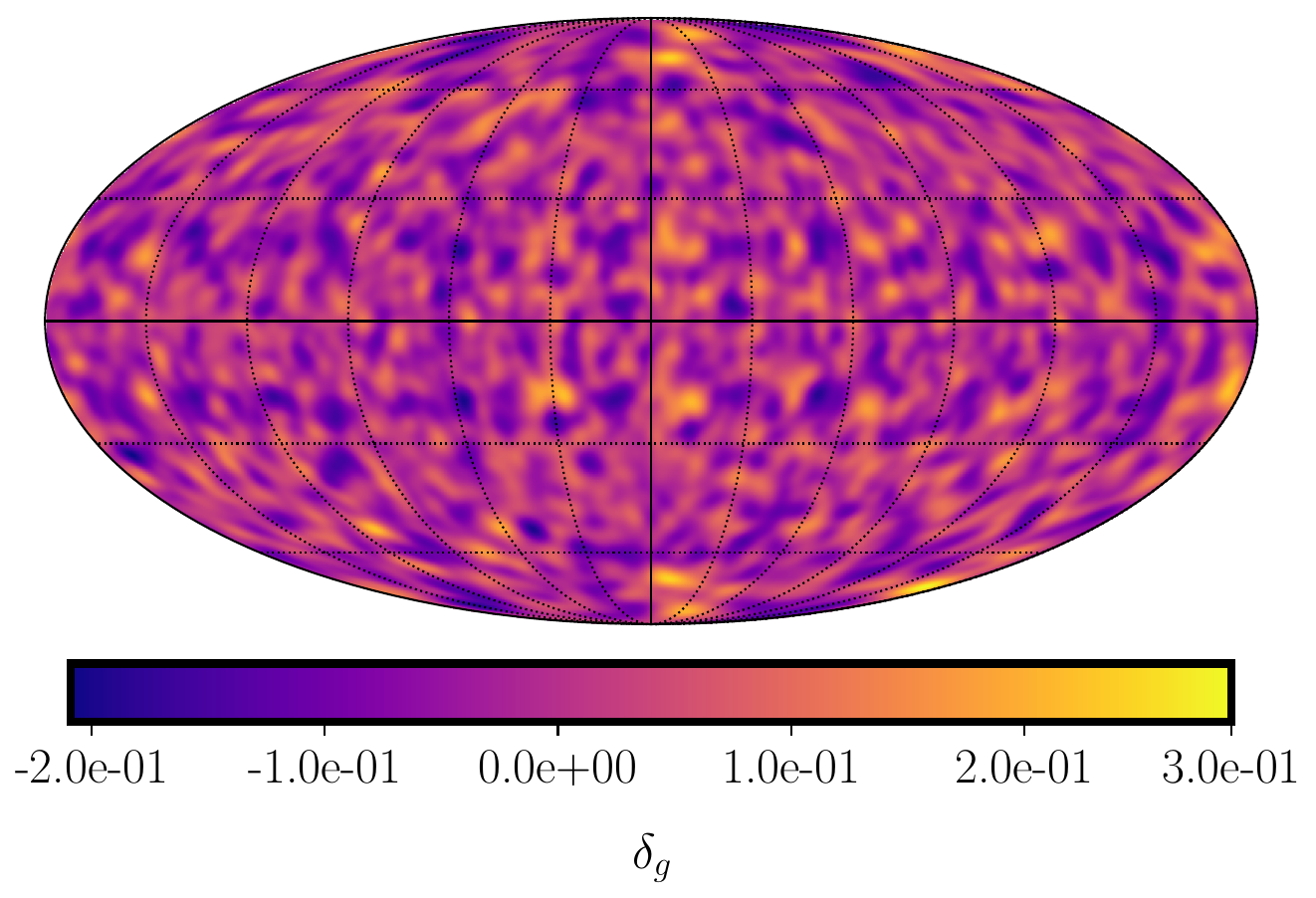}
        \caption{Galaxy clustering map.}
    \end{subfigure}
    
    \vspace{1em}
    
    \begin{subfigure}[b]{0.45\textwidth}
        \centering
        \includegraphics[width=\textwidth]{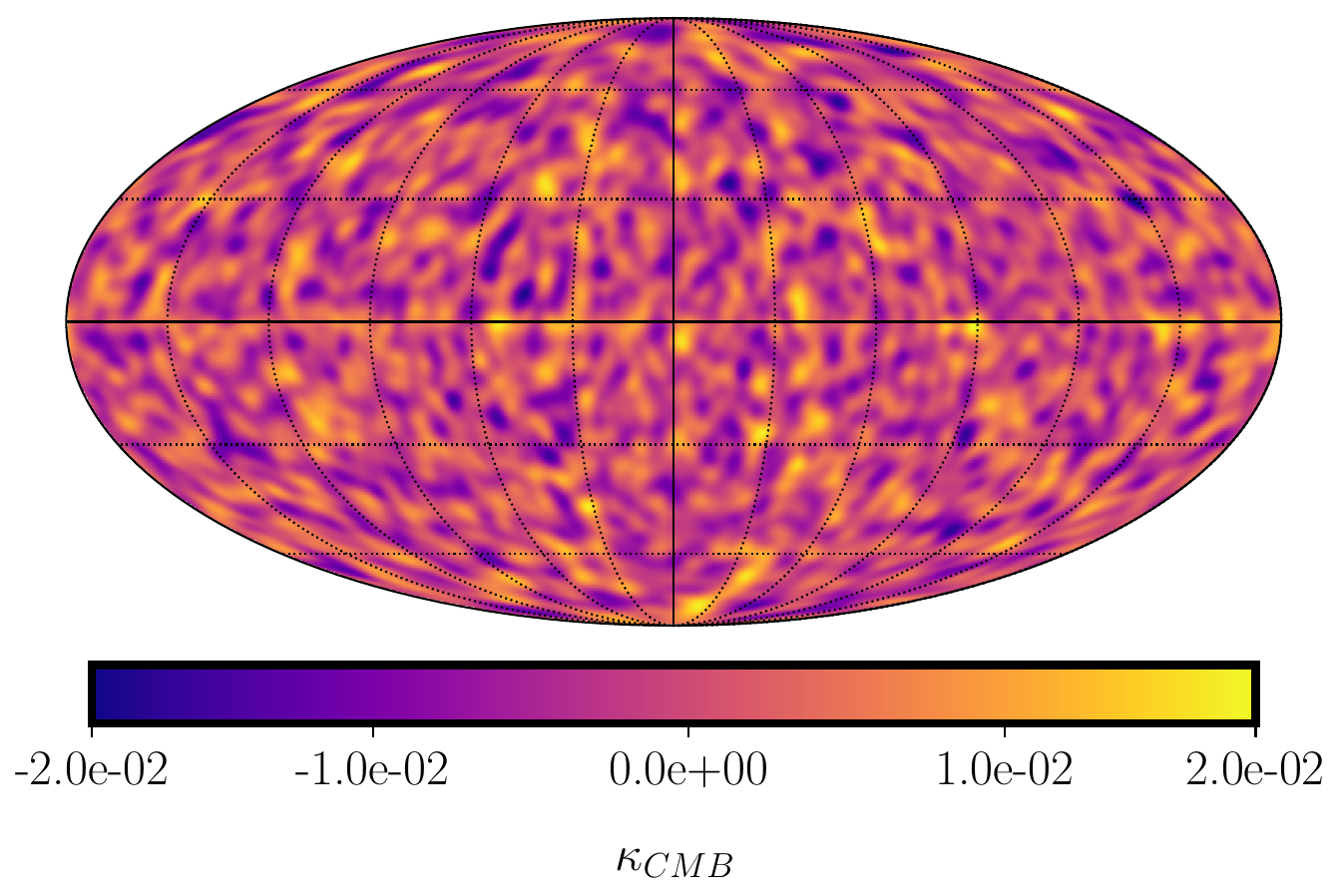}
        \caption{CMB lensing map.}
    \end{subfigure}
    \hfill
    \begin{subfigure}[b]{0.45\textwidth}
        \centering
        \includegraphics[width=\textwidth]{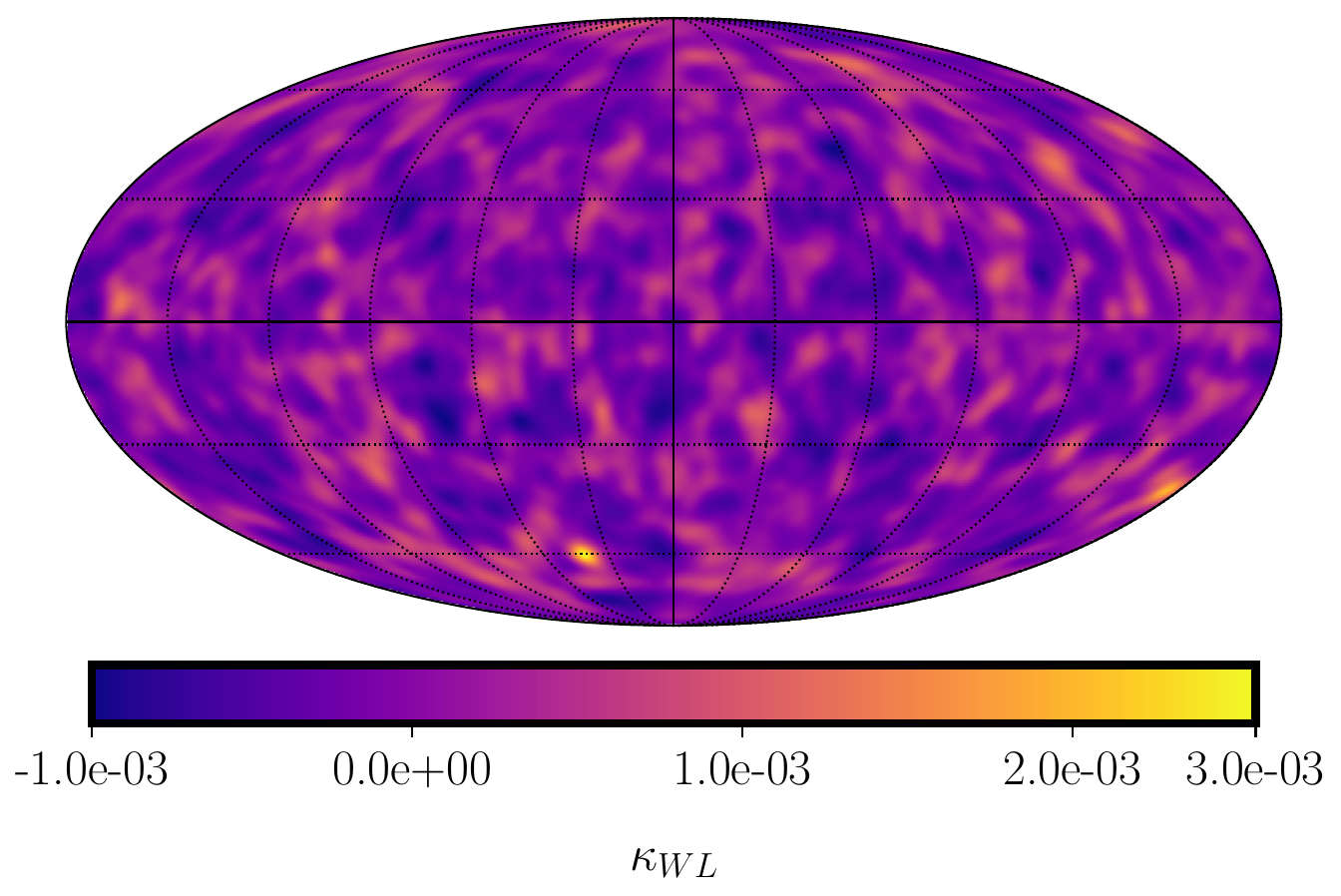}
        \caption{Convergence $\kappa$ map.}
    \end{subfigure}
    
    \caption{Example maps produced by post-processing the \cosmogrid\ with the \ufalcon\ software. For the clustering map we use the redshift distribution from the BOSS CMASS selection~\cite{BOSS:2016wmc} and a corresponding linear bias value of $b=2.086$, and for the shear map we use the first redshift bin from KiDS-1000~\cite{KiDS:2020suj}. All maps plotted above are dimensionless.\label{fig:ufalcon_maps}}
\end{figure}

To process the simulations described above into cosmological signal maps we created \ufalcon{} (\textbf{U}ltra \textbf{Fa}st \textbf{l}igh\textbf{con}e), an updated version of the \texttt{UFalcon} code, first presented in~\cite{Sgier:2018soj, Sgier:2020das}. The source code and documentation (including usage and installation instructions) for the package are available at the following webpage: \url{https://cosmology.ethz.ch/research/software-lab/UFalcon.html}. The updated code is configured to output maps of galaxy weak lensing, galaxy clustering, CMB lensing, and the ISW temperature perturbation. Whilst \texttt{UFalcon} was originally configured to process N-body code snapshot output, the updated version directly handles lightcone output, enabling processing of the \cosmogrid{}. Processing a \cosmogrid{} lightcone into any of the signal maps takes O($\sim$~minutes) on a Mac M1 machine. In the following subsections, we briefly describe the procedures implemented in \ufalcon{} to produce full-sky realizations of each of the above probes. We produce all maps in the pipeline at a fiducial \texttt{HEALPix}~\cite{Gorski:2004by} resolution of \texttt{nside}=1024. Some example maps generated using \ufalcon{} are shown in Fig.~\ref{fig:ufalcon_maps}.

\subsubsection{Galaxy weak lensing}
Galaxy weak lensing (WL) refers to the distortion of the observed shapes of galaxies due to the gravitational deflection of light as it passes through intervening mass distributions. We can indirectly measure the dark matter distribution via analysis of the coherent alignment of galaxy shapes due to gravitational lensing (see ~\cite{Refregier:2003ct,Kilbinger:2014cea} for reviews). In this context, galaxy shear refers to the differential stretching of the observed shape of a galaxy, while convergence describes its dilation or contraction.

In the Born approximation, which ignores the fact that light rays do not follow straight lines, the convergence $\kappa$ can be related to the line-of-sight projection of the dark matter field $\delta$ with a kernel as~\cite{Kilbinger:2014cea}:
\begin{equation}\label{eqt:wl_integral}
\kappa(\mathbf{\theta}) = \frac{3H_0^2\Omega_m}{2c^2}\int_0^{z_s} \frac{g(z, z_s)}{a(z)E(z)}\delta(\chi(z)\mathbf{\theta}, z) \mathrm{d}z, 
\end{equation}
where $\Omega_m$ is the matter density parameter, $z_s$ represents the limiting redshift of the galaxy sample, $H_0$ is the Hubble constant, $E(z)=H(z)/H_0$ is the dimensionless Hubble parameter, $c$ is the speed of light, and $g(z,z_s)$ is the galaxy lensing radial function. The lensing radial function is defined as~\cite{Kilbinger:2014cea}:
\begin{equation}
    g(z,z_s) = \int_z^{z_s} \mathrm{d}z' n(z') \frac{\mathcal{D}(z')\mathcal{D}(z',z_s)}{\mathcal{D}(z_s)},
\end{equation}
where $\mathcal{D}(z)=\frac{H_0}{c}\chi(z)$ is the dimensionless comoving distance at redshift $z$, $\mathcal{D}(z,z')=\mathcal{D}(z')-\mathcal{D}(z)$ and $n(z)$ is the redshift distribution of source galaxies normalized such that $\int n(z')\mathrm{d}z'=1$. 

To compute a map from the \cosmogrid{} lightcones, we approximate the integral in equation~\ref{eqt:wl_integral} as a sum over the redshift shells that comprise a lightcone. Using the relative number of simulation particles in a given pixel as a proxy for the underlying dark matter contrast, the convergence field can be written (up to an overall constant) as~\cite{Teyssier:2008zd,Sgier:2018soj}:

\begin{equation}
    \begin{split}\label{equation:wl_from_sims}
        \kappa(\theta_{\text{pix}}) \approx \frac{3}{2} \Omega_m \left(\frac{H_0}{c}\right)^3 \frac{N_{\text{pix}}V_{\text{sim}}}{4 \pi N_p}\sum_b W^{\kappa}_b(z_b) \frac{n_p(\theta_{\text{pix}})}{\mathcal{D}^2(z_b)},\\
        W^{\kappa}_b(z_b) =\frac{\int_{\Delta z_b} \frac{\mathrm{d}z}{E(z)} \frac{1}{a(z)} \int_z^{z_s} \mathrm{d}z'n(z') \frac{\mathcal{D}(z)\mathcal{D}(z,z')}{\mathcal{D}(z')}}{\int_{\Delta z_b} \frac{\mathrm{d}z}{E(z)} \int_{z_0}^{z_s} \mathrm{d}z' n(z')}.
    \end{split}
\end{equation}

In this equation, $N_{\text{pix}}$ is the total number of pixels in the map, $n_p$ is the number of particles at the pixel position $\theta_{\text{pix}}$, $N_p$ is the total number of particles, $V_{\text{sim}}$ is the volume of the simulation box and $\Delta z_b$ represents the redshift width of the shell. Note that the user must specify the input cosmological parameters and the redshift distribution function $n(z)$ to compute these maps.

Using equation~\ref{equation:wl_from_sims}, we are able to compute weak lensing convergence maps from an N-body simulation lightcone. However, in survey data, we measure the spin-2 weak lensing \textit{shear} field by analyzing galaxy shapes. We can relate this to the convergence field via~\cite{Wallis:2017lwt}:
\begin{equation}
    \begin{split}
        _2\gamma_{\ell m} = \frac{-1}{(\ell)(\ell +1)} \sqrt{\frac{(\ell+2)!}{(\ell-2)!}} \kappa_{\ell m},
    \end{split}
\end{equation} 
where $_2\gamma_{\ell m}$ are the spin-2 spherical harmonic coefficients of the weak lensing shear observable. This relation is integrated into the \ufalcon{} code and the user can specify if they want convergence or shear output. 

\subsubsection{Galaxy clustering}
Galaxy clustering refers to the non-random spatial distribution of galaxies in the universe which act as \textit{biased tracers} of the underlying dark matter field (see ~\cite{Desjacques:2016bnm} for a recent review). In \ufalcon{} galaxy clustering is modeled using the linear bias approach such that the galaxy over-density $\delta_g$ is related to the underlying dark matter over-density $\delta$ via:
\begin{equation}
    \begin{split}
        \delta_g = b(z) \delta.
    \end{split}
\end{equation}
We can compute the projected galaxy field as an integral over the dark matter distribution: 
\begin{equation}
\delta_g(\mathbf{\theta}) = \int_0^{z_s} b(z) n_g(z) \delta(\chi(z)\mathbf{\theta}, z) \mathrm{d}z, 
\end{equation}
where $n_g(z)$ represents the normalized redshift distribution of the galaxy survey. We can then make an analogy to equation \ref{equation:wl_from_sims}, and approximate this integral as a weighted sum over lightcone slices:
\begin{equation}
    \begin{split}
        \delta_g(\theta_{\text{pix}}) \approx \left(\frac{H_0}{c}\right)^2 \frac{N_{\text{pix}}V_{\text{sim}}}{4 \pi N_p}\sum_b W^{\delta_g}_b(z_b) \frac{n_p(\theta_{\text{pix}})}{\mathcal{D}^2(z_b)},\\
        W^{\delta_g}_b(z_b) =\frac{\int_{\Delta z_b} \frac{\mathrm{d}z}{E(z)}H(z)b(z)n_g(z)}{\int_{\Delta z_b} \frac{\mathrm{d}z}{E(z)}}.
    \end{split}
\end{equation}

\subsubsection{CMB weak lensing}
Weak gravitational lensing of the CMB refers to the deflection of CMB photons as they travel through the large-scale structure of the Universe. This induces a signature correlation in CMB temperature and polarization measurements which can be used to reconstruct the lensing potential. This is a probe of the intervening matter between the surface of last scattering where the CMB photons are emitted and the observer (see e.g. ~\cite{Lewis:2006fu} for a review). 

We implement the CMB weak lensing probe in \ufalcon{} in an analogous manner to galaxy weak lensing. In this case, the redshift distribution of the source can be effectively modeled as a Dirac delta-function $n(z)=\delta_D(z - z_{\ast})$ where $z_{\ast} \sim 1100$ is the redshift of recombination, and $\delta_D(z)$ is the Dirac delta function. Inserting $n(z)=\delta_D(z - z_{\ast})$ in the galaxy weak lensing window function of equation~\ref{equation:wl_from_sims} and integrating this out leads to a simplified expression for the CMB lensing window function:
\begin{equation}
    \begin{split}
        W^{\kappa_{\text{CMB}}}_b(z_b) =\frac{\int_{\Delta z_b} \frac{\mathrm{d}z}{E(z)} \frac{\mathcal{D}(z)\mathcal{D}(z,z_{\ast})}{\mathcal{D}(z_{\ast})} \frac{1}{a(z)}}{\int_{\Delta z_b} \frac{\mathrm{d}z}{E(z)}}.
    \end{split}
\end{equation}

N-body simulation lightcones typically contain shells up to a limiting redshift $z_{\text{max}}$. This means the CMB lensing maps produced using \texttt{UFalcon2} include only the contribution from $0<z<z_{\text{max}}$. For sufficiently high $z_{\text{max}}$ we can model the remaining CMB weak lensing signal as a Gaussian random field. We can then supplement the \ufalcon{}-derived CMB lensing map with a Gaussian realization for the $z_{\text{max}}<z<z_{\ast}$ part (note we assume the radial correlation between this additional contribution and the $0<z<z_{\text{max}}$ piece is negligible)\footnote{The Gaussian realization is produced using the \texttt{synfast} routine from \texttt{healpy}~\cite{Zonca:2019vzt} with a theoretical $C_\ell$ that includes only contributions from this redshift range. To generate the required $C_\ell$, we modified the \class{} code according to \url{https://github.com/lesgourg/class_public/issues/33}.}.

\subsubsection{Integrated Sachs-Wolfe Effect}\label{subsubsec:isw}

The ISW effect \cite{SachsWolfe1967, ReesSciama1968} arises from the time evolution of the gravitational potential, which leads to an imprint on the CMB at large physical scales ($\ell < 200$). The ISW signal can be detected via cross-correlations with large-scale structure tracers and provides a powerful probe of the late-time dark energy evolution at low redshifts.

The previous computation of the ISW signal in \texttt{UFalcon} relied on snapshots of N-body simulations to compute the time dependence of the gravitational potentials \cite{Sgier:2020das}. In the new release, we provide an alternative approach that works on the lightcone, enabling cross-probe compatibility and reduced computational requirements. Our pipeline closely follows the method presented in Ref.~\cite{Naidoo:2021ylw}.

The ISW effect leads to a change in the CMB photon temperature according to:

\begin{equation}\label{eq:isw-temperature-effect}
    \frac{\Delta T_{\text{ISW}} (\hat{n})}{T_{\text{CMB}}} = \frac{2}{c^3} \int_{0}^{r_*} dr \frac{d\Phi}{dt} (r\hat{n}, \eta(r)) \cdot a(r),
\end{equation}

with $r_*$ the comoving distance to the surface of last scattering. We assume no anisotropic stress and ignore non-linear effects such as the Rees-Sciama effect \cite{ReesSciama1968}, which have a negligible impact at the low redshifts and large scales analyzed in this work~\cite{Seljak:1995eu}. In the linear regime, the time derivative of the potential $\Phi$ can be expressed in terms of the Hubble parameter and the linear growth factor:

\begin{equation}\label{eq:isw-linear-approximation}
    \frac{\Delta T_{\text{ISW}} (\hat{n})}{T_{\text{CMB}}} = \frac{2}{c^3} \int_{0}^{r_*} dr \ H(a(r)) [f(a(r)) - 1] \cdot \Phi(r \cdot \hat{n}, \eta(r)) \cdot a(r),
\end{equation}

with $f(a) = \frac{d(\ln(D(a)))}{d(\ln(a))}$ the linear growth rate, $D(a)$ the linear growth factor, and $H(a)$ the Hubble parameter, all computed at scale factor $a$.

As a final step to enable computations on the lightcone, we rewrite Equation \ref{eq:isw-linear-approximation} in terms of matter contrast $\delta$ instead of the potential $\Phi$. Motivated by Poisson's equation, we apply a Spherical Fourier-Bessel (SFB) decomposition of $\Phi$ and $\delta$, since the SFB basis functions are the eigenfunctions of the Laplacian operator on a spherical manifold. The SFB decomposition has been successfully applied in previous studies of the ISW effect (see, e.g. \cite{Shapiro_2012}) and in forecasts of power spectrum measurements from upcoming galaxy surveys (see, e.g., \cite{Grasshorn_Gebhardt_2021} and references therein). We describe the key steps in the SFB decomposition applied in \ufalcon, and refer the reader to \cite{Shapiro_2012} (see their section 3.1) for more details.

The SFB decomposition of a real-valued field defined on a finite region of 3D space is given by:

\begin{equation}\label{eq:sfb_decomposition}
    \begin{split}
        &\zeta (r, \theta, \phi) = \sum_{\ell=0}^{\infty} \sum_{m=-\ell}^{\ell} \sum_{n=1}^{\infty} \zeta_{\ell m n} R_{\ell n} (r) Y_{\ell m} (\theta, \phi), \\
        &R_{\ell n} (r) = \frac{j_{\ell} (k_{\ell n} \cdot r)}{\sqrt{N_{\ell n}}},
    \end{split}
\end{equation}

with $k_{\ell n} = q_{\ell n} / r(z_{\text{max}})$, $z_{\text{max}}$ the highest redshift used in the analysis, $q_{\ell n}$ the $n^{\mathrm{th}}$ root of $j_{\ell} (x)$, and $N_{\ell n} = [r(z_{\text{max}})]^3 \cdot [j_{\ell + 1} (k_{\ell n} \cdot r(z_{\text{max}}))]^2 / 2$ a normalization factor. The radial basis functions $R_{ln}(r)$ used in equation \ref{eq:sfb_decomposition} hold for the decomposition of functions defined on a finite 3D space \cite{SFB_IEEE}. In this work, the finite volume is a sphere of radius $r(z_{\mathrm{max}})$, motivated by the finite size of the simulation boxes. We note that this finite function domain additionally imposes boundary conditions on the basis functions, which can be derived via Sturm-Liouville theory (see section 1 in \cite{SFB_IEEE}). In this work, we use the zero-value boundary conditions for the SFB radial basis functions, which correspond to the expressions for $k_{ln}$ and $N_{ln}$ used in equation \ref{eq:sfb_decomposition}.

We apply the SFB decomposition to Poisson's equation in the linear regime, to obtain a relationship between the coefficients $\Phi_{\ell m n}$ and $\delta_{\ell m n}$. Furthermore, we decompose the ISW temperature anisotropies from equation \ref{eq:isw-linear-approximation} into spherical harmonics, obtaining:

\begin{equation}\label{eq:alm_and_tisw}
    \frac{\Delta T_{ISW} (\theta, \phi)}{T_{CMB}} = \sum_{\ell = 1}^{\infty} \sum_{m = - \ell}^{\ell} a_{\ell m} Y_{\ell m} (\theta, \phi)
\end{equation}

Combining the results of these decompositions, we obtain the final result:

\begin{equation}\label{eq:isw-final}
    a^{ISW}_{\ell m} = \frac{3 H_0^2 \Omega_{m, 0}}{c^3} \sum_{n=1}^{n_{\text{max}}} \frac{\delta_{\ell m n}}{k^2_{\ell n}} \int_{0}^{r(z_{\text{max}})} \mathcal{D}(a(r)) H(a(r)) [1 - f(a(r))] R_{\ell n} (r) dr.
\end{equation}

We note that this SFB algorithm works within a range of physical scales (denoted by $k_{\text{min}}$ and $k_{\text{max}}$ in the pipeline implementation). These can be overwritten by the user, but are automatically set to reasonable default values by \ufalcon{} (related to the physical size of the simulation box and the non-linear physical scale at the relevant redshifts under analysis).

\section{Theory predictions}\label{sec:theory_model_emu} 
In this section, we describe the theoretical predictions in our pipeline, which form a key component of the eventual parameter estimation. We begin by describing how the different angular power spectra $C_\ell$s are modeled theoretically in $\S$\ref{subsec:theory_modelling}, then how these predictions are computed by a theory code in $\S$\ref{subsec:theory_code}, before finally describing how we build a neural network emulator to rapidly output the predictions for each probe in $\S$\ref{subsec:emulator}. 

\subsection{Modeling} \label{subsec:theory_modelling}

The theoretical modeling of two-point functions in this pipeline generally separates into two regimes: the LSS probes (WL, clustering, and their respective cross-correlations) and the CMB probes (primary temperature/polarization and CMB lensing).

To compute the LSS correlations we make use of the Limber approximation~\cite{Limber:1954zz, Simon:2006gm, Kaiser:1991qi, Kaiser:1996tp} which simplifies the computation of the $C_\ell$s. This is generally accurate for $\ell \gtrsim 20$~\cite{LoVerde:2008re}. In this approximation, the angular power spectrum can be written as:
\begin{equation}
    \begin{split}
        C_{\ell}^{XY} = \int \frac{d\chi}{\chi^2} P\left(k = \frac{\ell + \frac{1}{2}}{\chi}, z(\chi)\right) \left[W^{X}\left(\frac{\ell+ \frac{1}{2}}{\chi}, z(\chi)\right) W^{Y}\left(\frac{\ell + \frac{1}{2}}{\chi}, z(\chi)\right)\right],
    \end{split}
\end{equation}
where $P(k)$ is the matter power spectrum, $\chi$ is the comoving distance, $z(\chi)$ is the redshift and $W^{X/Y}$ are window functions specific to the probes in the pipeline ($X,Y \in \{\gamma, \delta_g, \kappa^{\text{CMB}}, \Delta T_{\text{ISW}}\}$). In the following subsection, we describe the window functions for the different LSS $C_\ell$s computed in the pipeline as well as any systematics nuisance parameters used in their modeling.

\subsubsection{Galaxy weak lensing}
The window function for the WL shear signal can be derived by considering the deflection of light rays due to the presence of a gravitational potential~\cite{Kilbinger:2014cea}. In the Born approximation, this takes the following form (see e.g. ~\cite{Tarsitano:2020ddh}):
\begin{equation}
    W^{\gamma}(\chi) = \frac{3 H_0^2 \Omega_m}{2c^2} \frac{\chi}{a(\chi)} \int^{\chi_{\text{lim}}}_{\chi} d \chi' n(\chi') \frac{\chi'-\chi}{\chi'}, \\
\end{equation}
where $n(\chi)$ represents the photometric redshift distribution of the source galaxies (normalized to unity) and $\chi$ is the comoving distance\footnote{Note in this pipeline we have assumed a flat Universe when computing comoving distances.}. 

We add an intrinsic alignment (IA) component to the pure WL cosmological signal. This is a correlation in the shape of galaxies, in addition to any correlation induced by gravitational lensing, due to local tidal forces which tend to align galaxies (see ~\cite{Kiessling:2015sma, Kirk:2015nma, Joachimi:2015mma} for reviews). The total observed signal is computed as the sum of a cosmological shear part $C_{\ell, GG}$, an IA piece $C_{\ell, II}$ and their cross-correlation $C_{\ell, GI}$. We use the non-linear alignment model (NLA) to model the IA contribution~\cite{Bridle:2007ft}, for which the auto- and cross-correlation contributions can be computed in the Limber framework with the following window function: 
\begin{equation}
        W^{IA}(\chi) = A_\mathrm{IA}(z(\chi))n(\chi),
\end{equation}
where $A_\mathrm{IA}$ is the NLA amplitude parameter which we assume to be redshift-independent in this work (see~\cite{KiDS:2020suj}). We also include a $\Delta_z$ parameter for each WL redshift bin which can shift the photometric redshift distribution to account for redshift miscalibration. 

\subsubsection{Galaxy clustering}
We model galaxy clustering using the scale-independent linear bias approach~\cite{Bardeen:1985tr, Kaiser:1984sw}. The window function for this model reads:
\begin{equation}
    W^{\delta_g}(\chi) = n_g(\chi)b(\chi), \\
\end{equation}
where $b(\chi)$ is the redshift dependent linear bias. We allow for one free bias parameter per redshift bin in our analysis. We do not include any parameters to account for redshift distribution systematics. These are assumed to be negligible given we model spectroscopic surveys for galaxy clustering in this pipeline. Furthermore, we do not include the effect of redshift-space distortions (RSDs).

\subsubsection{CMB lensing cross-correlations}
The CMB lensing probe has a window function that can be derived analogously to that of the WL case but with a single source redshift at $z=z_{\ast}$. The window function thus takes the following form (see e.g. ~\cite{HerschelATLAS:2014txv}):
\begin{equation}
    W^{\kappa^{CMB}}(\chi) = \frac{3 H_0^2 \Omega_m}{2c} \frac{\chi}{a} \frac{\chi^* - \chi}{\chi^*},
\end{equation}
where $\chi*$ represents the comoving distance to the last scattering surface. We note that the Limber approximation is used only in the prediction of the CMB-lensing cross-correlations in the pipeline whilst the auto-correlation is computed without making this approximation.

\subsubsection{ISW cross-correlations}
For the ISW cross-correlation $C_\ell$s, we make an analogy to the Limber approximation following~\cite{Nicola:2016eua}:
\begin{equation}
    C^{TX} = \frac{3 \Omega_m H_0^2}{c^2} \Bigl(\frac{1}{(\ell+1/2)^2}\Bigr) \int^{z^*}_0 \mathrm{d}z W^X(\chi(z)) \frac{d}{\mathrm{d}z}[D(z)(1+z)] D(z) \times P^{lin}_{\delta \delta}(k=\frac{\ell + 1/2}{\chi(z)},0),
\end{equation}
where $z^*$ is the redshift at recombination and $W^{X}$ is as before the window function of an LSS probe in the pipeline.

\subsubsection{CMB TTTEEE and lensing auto-correlation}
For the CMB probes ($C_\ell^{TT}, C_\ell^{TE}, C_\ell^{EE}$ and $C_\ell^{\kappa \kappa}$), we instead compute the angular power spectra using the \class\ code directly~\cite{Blas:2011rf}. This computation does not rely on the Limber approximation and hence we can safely include larger scales ($\ell < 20$) for these spectra in our analysis.

\subsection{Theory code}\label{subsec:theory_code}
There exist many available codes to compute the non-linear matter power spectrum required for the Limber integrals, and the computation of the CMB $C_\ell$s given a set of cosmological parameters, $\vec{\theta}$ (see e.g. \class~\cite{Blas:2011rf} and \texttt{PyCosmo}~\cite{Refregier:2017seh, Tarsitano:2020ddh, Moser:2021rej}). In our pipeline, we adopt a hybrid approach. We use \class\ as our primary engine to compute the matter power spectrum and the CMB $C_\ell$s. We use the native implementation of \texttt{HMcode-2016}~\cite{Mead:2016zqy} to compute the non-linear contribution to the matter power spectrum. This spectrum is then passed into \texttt{PyCosmo}, which handles the Limber integrals for the LSS probes in the pipeline~\cite{Tarsitano:2020ddh}. 

\subsection{Emulator}\label{subsec:emulator} 

Emulators are neural networks that approximate the output of computationally intensive functions. This is done by representing the mapping from input parameters to function output as a series of non-linear transfer function operations. These are increasingly used in cosmology due to the significant speed gains over traditional methods (see ~\cite{Petri:2015ura,Bolliet:2023sst,Gong:2023nzy, Fischbacher:2022gua, Piras:2023aub,SpurioMancini:2021ppk, Euclid:2018mlb, Angulo:2020vky}).
Another advantage is that emulators are by default fully differentiable enabling robust numerical predictions of $\mathrm{d}C_\ell/\mathrm{d}\Vec{\theta}$ using automatic differentiation (AD).

In this pipeline, we train emulators to learn the mapping between input cosmological parameters and the output spectra of the theory code described in $\S$\ref{subsec:theory_code}. There are 43 distinct spectra to emulate (including all possible auto- and cross- correlation $C_\ell$s across all redshift bins of the LSS surveys) which can be interpreted as distinct outputs per input parameter set. Instead of training a separate network for each spectrum, we managed to train just six models to cover the entire range of $C_\ell$s whereby some of the networks are made to output concatenated sets of multiple spectra. This number was found as a compromise between the implementation complexity of training many separate models and the accuracy requirement for each emulated spectrum (see $\S$\ref{sec:emu} for further details). Although each network in the pipeline is created separately, we used a common training data generation framework and identical network architecture for each.

\paragraph{Training data} To produce the training data we sample roughly half a million points in parameter space using a Latin Hypercube (LHC) sampling. We train over the parameters of the $\nu \Lambda CDM$ model and nuisance parameters associated with the target $C_\ell$ (see $\S$\ref{subsec:emu_dist} for details on the exact parameters and priors used for each network).
At each sampled point of this parameter space, we use our theory code to compute the relevant spectra (the target output of the network). Note we fix the redshift distributions to the respective distributions of the surveys we model for the galaxy weak lensing and clustering spectra (see description of surveys in $\S$\ref{sec:survey_data}). Instead of training the models directly on the raw $C_\ell$, we further compute $\log(C_\ell)$ where possible. This helps to reduce the dynamic range of the training data and vastly improves the emulator fidelity. In the case of $C_\ell^{TE}$, taking the logarithm is not possible as this spectrum contains negative values across the entire parameter space. In this case, we follow the approach adopted in Ref.~\cite{SpurioMancini:2021ppk}: we train on the 512 highest variance PCA modes of the TE training data which similarly helps to reduce the dynamic range and ameliorates the final model accuracy. For $C_\ell^{\kappa \kappa}$, we follow Ref.~\cite{SpurioMancini:2021ppk} in also training on 512 PCA components as we similarly find this approach gives slightly better performance compared to training on $\log(C_\ell^{\kappa \kappa})$.
For a subset of the LSS $C_\ell$s, there are also negative values in the training data (for example WL cross-correlation $C_\ell$s with large IA amplitudes). In these corner cases, we add an ``offset'' vector, $\Vec{b}$, to these spectra to enforce positivity before taking the logarithm. We choose the offset vector to be twice the absolute value of the minimum of the concatenated vector of $C_\ell$s across the emulator training set i.e. $\Vec{b}= 2\cdot|\min(C_\ell)|$. We note that we choose $\omega_{\text{cdm}}$ as opposed to $\Omega_m$ in our parameterization of the cosmological model as our initial attempt with the latter consistently gave insufficient accuracy for the $C_\ell^{TE}$ spectrum.\footnote{We hypothesize that this is due to changes in $\omega_{\text{cdm}}$ producing a smoother response in the TE spectrum compared to $\Omega_m$ which makes the mapping from parameters to spectrum easier for a network to learn. See discussion at \url{https://github.com/alessiospuriomancini/cosmopower/issues/17}.}

\paragraph{Architecture} For each emulator in this pipeline, we use the architecture outlined in Ref.~\cite{SpurioMancini:2021ppk}\footnote{Available at \url{https://github.com/alessiospuriomancini/cosmopower}}. This consists of a neural network, implemented in the \texttt{Tensorflow} framework~\cite{tensorflow2015-whitepaper}, with 4 hidden layers of 512 neurons each. The activation function for each layer with input $\Vec{x}$ is: 
\begin{equation} \label{eqt:activation}
    f(\vec{x}) = \left( \Vec{\gamma} + (1 + \exp(- \Vec{\beta} \cdot \Vec{x}))^{-1} \cdot (1-\Vec{\gamma}) \right) \cdot \Vec{x},
\end{equation}
where the free parameters $\Vec{\beta}$ and $\Vec{\gamma}$ are trained in addition to the weights in each layer. To train the models, we use the gradient descent optimizer \texttt{ADAM}~\cite{adamxyz} with default parameters. We use a batch size of 1024 and train for $\sim 200$ epochs which takes $\sim$ 3 hours on a Mac M1 machine. We check across 60000 test points that the median relative absolute deviation between the emulated and true spectra is less than 0.1\% for all of our models. For further details and emulator validation, the reader is referred to Appendix~\ref{sec:emu}.

\section{Survey modeling}\label{sec:survey_data}

\begin{figure}[htbp!]
\centering
\includegraphics[scale=0.16]{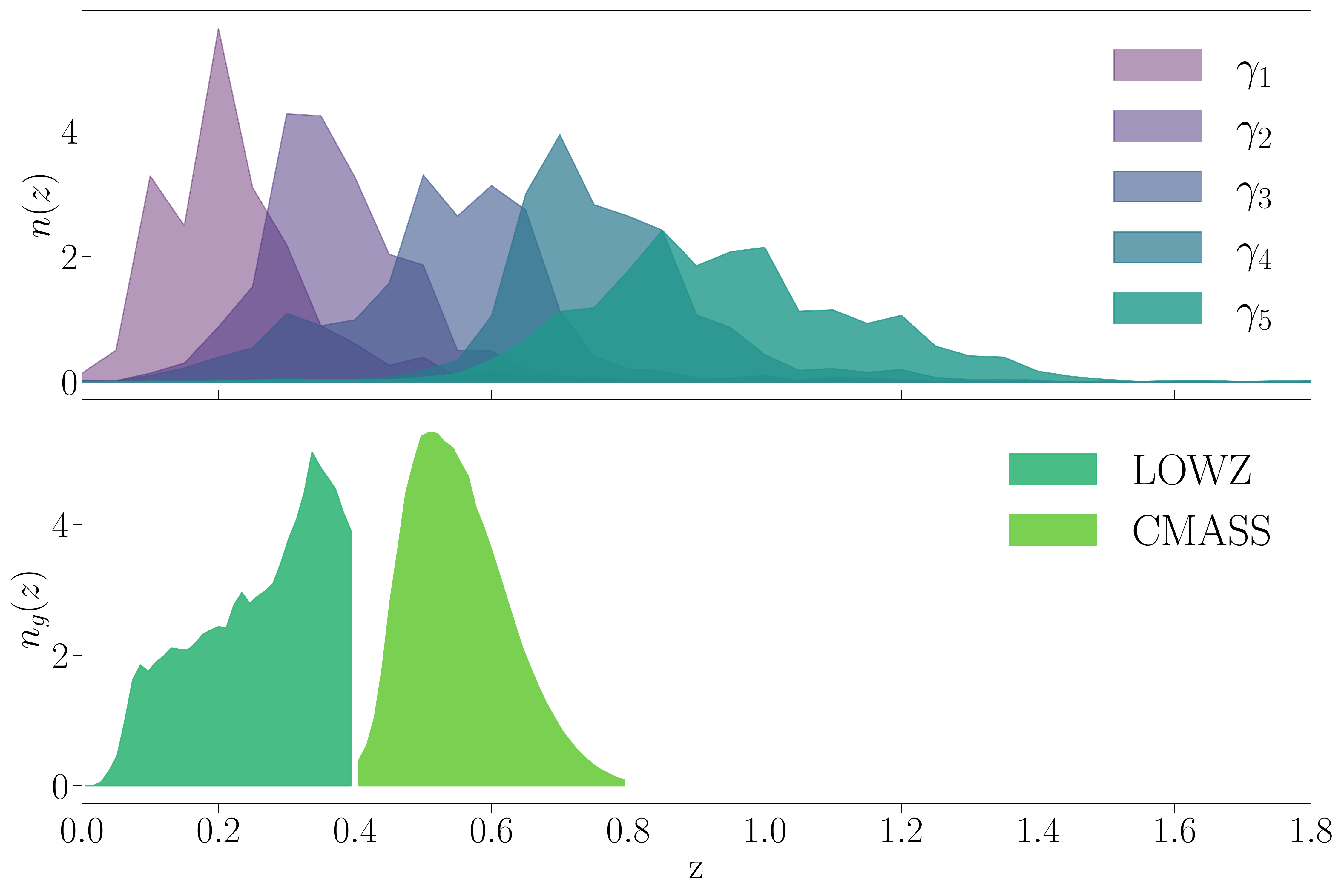}
\caption{The redshift distributions for the KiDS-1000~\cite{Heymans:2020gsg, Hildebrandt:2018yau, Hildebrandt:2020rno} and BOSS-DR12~\cite{BOSS:2016wmc} modeling. \label{fig:redshift}}
\end{figure}

To obtain forecasted constraints for mock data we now choose three specific surveys to model. Specifically, we include mock observations of KiDS-1000 WL data~\cite{Hildebrandt:2018yau,KiDS:2020suj}, BOSS galaxy clustering data~\cite{BOSS:2016wmc} and \textit{Planck} CMB data (both primary TTTEEE and reconstructed lensing)~\cite{Planck:2019evm, Planck:2018lbu}. 

We analyze the LSS probes and their cross-correlations using map-level mock observations. In the following, we explain in turn how we create noisy, masked mock observations for each of the LSS surveys. We then describe how we measure the auto- and cross- correlation power spectra and account for the survey mask. Finally, we describe the scale-cuts chosen for this analysis. 

The \textit{Planck} CMB correlations $C_\ell^{TT},C_\ell^{TE},C_\ell^{EE}$ and $C_\ell^{\kappa \kappa}$ are instead included in the framework at the likelihood level using official data products (e.g. covariance matrices) and a mock data vector drawn from the same underlying cosmology as the simulations. We, therefore, delay the description of these until we describe our combined likelihood in $\S$\ref{sec:combined_probes_like}.

\subsection{Map level modeling}\label{subsec:map_level_model}

\subsubsection{Coordinate system}
Given that the LSS data from BOSS DR12 and KiDS-1000 is supplied in celestial coordinates, whilst \textit{Planck} data is in galactic coordinates we must choose a common coordinate choice for each data set. We choose to place all data in galactic coordinates and achieve this by a pre-processing step using the \texttt{rotator} method from \texttt{healpy} to change the coordinates of the observed galaxy positions in the LSS surveys at the catalog level.

\subsubsection{KiDS-1000}
We create mock observations to mimic the fourth data release~\cite{Kuijken:2019gsa} of the Kilo-Degree Survey (KiDS). This is a public survey run by the European Southern Observatory (ESO)\footnote{\url{https://kids.strw.leidenuniv.nl/}}, which aims to measure the correlations in distortion of galaxy shapes due to weak gravitational lensing. The photometric redshift of each galaxy is measured using 9 bands including the optical \textit{ugri} as well as the near-infrared ZYJH$K_s$. The addition of the near-infrared bands is possible thanks to the KiDS partner survey VIKING (the VISTA Kilo-degree INfrared Galaxy survey~\cite{edgexyz}). The data in the fourth release cover $1006$ $\mathrm{deg}^2$ on the sky and contain $~\sim 2\times10^7$ individual galaxies across all of the photometric redshift bins for which shape measurements are made~\cite{Giblin:2020quj}. The data are compiled in the galaxy catalog which is a database containing the estimated photometric redshift, the observed ellipticity, and the position of each object in the survey. The fiducial cosmological parameter constraints from 2-point statistics for this data release are presented in~\cite{KiDS:2020suj}, and there have since been many studies including different (higher-order) statistics as well as cross-correlations with other surveys (see ~\cite{Fluri:2022rvb, Troster:2021gsz,Heymans:2020gsg}). For the present analysis, we focus on the two-point $C_\ell$ statistic and follow a similar methodology to~\cite{Troster:2021gsz} in analyzing the data using the pseudo-$C_\ell$ approach. In particular, we subdivide the survey galaxies into 5 distinct redshift bins and use the associated photometric redshift distribution functions~\cite{Hildebrandt:2020rno} ($\gamma_1$ to $\gamma_5$ in Fig.~\ref{fig:redshift}).

To produce map-level mock observations of the KiDS-1000 survey, we first use \ufalcon{} to produce full-sky cosmological signal maps for each photometric redshift bin. To these signal maps, we add a shape noise realization. To produce this we take the catalog of galaxies from the KiDS-1000 data release and apply a random rotation to each measured ellipticity by an angle drawn from the range $\phi \in [0,2\pi)$. This cancels the correlation between galaxy shapes due to lensing and leaves a random shape noise realization. We bin the resulting rotated galaxy ellipticities into pixels of an \texttt{nside}=1024 map to generate a noise realization for each shear component. Finally, we create a survey mask using the `lensing weights' derived from \texttt{lensfit} following Ref.~\cite{Troster:2021gsz}\footnote{\url{https://github.com/tilmantroester/KiDS-1000xtSZ}}. For this, we bin the \texttt{lensfit} weights into pixels of a \texttt{HEALPix} map to produce a mask that weights the shear pixel value according to the shape measurement uncertainty. 

\subsubsection{BOSS DR12} 
We include a mock data analysis of the BOSS $12^{th}$ and final public data release\footnote{\url{https://data.sdss.org/sas/dr12/boss/}} (hereafter BOSS DR12). BOSS is part of the Sloan Digital Sky Survey (SDSS) III project which took measurements using a 2.5-meter optical imaging telescope. This data release consists of two spectroscopic galaxy catalogs (LOWZ and CMASS). The galaxies within these samples are targeted by applying color-color and color-magnitude cuts to the SDSS photometric catalog~\cite{Reid:2015gra,SDSS-III:2015hof}.

We use two broad redshift bins of $0<z<0.4$ and $0.4<z<0.8$ corresponding to the LOWZ and CMASS cuts in the BOSS fiducial analysis (see Fig.~\ref{fig:redshift}). We produce galaxy over-density maps using \ufalcon{} for each redshift distribution. To create noisy mock realizations, we then add a random shot-noise realization produced by randomly shuffling the positions of the BOSS DR12 catalog galaxies within the survey mask following the procedure outlined in Ref.~\cite{Nicola:2016eua}. This breaks any spatial correlations between the galaxy positions and leaves a random shot-noise realization.

We compute the survey mask from the acceptance and veto masks provided by the BOSS collaboration. These are made available in the \texttt{MANGLE} format~\cite{Swanson:2007aj, Hamilton:2003ea}. The acceptance mask for a given region is defined as:
\begin{equation}
    C = \frac{N_{\text{observed}}}{N_{\text{targeted}}},
\end{equation}
i.e. the ratio of the number of observed galaxies with measured redshift to the number of targeted galaxies. The veto masks are instead binary maps that mask areas affected by observational factors~\cite{Reid:2015gra}. We follow the approach taken in Ref.~\cite{Loureiro:2018qva} in creating a binary mask from these: first, we convert the acceptance mask to a high resolution \texttt{nside}=16384 \texttt{HEALPix} map and subsequently apply the binary veto masks at this resolution. We then downgrade this combined map to the resolution used in our analysis (\texttt{nside}=1024). Finally, we produce a binary mask by assigning 1 to pixels where the resultant low-resolution map is greater than 0.75 and 0 otherwise following Ref.~\cite{Doux:2017tsv}.  

\subsubsection{\textit{Planck} CMB lensing cross-correlations \label{subsubsec:planck_model}} 
The \textit{Planck} project, a public survey from the European Space Agency, has provided unprecedented nearly full-sky observations of the CMB and its anisotropies~\cite{Planck:2018vyg}. In this pipeline, we create mock observations to simulate the third full data release (PR3)~\cite{Planck:2018nkj}. The gravitational lensing of the CMB, caused by the deflection of the CMB photons by large-scale structure at late-times, was detected by the \textit{Planck} team at over $40 \sigma$~\cite{Planck:2018lbu}. 

In this pipeline, the CMB primary $C_\ell^{TT}, C_\ell^{TE}$ and, $ C_\ell^{EE}$ as well as the lensing auto-correlation $C_\ell^{\kappa \kappa}$ are modeled at the likelihood level (see $\S$\ref{subsubsec:CMB_like}). However, for the cross-correlations of \textit{Planck} CMB lensing, we follow a procedure similar to the LSS probes whereby we create map-level mock observations. To achieve this, we use \ufalcon{} to produce a CMB lensing signal realization. To this, we add a CMB lensing noise realization. This is produced by subtracting the input cosmological signal as well as the mean field bias estimate from the \textit{Planck} Full Focal Plane (FFP10) CMB lensing simulations available from the Planck Legacy Archive\footnote{\url{https://pla.esac.esa.int/home}}. 

Finally, we apply the \textit{Planck} lensing survey mask. We downgrade this mask to our fiducial resolution of \texttt{nside}=1024 following the procedure outlined in \citep{Planck:2019evm} and further apodize the mask using a Gaussian smoothing kernel of FWHM $1.0$ deg. This apodization scale is chosen as we find there is significant power leakage from noise-dominated small-scale modes until this scale~\cite{Krolewski:2021yqy, Garcia-Garcia:2021unp}.

\subsubsection{\textit{Planck} ISW cross-correlations} 
We make mock observations of the ISW signal on the map-level utilizing the new routine implemented in \ufalcon{} (see $\S$\ref{subsubsec:isw}). To create a noisy ISW realization we add one \textit{Planck} temperature FFP10 end-to-end simulation to each of the maps produced by \ufalcon{}. The simulations contain both a CMB signal and noise component. Adding these to our ISW signal maps is hence an approximation since we are double-counting the ISW signal component. We validate this approximation for our use-case by checking that a covariance matrix built using these noisy ISW realizations has deviations in the diagonal elements of no more than 10\% compared to an analytically derived Gaussian ISW covariance.

We use the mask provided by the \textit{Planck} team again downgrading to our fiducial resolution of \texttt{nside}=1024 and apodizing using a Gaussian smoothing kernel of FWHM $1.0$ deg following Ref.~\cite{Krolewski:2021znk}. 

\subsection{Measuring spherical harmonic power spectra}\label{subsec:masking}
Once we have produced the masked, noisy map-level mock observations for each of the LSS probes in the pipeline, we can measure the associated auto- and cross- correlation power spectra. The spherical harmonic power spectrum or $C_\ell$ is defined as the ensemble average of the products of the spherical harmonic coefficients of two maps $\psi_{lm}$ and $\phi_{lm}$: 
\begin{equation} \label{eqn:cl_definition}
    \langle \psi_{\ell m} \phi_{\ell' m'} \rangle = \delta_{mm'} \delta_{\ell \ell'} C_\ell^{\psi \phi}.
\end{equation}
We can estimate the spherical harmonic power spectra from our maps by computing the spherical harmonic decomposition and using the following estimator: 
\begin{equation}
    C_\ell = \frac{1}{2\ell+1} \sum_m \psi_{\ell m} \phi_{\ell m}.
\end{equation}
However, we must account for the fact that the data is recorded over the partial sky when comparing the measured $C_\ell$s to theoretical predictions. We can generally relate the \textit{ensemble average} of the $C_\ell$s measured on the cut-sky to the full-sky predictions via a mode coupling matrix:
\begin{equation} \label{equation:psuedo_cl}
    \begin{split}
        <\Tilde{C}^{XY}_\ell> = \sum_{\ell '} M_{\ell \ell'} C_{\ell '},
    \end{split}
\end{equation}
where $<\Tilde{C}^{XY}_\ell>$ is the ensemble average of the cut-sky measurement, $M_{\ell \ell'}$ is the mode coupling matrix, and $C_{\ell}$ is the full-sky prediction. Producing an unbiased estimate of the full-sky $C_\ell$ (to compare with theory) then amounts to solving the inverse of equation~\ref{equation:psuedo_cl}. This is in general not possible given the implicit loss of information associated with measuring $C_\ell$s on the masked sky. To make progress, we use the MASTER algorithm~\cite{Hivon:2001jp} implemented in the code \texttt{NaMaster}~\cite{Alonso:2018jzx}. This enables an efficient solution to the inverse problem by assuming that the true angular power spectrum is piecewise constant over some bins. It is then possible to invert the associated \textit{binned} mode coupling matrix and produce an unbiased ``bandpower'' estimate of the full-sky $C_\ell$ for each bin. 

Since the theoretical Cls are not, in reality, constant over the $\ell$-range of a given bin, we must make a correction to these before comparing them with the bandpowers of the data. To achieve this we first couple the theoretical $C_\ell$s with the original mode-coupling matrix ($M_{\ell \ell'}$). Then we bin the resulting coupled $C_\ell$s into the same bandpowers as the corresponding data we wish to compare with and decouple the result with the \textit{binned} mode-coupling matrix~\cite{Alonso:2018jzx}. This process accounts for the (small) bias induced by the assumption of step-wise full-sky Cls. Note for the cross-correlations we use the intersection of the two associated survey masks when measuring the cross $C_\ell$.

\begin{figure*}[htbp!]
\centering
\includegraphics[scale=0.14]{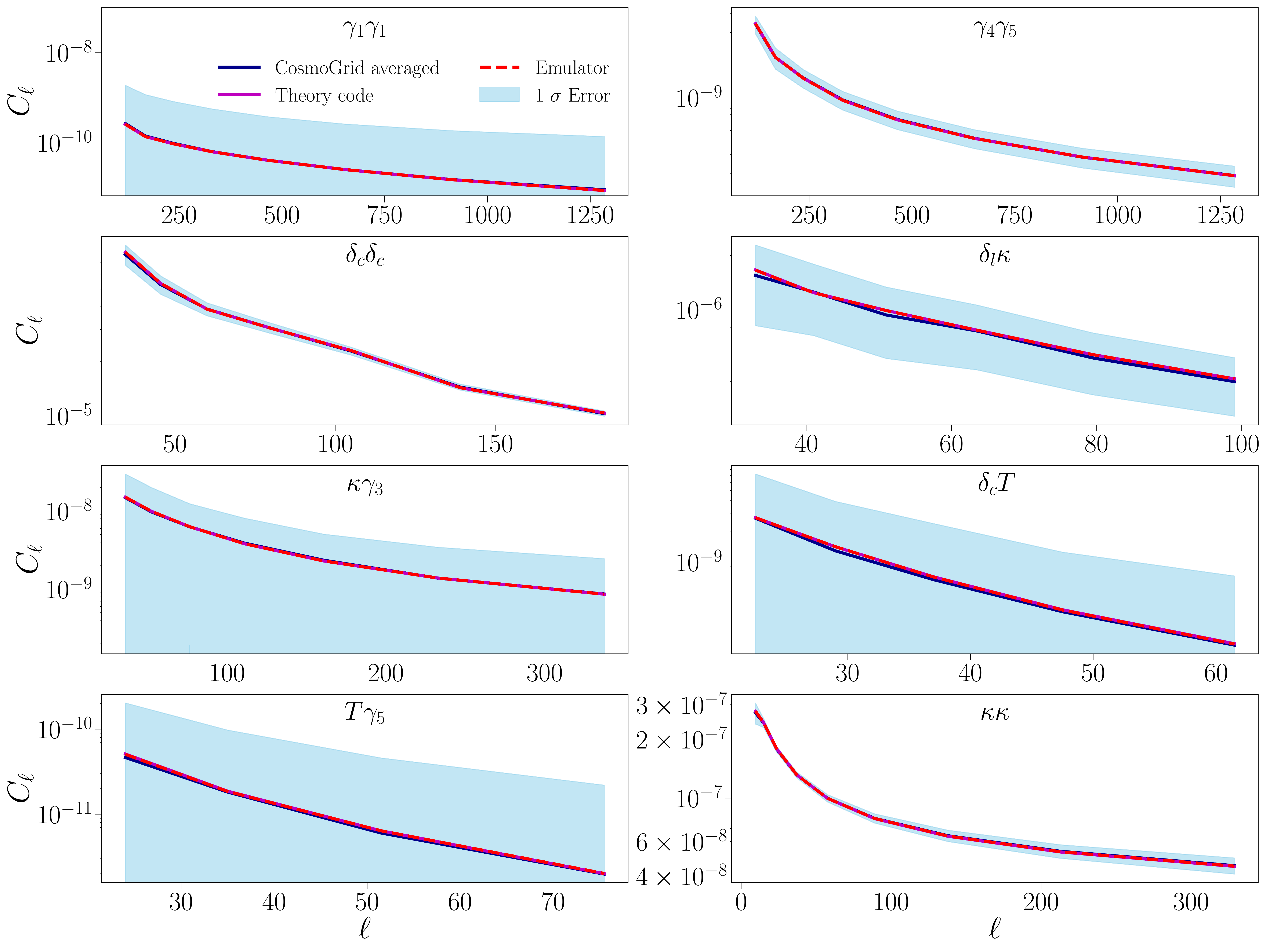}
\caption{A comparison of the mean of 200 \ufalcon{}-derived mock (noiseless) signal observations at the fiducial cosmology $C_\ell$s with theoretical predictions from the base theory code and the multiprobe emulator for a selection of spectra. Here $\delta_l$ and $\delta_c$ represent LOWZ and CMASS galaxy clustering respectively. The $C_\ell$s measured from the mock observations are mask-deconvolved and the theory predictions are treated to include the effects of mask deconvolution following the procedure outlined in~\ref{subsec:masking}. The $1-\sigma$ errorbars are determined from the square root of the diagonal of the multiprobe covariance matrix described in $\S$\ref{subsec:covariance_matrix}.}\label{fig:ufalc-theory-comp}
\end{figure*}

An important cross-check of the pipeline so far is made at this point in Fig.~\ref{fig:ufalc-theory-comp}. Here we compare the mean $C_\ell$ of 200 noiseless masked map-level mock observations against the theoretical predictions at the simulation fiducial cosmology from the theory code and the emulator including the effects of mask deconvolution. Whilst Fig.~\ref{fig:ufalc-theory-comp} shows only a selection of the 43 total spectra in the pipeline, we checked that for each spectrum the deviation between the average of 200 simulated spectra and the emulated spectrum is less than $0.1\sigma$ across all bandpowers. The $1-\sigma$ errorbar is determined from the square root of the diagonal of the multiprobe covariance matrix (described in $\S$\ref{subsec:covariance_matrix}). 

\subsection{Scale cuts}
Scale-cuts generally refer to the restriction of multipole range over which we perform cosmological analysis to the region where we trust the underlying theoretical model. For the CMB probes modeled at the likelihood level (see $\S$\ref{subsubsec:CMB_like}), we take the scale-cuts and binning provided by the official analyses. We instead choose the scale-cuts wherever we independently implement the likelihoods (for example for the cross-correlations of the LSS probes). In this section, we describe the motivation behind the scale-cuts in this pipeline for each auto- and cross- correlation $C_\ell$ (see Table~\ref{tab:scale_cuts} for an overview).

\paragraph{KiDS-1000} For the KiDS-1000 auto-correlations, we choose the same scale cuts ($100<\ell<1500$) as used in~\cite{Troster:2021gsz}. These scales represent the range over which the KiDS-1000 methodology is validated. Higher $\ell$ modes are discarded due to complications in modeling the non-linear matter power spectrum and baryonic processes on small scales.

\paragraph{BOSS DR12} We restrict the analysis of galaxy clustering (both auto- and cross- correlations) to large scales where the linear bias model implemented in the pipeline is accurate. Specifically, we choose $\ell_{\text{max}}$ such that the Fourier modes entering the Limber integral satisfy $k<0.1 \mathrm{Mpc}^{-1}$ following~\cite{Doux:2017tsv}. For the low-$\ell$ cut, we choose a conservative $l_{\text{min}}=30$ for the autocorrelations since, for lower $\ell$ values, RSDs become important~\cite{SDSS:2006egz} and the Limber approximation is no longer accurate~\cite{LoVerde:2008re}. 

\paragraph{KiDS-1000 cross-correlations with BOSS DR12} We generally follow the principle that the cross-correlation scale-cuts are given the two most conservative cuts applied to the respective auto-correlations (i.e., in this case, the high-$\ell$ cut imposed by the linear bias modeling of BOSS). Furthermore, we follow Ref.~\cite{Heymans:2020gsg} in only including cross-correlation bins where there is not a significant overlap between the source galaxies of KiDS-1000 and the galaxies in the BOSS DR12 catalog. This is because in this regime the IA terms become important and it is not clear that the NLA model works sufficiently well to produce unbiased constraints. In particular, we choose to only include cross-correlations for $C_\ell^{\gamma_3 \delta_{\text{lowz}}},C_\ell^{\gamma_4 \delta_{\text{lowz}}}, C_\ell^{\gamma_5 \delta_{\text{lowz}}}, C_\ell^{\gamma_4 \delta_{\text{cmass}}}$ and $ C_\ell^{\gamma_5 \delta_{\text{cmass}}}$ (see Fig.~\ref{fig:redshift} for the respective redshift distributions).

\paragraph{\textit{Planck} CMB lensing cross-correlations} For the \textit{Planck} CMB lensing cross-correlations, we consider multipoles in the range $30<\ell<400$. The low-$\ell$ cut is motivated by the use of the Limber approximation used for the theoretical predictions which can be inaccurate for lower $\ell$'s~\cite{LoVerde:2008re}. For high $\ell$'s, we use the conservative $\ell$ cut from the \textit{Planck} 2018 lensing analysis of $\ell_{\text{max}}=400$. This was chosen by the \textit{Planck} team due to potential uncertainties in the lensing bias terms becoming increasingly important at smaller scales.

\paragraph{\textit{Planck} ISW cross-correlations} Whilst for the other probes we choose a conservative $\ell_{\mathrm{min}}=30$ due to the Limber approximation, we choose a more aggressive scale-cut of $\ell_{\text{min}}=20$ for the ISW cross-correlations. This helps to boost the ISW signal-to-noise ratio (S/N) which peaks at large scales. We expect any bias due to the inaccuracy of the Limber approximation in this regime to be small, especially given the ISW cross-correlations are typically low S/N measurements compared to the other probes in the analysis. We do not include the $T\gamma_1, T\gamma_2$ cross-correlations in the analysis as we found making the simulation-based estimates of these cross-correlations unstable due to the very low S/N. We leave the possibility of using beyond-Limber techniques~\cite{Fang:2019xat,LSSTDarkEnergyScience:2022lno}, to push the analysis of the ISW to larger scales and thereby further increase the S/N, to future work.

\begin{table}[h]
\centering
\caption{The $\ell$-cuts and multipole binning for the auto- and cross- $C_\ell$s in this analysis. The CMB $C_\ell^{TT},C_\ell^{TE},C_\ell^{EE}$ and $C_\ell^{\kappa \kappa}$ are implemented at the likelihood level using \textit{Planck} official data products. We, therefore, inherit the scale-cuts and binning from these analyses~\cite{Planck:2019nip,Planck:2018lbu}. Note the number of bins for these probes is not shown as these use binning schemes that are in general more complex than the simple logarithmic binning used for the LSS probes.\label{tab:scale_cuts}}
\begin{tabular}
{p{0.2\linewidth}p{0.25\linewidth}p{0.25\linewidth}p{0.25\linewidth}}
\hline \hline

&Dataset & $\ell$-range & No. log-bins \\
\hline
&TT, EE low-$\ell$ & 2--30 & --- \\
&TTTEEE (high-$\ell$) & 30--2508 & --- \\
&$\delta_l \delta_l$ & 30--110 & 5 \\
Auto-correlations &$\delta_c \delta_c$ & 30--210 & 7 \\
&$\gamma \gamma$ & 100--1500 & 8 \\
&$\kappa \kappa$ & 8--400 & 9 \\
\hline
&$\delta_l \gamma$ & 30--110 & 6 \\
&$\delta_c \gamma$ & 30--210 & 6 \\
&$\kappa \gamma$ & 30--400 & 7 \\
&$\delta_l \kappa$ & 30--110 & 6 \\
Cross-correlations &$\delta_c \kappa$ & 30--210 & 6 \\
&T$\kappa$ & 20--120 & 6 \\
&T$\gamma$ & 20--90 & 4 \\
&T$\delta_l$ & 20--70 & 5 \\
&T$\delta_c$ & 20--70 & 5 \\
\hline \hline
\end{tabular}
\end{table}

\section{Inference} \label{sec:combined_probes_like} 

\subsection{Likelihood}
In this section, we describe how we evaluate the likelihood in the combined probes framework. We begin by describing the CMB probes that are included in the analysis at the 2-point level (i.e. not through the map-making procedure described in $\S$\ref{subsec:map_level_model}). We then detail how the mock data vector and covariance matrix are derived for each of the $C_\ell$s in the pipeline before combining to arrive at the likelihood.

\subsubsection{CMB probes}\label{subsubsec:CMB_like}
In this pipeline, $C_\ell^{TT}, C_\ell^{TE}, C_\ell^{EE}$ and $C_\ell^{\kappa \kappa}$ are modeled at the 2-point level. We follow this approach primarily because the \cosmogrid{} simulations go up to $z=3.5$ and hence cannot produce maps for the higher redshift probes\footnote{Note an alternative map-level approach would be to use the \texttt{synfast} routine from \texttt{healpy} to produce Gaussian random map-level realizations for these probes (see~\cite{Nicola:2016eua, Sgier:2021bzf}). We decided against using this approach as the scale-cuts required for this approach are more restrictive (meaning less constraining power) due to the presence of high-$\ell$ foregrounds in the component-separated data maps (see Ref.~\cite{Nicola:2016eua} for further details).}.

\paragraph{CMB primary} For the high-$\ell$ ($\ell>30$) part of the CMB primary TTTEEE likelihood, we follow the Python implementation of the \texttt{Plik\_lite} likelihood, \texttt{planck-lite-py}\footnote{\url{https://github.com/heatherprince/planck-lite-py}}, described in Ref.~\cite{Prince:2019hse}. This likelihood is already marginalized over the \textit{Planck} foreground parameters following the procedure first outlined in Ref.~\cite{Dunkley:2013vu}. The marginalization simplifies the likelihood significantly: this uses just one nuisance parameter, $A_{\text{planck}}$, which accounts for the uncertainty in the overall \textit{Planck} calibration. For the low-$\ell$ ($\ell<30$) TT and EE likelihood we use the log-normal binned likelihood, \texttt{planck-low-py} presented in~\cite{Prince:2021fdv}\footnote{\url{https://github.com/heatherprince/planck-low-py}}. Ref.~\cite{Prince:2021fdv} showed that $D^{TT/EE}_\ell = \ell (\ell+1)C^{TT/EE}_\ell/2 \pi$, once binned into 2 and 3 bins for TT and EE respectively, closely follows an offset-lognormal distribution. This means that one can write a Gaussian likelihood in $x=\ln(D_{\text{bin}})$ as: 
\begin{equation}
    \mathcal{L}(x) = \frac{1}{2 \pi \sigma (x-x_0)}e^{-(\ln(x-x_0)-\mu)^2/(2 \sigma^2)},
\end{equation}
where $x_0$ is the offset parameter and $\mu$ and $\sigma$ are the mean and width of the Gaussian distribution respectively. We take the values for these parameters found in Ref.~\cite{Prince:2021fdv} for the \textit{Planck} PR3 low-$\ell$ likelihood. We validate our implementation of the \texttt{Plik\_lite} likelihood and the low-$\ell$ lognormal bins by comparing to the official \textit{Planck} chains from the full likelihood and find excellent agreement (see $\S$ ~\ref{subsec:valid_planck_alone}). 

\paragraph{CMB lensing auto-correlations} Our CMB lensing analysis follows the usual quadratic-estimator approach~\cite{Lewis:2006fu, Hu:2001kj}. Generally, when one measures the lensing potential using a quadratic estimator, the estimated $C_\ell$s are contaminated with biases which must be subtracted to extract the unbiased $C_L^{\phi \phi}$. Specifically, one must remove:
\begin{itemize}
    \item The $N0$ bias. This is a 0th-order bias that accounts for the disconnected signal expected even in the absence of lensing. This can be generally assumed to be fixed with respect to varying cosmology and is typically estimated from simulations.
    \item The $N1$ bias. This enters at first-order and is the non-Gaussian secondary contraction of the trispectrum. This is generally cosmology-dependent. 
    \item Point source correction. This is due to the non-Gaussian signal produced by unresolved point sources in the data and can be assumed fixed and computed from the data.
\end{itemize}
In this analysis, since we use mock data, we generally do not have to consider the removal of terms that are fixed with respect to cosmological parameters ($N0$ and point source correction). However, since the computation of the cosmology-dependent N1 bias relies on sampling the CMB primary $C_\ell$s at each point in parameter space, this can impact the shape and width of the contours. We hence include the effect of its removal in our likelihood to give realistic forecast constraints~\cite{Planck:2015mym}. To do this we implement the ``CMB marginalized'' likelihood supplied by the \textit{Planck} team~\cite{Planck:2018lbu}. This uses a minimum variance (MV) reconstruction of the lensing potential based on measurements of both the temperature and polarization data. In this likelihood, the dependence on the CMB primary spectra is marginalized over to obtain a likelihood that depends on cosmological parameters only through the lensing potential power spectrum $C_L^{\phi \phi}$. This procedure boils down to adding a term to the original covariance matrix and applying a linear correction to the theoretical prediction of this quantity. This approach was introduced in the regime of ``lensing-only'' analyses (i.e. not including the CMB primary $C_\ell^{TT},C_\ell^{TE}, C_\ell^{EE}$), and validated only in this case~\cite{Planck:2018lbu}. We, therefore, checked that using this likelihood in the case where we include CMB primary spectra does not introduce any biases compared to the official \textit{Planck} TTTEEE+ lensing analysis, which is non-marginalized (see $\S$\ref{subsec:valid_planck_lensing} for further details).

\subsubsection{Mock data vector}
In this analysis, we use two different mock data vector cosmologies. The baseline case is the fiducial cosmology of the \cosmogrid, shown in Table~\ref{tab:fidcosmo}. For $\tau_{\text{reio}}$ and the nuisance parameters in the pipeline (which are not included in the simulation cosmology), we choose the baseline values shown in Table~\ref{tab:fidnuisance}. The second mock cosmology is termed the ``high-$M_\nu$'' case which has $M_\nu=0.15~\mathrm{eV}$ and therefore a raised neutrino energy density $\Omega_\nu$. The increase in $\Omega_\nu$ is compensated by a decrease in $\Omega_{\text{cdm}}$ such that the overall matter density $\Omega_m = \Omega_{\text{cdm}} + \Omega_\nu + \Omega_b$ is kept the same and all other parameters are identical to the baseline case. We use these two cosmologies to explore the neutrino mass constraints in two settings: (1) where the constraint looks like an upper limit (the baseline case) and (2) where there is a possibility of a Gaussian constraint (the \highmnu{}). We additionally vary the $A_L$ parameter for some of the mock observations away from the baseline value of $A_L=1$.

\begin{table}[h]
\centering
\begin{tabular}{cc}
    Parameter & Value \\
    \hline
    \hline
    $\tau_{\text{reio}}$ & 0.055 \\
    $A_\mathrm{IA}$ & 0.0 \\
    $\Delta_{z1}-\Delta_{z5}$ & 0.0 \\
    bias lowz & 1.837 \\
    bias cmass & 2.086 \\
    $A_{\text{planck}}$ & 1.0 \\
    \hline
\end{tabular}
\caption{Optical depth and nuisance parameter choices for the mock data vectors produced in this pipeline. Note that for the galaxy biases we choose the best-fit values corresponding to~\cite{Doux:2017tsv}. \label{tab:fidnuisance}}
\end{table} 

We follow the procedure outlined in $\S$\ref{sec:survey_data} to create mock observations for each of the LSS and associated cross-correlation $C_\ell$s in the pipeline. For the CMB primary and CMB lensing mock signals, we instead use a $C_\ell$ computed by our theory code described in $\S$\ref{subsec:theory_code}. The $C_\ell$ is then binned using the same procedure from the official analysis pipelines we have re-implemented. We compute our results using mock data vectors that do not additionally include a survey-specific noise draw, though survey noise is included in the covariance matrix (see $\S$\ref{subsec:covariance_matrix})\footnote{We checked that in the baseline case of $M_\nu=0.06\mathrm{eV}$, the average of 20 MCMC chains produced using noisy input mock data vectors produced nearly identical results to those presented in this work indicating there is little impact of noise bias.}.

\subsubsection{Simulation-based covariance matrix}\label{subsec:covariance_matrix} 

\begin{figure*}[htbp!]
\centering
\includegraphics[width=0.8\paperwidth]{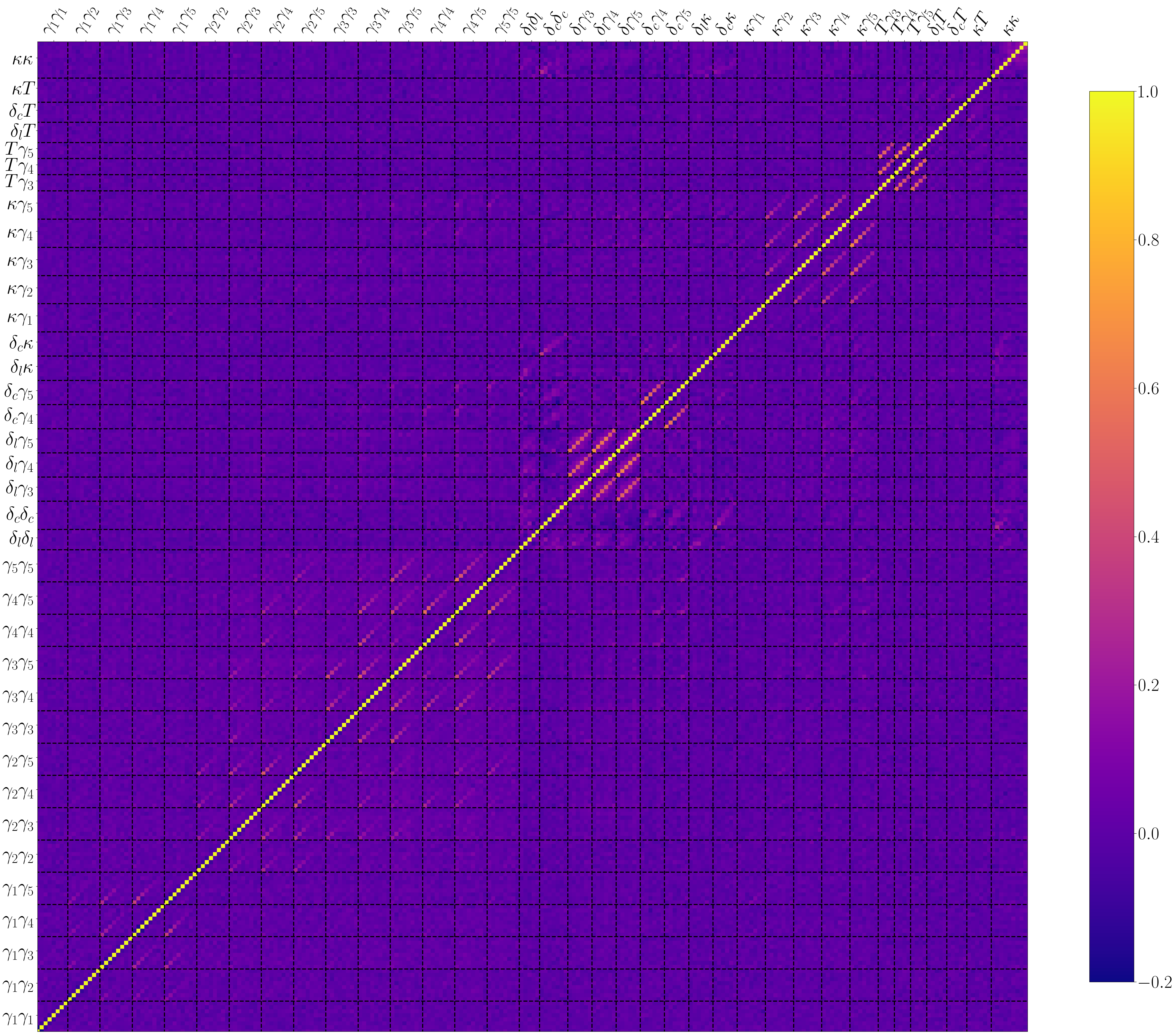}
\caption{The correlation matrix for the LSS probes (including CMB-$\kappa$ and the associated cross-correlations). This is derived from 2000 quasi-random mock realizations of fiducial cosmology. Each mock realization consists of a set of noisy mask-deconvolved $C_\ell$s which are derived from the same underlying dark matter simulation. This means the signals have the correct cross-correlation power. \label{fig:corr}}
\end{figure*}

We estimate the covariance of the LSS probes in the pipeline from a set of 2000 correlated mock realizations. These are generated by applying both a survey mask and a survey-specific noise realization to the \ufalcon{}-generated cosmological signal maps (as described in~\ref{sec:survey_data}). We add 10 random noise realizations for each of the 200 fiducial signal realizations to produce 2000 pseudo-random realizations. We use the \texttt{NaMaster} code to estimate the binned mask-deconvolved $C_\ell$s for each set of correlated mock observations from which we can estimate the LSS block of the covariance matrix. We add the covariance for $C_\ell^{\kappa \kappa}$ to the LSS covariance matrix as follows. We use the CMB lensing mocks to estimate the $C_\ell^{\kappa \kappa}-C_\ell^{\kappa X}$ and $C_\ell^{\kappa \kappa}-C_\ell^{XY}$-type (where $X$ and $Y$ stand for any of the other LSS probes) terms of the covariance matrix. We then insert the `CMB marginalized' official \textit{Planck} covariance matrix into the $C_\ell^{\kappa \kappa}-C_\ell^{\kappa \kappa}$ sub-block of the covariance matrix. The full LSS combined probes correlation matrix (including the $C_\ell^{\kappa \kappa}$ contributions) is shown in Fig~\ref{fig:corr}. 

We checked that the obtained covariance matrix is converged by analyzing the fractional change in the diagonal elements as a function of the number of realizations used to compute the covariance. We found that above $\sim 600$ realizations, the mean change is at the sub-percent level (see $\S$\ref{subsec:covmat_conv}). We checked that the diagonal power of our covariance matrix built from simulations matches within 10\% of the Gaussian analytic covariance matrix built using the \texttt{NaMaster} framework and further that using this Gaussian covariance as opposed to our simulation-based covariance produces only a small shift in the derived contours(see $\S$\ref{subsec:compare_to_gauss}).

\subsubsection{Combining the CMB and LSS likelihoods}\label{subsec:multiprobe_likelihood}

In this pipeline, we treat the CMB primary contribution to likelihood as independent of the LSS likelihood. This is equivalent to summing the individual contributions at the level of the log-likelihood ($\log(\mathcal{L})$):
\begin{equation}
    \begin{split}
        \log(\mathcal{L}_{\text{tot}}(\Vec{\theta})) = \log(\mathcal{L}_{\text{LSS}}(\Vec{\theta})) + \log(\mathcal{L}_{\text{CMB}}(\Vec{\theta})).
    \end{split}
\end{equation}

We, therefore, lose the cross-covariance between the CMB correlations (TTTEEE) and the other spectra in our pipeline (meaning we slightly underestimate the error bars in constraints). We expect this to have a negligible impact given that the amplitude of the cross-correlation with LSS tracers is negligible compared to the CMB primary $C_\ell^{TT}, C_\ell^{TE}$ and $C_\ell^{EE}$ (see e.g. ~\cite{Schmittfull:2013uea}, though this statement may need to be re-addressed for future surveys~\cite{Peloton:2016kbw}). For the CMB lensing auto-correlation ($C_\ell^{\kappa \kappa}$), where these cross-terms are important, we do include the cross-covariance with the other probes in the simulation-based covariance matrix (see $\S$\ref{subsec:covariance_matrix}).

For the LSS probes (including the ISW and CMB lensing cross-correlations) and $C_\ell^{\kappa \kappa}$, we compute the likelihood using our full covariance matrix and the mock data vector assuming the Gaussian likelihood approximation: 
\begin{equation}
    \ln(\mathcal{L(\theta)}) \propto (\vec{D}-\vec{f}(\theta))^T C^{-1} (\vec{D}-\vec{f}(\theta)),
\end{equation}
where $\vec{D}$ is the mock data vector, $\vec{f}(\theta)$ is the theory prediction (from the emulator) and $C^{-1}$ is the inverse covariance matrix. The naïve estimate of the inverse covariance matrix from a finite number of simulations is typically biased to be too large (i.e. we underestimate the size of confidence regions)~\cite{Hartlap:2006kj}. We hence de-bias our estimate by multiplying the naïve inverse by a factor: 
\begin{equation}\label{eqt:hartlap}
    C^{-1} \to \frac{n-p-2}{n-1} C^{-1}, 
\end{equation}
where $n=2000$ is the number of quasi-independent realizations used to estimate the covariance and $p$ is the number of data points (which is set-up dependent but takes a maximum value of $p=243$ when all possible LSS correlations are included).

\subsection{Data Compression} \label{subsec:data_compress}

The need for effective data compression in cosmological analysis has been consistently emphasized, especially in the context of multi-probe studies~\cite{Tegmark:1996bz, Heavens:2017efz}. For example, compression is important in the case that the covariance is estimated from simulations where for a data vector of size $n$, typically $\sim n+3$ simulations are required to get an accurate estimate of covariance~\cite{Heavens:2017efz}. Even in the case where the covariance is analytically estimated, reducing the dimensionality via data compression can help to improve the numerical stability of the inverse of the covariance matrix. Compression also allows for less memory usage and computational cost in the inference step of analysis due to the reduced size of the covariance matrix \& data vector involved. 

We explore two distinct compression methods: (1) MOPED and (2) PCA. The compression algorithms are implemented at the stage of the pipeline just before inference takes place. The covariance matrix and mock data vector are compressed using the methods described below and passed into the inference pipeline. The associated compression vectors are also supplied so that the theoretical predictions can be compressed at each stage of the inference.

The MOPED~\cite{Heavens:1999am} algorithm is a compression technique that reduces the dimensionality of a dataset ($N_{\text{data}}$) to the number of parameters being constrained. This works by finding a decomposition of the data vector onto a set of `MOPED basis vectors' ($b_i$) which are chosen to maximize the sensitivity of the likelihood to each of the parameters being constrained. The set of compression vectors is found by weighting the derivative of the given summary statistic (here, the $C_\ell$s) with respect to the input cosmological parameters by the inverse of the covariance matrix and subsequently using Gram-Schmidt orthonormalization:

\begin{equation}
\begin{split}
Y_i &= b_i^T X,
\\
b_i &= \frac{C^{-1} \frac{\partial \Psi}{\partial \theta_i} - \sum_{q=1}^{i-1} \left(\frac{\partial \Psi}{\partial \theta_i}^T b_q\right)b_q}{\sqrt{\left(\frac{\partial \Psi^T}{\partial \theta_i} C^{-1} \frac{\partial \Psi}{\partial \theta_i}- \sum_{q=1}^{i-1} \left(\frac{\partial \Psi}{\partial \theta_i}^T b_q\right)^2\right)}},
\end{split}
\end{equation}
where $X$ is the uncompressed data vector, $Y_i$ is the compressed data-vector and $C$ and $\Psi$ are respectively the covariance matrix and theoretical $C_\ell$ at the fiducial cosmology. Ref.~\cite{Heavens:1999am} showed that MOPED compression is optimal and lossless for Gaussian-distributed data if the input covariance matrix is parameter-independent (see also~\cite{Heavens:2020spq}). The main computational challenge involved in applying this technique is to find accurate derivatives of the theoretical prediction for the data at the fiducial cosmology. We compute the derivatives using the AD afforded to us by using an emulator-based approach for the theoretical predictions. As in~\cite{Piras:2023aub}, we found that this approach was consistently more robust than the more rudimentary `finite differences' method (see also~\cite{Campagne:2023ter} for a recent demonstration of using AD for MOPED compression).  

The second technique, PCA, allows us to find a new set of coordinates that can efficiently represent the data by choosing the modes giving rise to the highest variance across a dataset. In the present case, we perform a PCA compression on the emulator training set to find the vectors that capture the most variance as we span the parameter space within the priors (following an approach similar to~\cite{Zurcher:2022clh, Fischbacher:2022gua}). PCA is a useful comparison to MOPED as one is free to keep as many PCA components ($N_{\text{PCA}}$) as desired. This means that in the limit of $N_{PCA} \to N_{data}$ we will always recover the full information content of a dataset and hence this scheme is guaranteed to be lossless for some choice of $N_{\text{PCA}}$.

\subsection{Fiducial model} \label{subsec:fid_model}

The fiducial model in this study is standard $\Lambda$CDM. When exploring neutrino mass constraints we additionally allow the $M_\nu$ parameter to vary. In this case, we model neutrino masses using three degenerate mass states given this is sufficiently accurate for the precision of current data~\cite{Vagnozzi:2017ovm, Giusarma:2016phn, Archidiacono:2016lnv}. In some cases, we also include a free lensing amplitude rescaling $A_L$ parameter. This phenomenological parameter, first introduced in~\cite{Calabrese:2008rt}, rescales the amplitude of the lensing potential in the calculation of the lensed CMB primary anisotropies (i.e. the TTTEEE spectra):
\begin{equation}\label{eqn:a_l_definition}
    C_\ell^{\phi \phi} \to A_L C_\ell^{\phi \phi}. 
\end{equation}
Note the lensing potential is not rescaled when computing the CMB lensing reconstruction. This is used as an internal consistency check of CMB data where the expected value is $A_L=1$. Deviations from $A_L=1$ could indicate internal systematics in CMB data that can have consequences for parameter constraints as will be discussed in more detail in $\S$\ref{subsubsec:a_l_mnu-results}. 

In addition to the extension parameters $M_\nu$ and $A_L$, we also vary several nuisance parameters specific to each of the probes:

\paragraph{Weak lensing} For the WL modeling, we allow for a free $\Delta_z$ shift in the mean redshift for each of the bins as well as a free IA amplitude ($A_{\mathrm{IA}}$). Since in this analysis, we rely on DM-only simulations, we make a simplification compared to the fiducial KiDS-1000 analysis by not modeling the effect of baryonic feedback, which can affect the matter power spectrum at scales included in our forecast analysis. Furthermore, when considering cross-correlations between weak lensing shear and other tracers we ignore the cross-correlation with the intrinsic alignment component. In this forecast, the underlying data vector has $A_{IA}=0$ so these terms will not be present and we expect this to be a low-level signal in the KiDS-1000 data~\cite{Heymans:2020gsg}. Finally, the KiDS-1000 block of the covariance matrix is increased by an additive term $C_m$ to account for the uncertainty in the shear calibration as described in Ref.~\cite{Joachimi:2020abi}. We do not model the impact of anisotropic redshift distributions (for both WL and galaxy clustering) which typically lead to sub-percent biases for the surveys considered in this work, though these could be important when analyzing data from future surveys~\cite{BaleatoLizancos:2023zpl}.

\paragraph{Galaxy clustering} For galaxy clustering, we allow for one linear bias parameter for each bin. Using the linear bias approach limits our analysis (for both auto- and cross- $C_\ell$s) to large, linear scales where the model is accurate~\cite{Doux:2017tsv}. We do not include any nuisance parameters that account for foreground contamination\footnote{In real data analysis, this approach would need to be verified by checking the impact of residual foregrounds on the auto- and cross- correlation signals of the data maps. This could be achieved, for example, by checking the cross-correlation against known foreground maps~\cite{Doux:2017tsv,Nicola:2016qrc}.}.

\paragraph{CMB} For the CMB analysis, we include the $A_{\text{planck}}$ nuisance parameter wherever we encounter a CMB temperature/polarization auto- or cross- correlation $C_\ell$ (including ISW cross-correlations). This parameter is defined via an overall rescaling at the map level, and hence at the $C_\ell$-level this becomes: 
\begin{align}
    C_\ell^{TT/EE/TE} \to C_\ell^{TT/EE/TE}/A_{\text{planck}}^2, \\
    C_\ell^{TX} \to C_\ell^{TX}/A_{\text{planck}},
\end{align}
where $X$ represents an LSS probe in the ISW cross-correlation. We include this in the likelihood. We do not vary any nuisance parameters for the CMB lensing part of the likelihood.

\subsection{Inference pipeline}\label{subsec:inference}
\begin{table}[h]
\centering
\begin{tabular}{ccc}
    parameter & prior min & prior max \\
    \hline
    \hline 
    \multicolumn{3}{c}{cosmology} \\
    \hline
    $\omega_b$ & 0.018 & 0.026 \\
    $\Omega_{m}$ & 0.15 & 0.8 \\
    $n_s$ & 0.85 & 1.05\\
    $h$ & 0.6 & 0.9 \\
    $\sigma_8$ & 0.4 & 1.3 \\
    $\tau_{\text{reio}}$ & 0.02 & 0.12\\    
    \hline 
    \hline 
    \multicolumn{3}{c}{nuisance} \\
    \hline
    $A_\mathrm{IA}$ & -6 & 6 \\
    bias lowz & 0.1 & 5 \\
    bias cmass & 0.1 & 5 \\
    $\Delta_{z1}-\Delta_{z5}$ & \multicolumn{2}{c}{$\mathcal{N}(\mu_z, \textbf{C}_z)$} \\
    $A_{\text{planck}}$ & \multicolumn{2}{c}{$\mathcal{N}(1,0.025)$} \\
    \hline 
    \hline
    \multicolumn{3}{c}{extensions} \\
    \hline
    $M_\nu$ & 0.001 & 0.4 \\
    $A_L$ & 0 & 5\\
    \hline
\end{tabular}
\caption{The parameters varied in the analysis and the associated priors.\label{tab:vary_params}}
\end{table}
To perform cosmological inference in this pipeline we use the \texttt{emcee} package~\cite{Foreman-Mackey:2012any}. The \texttt{emcee} algorithm uses an ensemble of walkers in parallel and capitalizes on the affine-invariant ensemble sampling technique to efficiently explore the parameter space, providing accurate and computationally cost-effective parameter space exploration. Using \texttt{emceee} combined with our emulators enables running a full MCMC analysis in O($\sim$~minutes).

The priors for the cosmological/nuisance parameters in the analysis runs are shown in Table~\ref{tab:vary_params}. We use the same priors for all results shown in the main text of this paper. These are typically uninformative, flat uniform priors, with only two exceptions. First, we follow~\cite{KiDS:2020suj} in placing a correlated Gaussian prior on the $\Delta_z$ shift nuisance parameters for which we use the covariance matrix provided in the KiDS-1000 data release. This encodes the correlated errors from the photometric redshift distribution determination procedure~\cite{KiDS:2020suj}. Second, following~\cite{Planck:2019nip} we place a tight Gaussian prior on the re-scaling parameter $A_{\text{planck}}$.

\section{Results}
\label{sec:results}
\subsection{Baseline $\Lambda$CDM constraints}

\begin{figure*}[htbp!]
\centering
\includegraphics[scale=0.28]{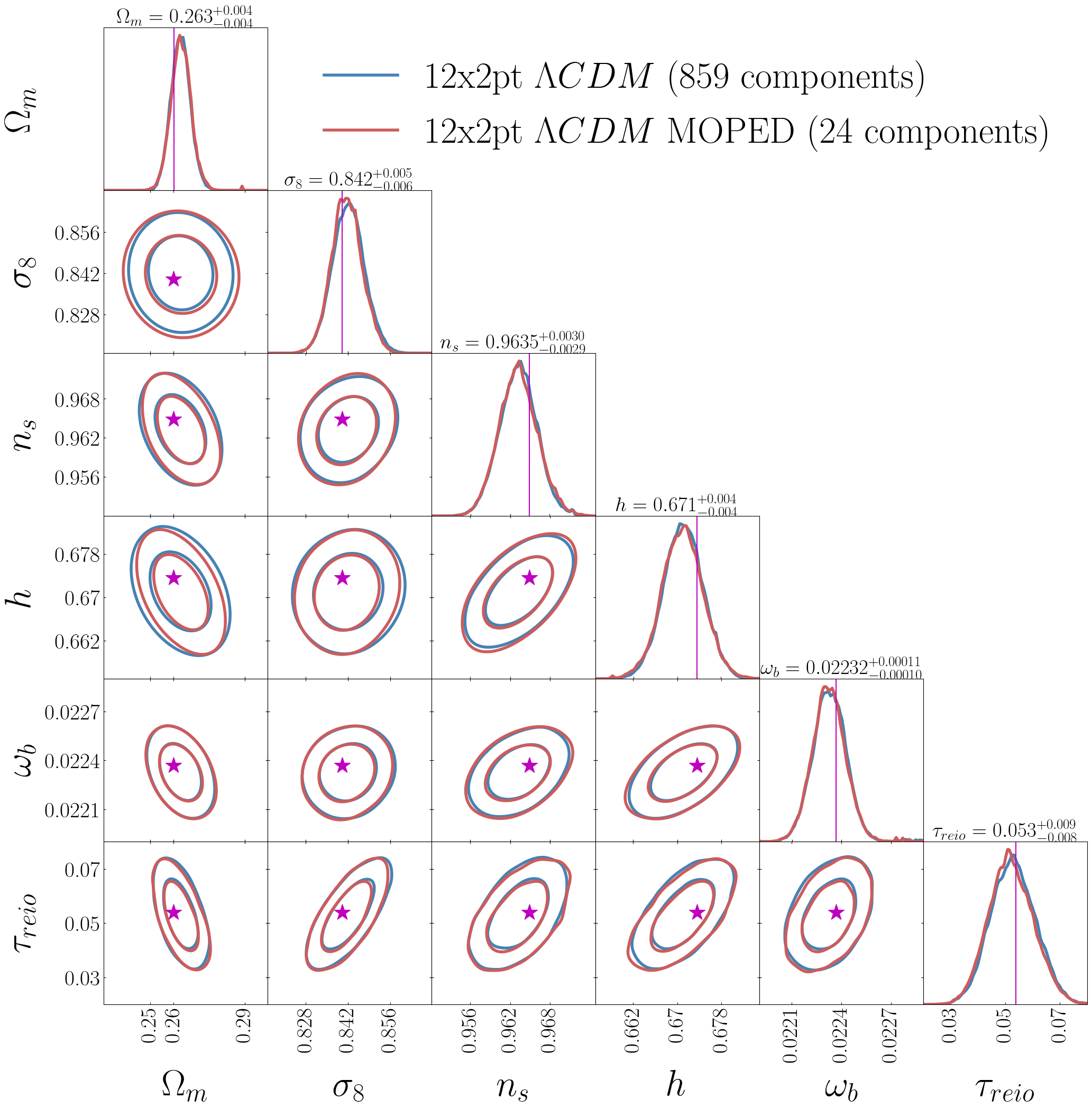}
\caption{The baseline $12 \times 2$pt $\Lambda$CDM results from the pipeline (blue) and comparison to MOPED compressed likelihood (red). The fiducial cosmology of the input mock data vector is shown in the purple stars and lines. The MOPED compressed data vector contains just 24 data points compared to the original 859 and preserves the information contained in the full data vector.\label{fig:full_moped}}
\end{figure*}

The baseline $\Lambda$CDM posterior contours are shown in Fig.~\ref{fig:full_moped}. These contours, shown in blue, represent the constraints from analyzing the full set of mock data at the baseline cosmology. This consists of a combination of CMB primary ($C_\ell^{TT},C_\ell^{TE},C_\ell^{EE}$), CMB lensing ($C_\ell^{\kappa \kappa}$), WL ($C_\ell^{\gamma \gamma}$) and galaxy clustering ($C_\ell^{\delta \delta}$) as well as the cross-correlations ($C_\ell^{T \kappa}, C_\ell^{T \gamma}, C_\ell^{T \delta}, C_\ell^{\delta \gamma}, C_\ell^{\kappa \gamma}, C_\ell^{\kappa \delta}$). Note we do not include any cross-correlations of CMB polarization with LSS tracers as these are too low significance to be detected in current data. We find that the pipeline produces unbiased contours in the $\Lambda$CDM setting: the fiducial model parameters (the purple stars) are within the $1\sigma$ marginalized contours. We also find this when combining smaller subsets of auto- and cross- correlations in the pipeline (e.g. a mock WL-only analysis). 

Whilst this full combination of data is somewhat optimistic given the current levels of tensions between datasets (see e.g.~\cite{Handley:2019wlz}), the constraining power of the $12 \times 2$pt combination is nevertheless interesting to compute. We find that this combination of data enables a determination of the Hubble constant at the 0.5\% level and $\Omega_m$ at the 1.5\% level improving the constraints on these parameters from the official \textit{Planck} CMB TTTEEE + CMB lensing analysis by $\sim 30\%$ in each case~\cite{Planck:2018vyg}. The increase in constraining power comes from the consistent addition of the LSS auto- and cross-correlation $C_\ell$s.

\subsection{Data compression for combined probes}

\begin{figure}
  \centering
  \begin{subfigure}[b]{0.47\textwidth}
    \includegraphics[width=\textwidth]{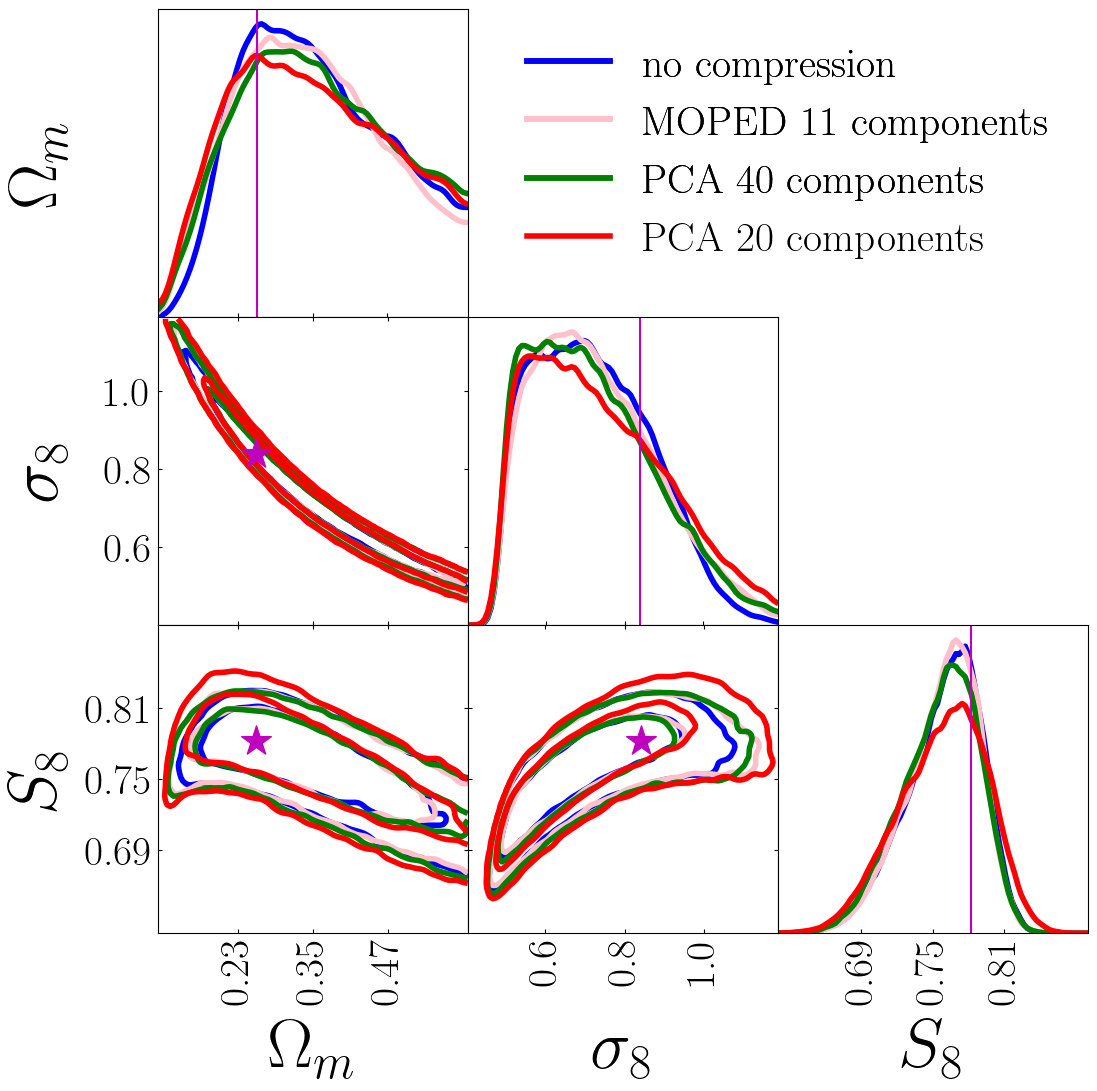}
    \caption{WL}
  \end{subfigure}
  \hfill
  \begin{subfigure}[b]{0.47\textwidth}
    \includegraphics[width=\textwidth]{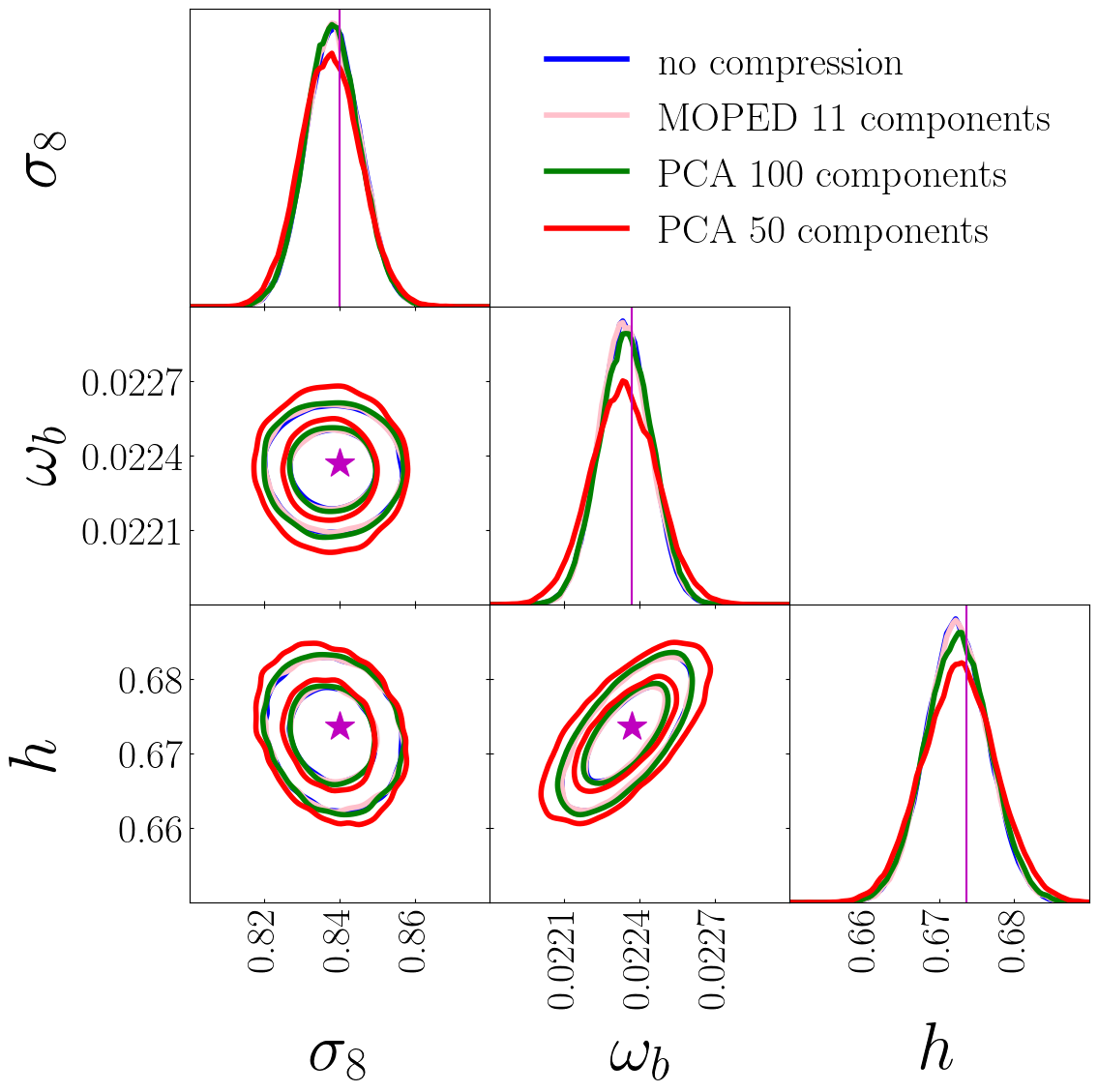}
    \caption{CMB primary TTTEEE}
  \end{subfigure}
  \caption{A comparison of the MOPED and PCA compression schemes: \textbf{left} KiDS-1000 WL, \textbf{right} \textit{Planck} CMB primary TTTEEE example.The fiducial cosmology of the input mock data vector is shown in the purple stars and lines. MOPED outperforms PCA in both cases in terms of reducing the size of the data vector whilst preserving the information in the likelihood and avoiding biases.}
  \label{fig:pca_moped_compare}
\end{figure}

In this subsection, we compare the PCA and MOPED compression algorithms. In Fig.~\ref{fig:pca_moped_compare} we show the comparison of the two techniques for a mock KiDS-1000 analysis (left) and a mock \textit{Planck} analysis (right). The uncompressed contours are shown in blue and represent the contours derived from the full data vector including all of the extractable information from the 2-point statistics. In red and green are the contours for the same setting but using PCA compression with different numbers of components. We find that $\sim 40$ and $\sim 100$ components are required respectively to faithfully reproduce the uncompressed contours (green contours in Fig.~\ref{fig:pca_moped_compare}) whilst fewer components lead to deviations (red contours in Fig.~\ref{fig:pca_moped_compare}). In contrast, the MOPED compression algorithm condenses the information to a data vector of size just 11 in both cases whilst faithfully reproducing the original contours. The compression to 11 data points with MOPED can be explained as follows. For the KiDS-1000 WL case, we have 5 cosmological ($\Lambda$CDM parameters but with fixed $\tau_{\text{reio}}$ as this is not constrained by LSS data) and 6 nuisance parameters ($A_\mathrm{IA}$ and 5 $\Delta z$ shifts). For the \textit{Planck} CMB primary case we compress the high-$\ell$ likelihood to just 6 data points (one for each of the 6 cosmological parameters of $\Lambda$CDM). We then add the 5 log-normal bins for the low-$\ell$ T/E likelihood giving overall 11 data points. 

In Fig.~\ref{fig:full_moped} we apply the MOPED compression scheme to the full combined probes set-up. The uncompressed baseline constraints use 859 data points for all of the CMB and LSS data auto and cross-correlations. In principle, both of the CMB primary ($C_\ell^{TT}, C_\ell^{TE}, C_\ell^{EE}$) part and the LSS part of the likelihood could be compressed together to produce a data vector equal the total count of variable parameters (plus 5 lognormal bins for the compressed low-$\ell$ T/E likelihood). We, however, choose to apply a separate compression of these components. This is necessary as we found that the combined covariance matrix of LSS and CMB is too large to be inverted without numerical instability which induces inaccuracies in the computation of the MOPED compression vectors and ultimately results in biased contours. 

Using the separate MOPED compression scheme described above, we demonstrate that the data vector can be drastically reduced from 859 to just 24 data points whilst preserving the information contained in the full combination of CMB + LSS. This consists of 13 data points for the LSS block where we vary 5 cosmological parameters and 8 nuisance, and 11 for the CMB block where we vary 6 cosmological parameters and add 5 log-normal bins for the compressed low-$\ell$ likelihood. We find that whilst there is a slight loss in constraining power in some of the 2D parameter projections (most notably in the $\Omega_m$-$\sigma_8$ plane), the 1D marginalized parameter constraints in the MOPED compressed scenario show excellent agreement with the uncompressed case with the mean and standard deviations of each parameter constraint agreeing to within 1\%. The compressed likelihood additionally enables a modest speed-up of 10-20\% for the 12x2pt analysis compared to the uncompressed case.

In addition to the results on mock data, we checked that the MOPED algorithm is also able to reproduce the full \textit{Planck} high-$\ell$ TTTEEE + low-$\ell$ T/E + lensing results with just 16 data points. Here we again perform a separate compression and have 11 data points for the CMB primary likelihood and 5 for the CMB lensing contribution. using the real survey data (see~\ref{subsec:valid_planck_lensing} for further details). We release, for the first time, this MOPED compressed \textit{Planck} PR3 CMB primary + CMB lensing likelihood publicly\footnote{\url{https://github.com/alexreevesy/planck_compressed}.}.

\subsection{Neutrino masses} 
\subsubsection{Constraining power}

\begin{figure}[htb]
    \centering
    \includegraphics{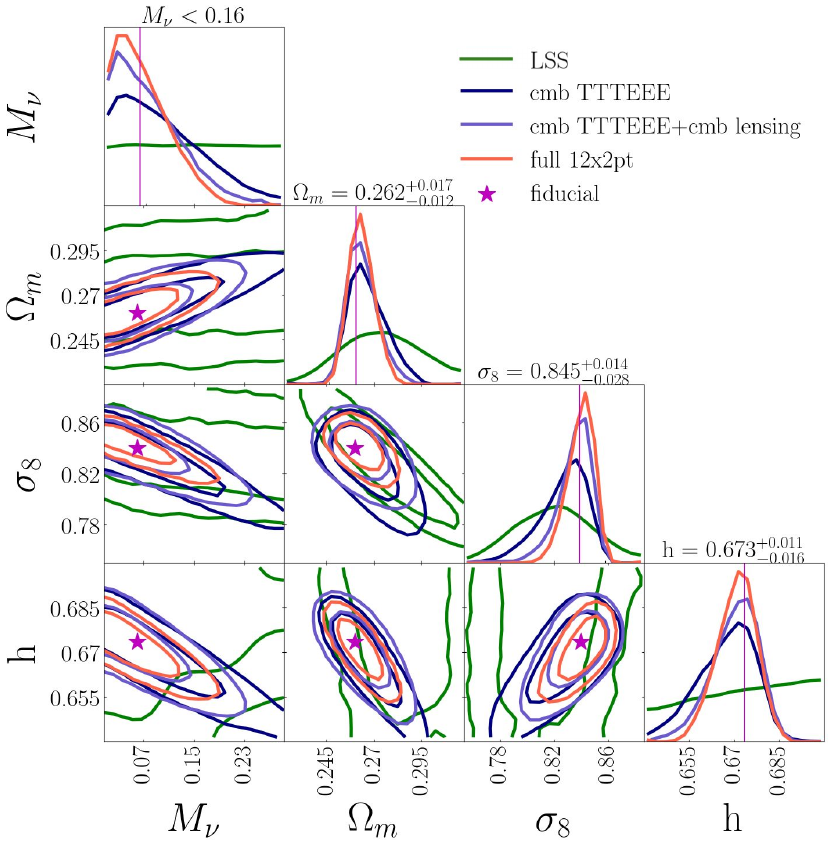}
    \caption{The constraining power of different combinations of data at the baseline cosmology ($M_\nu = 0.06 \mathrm{eV}$). The fiducial cosmology of the input mock data vector is shown in the purple stars and lines. The ``LSS'' line in green corresponds to the constraint from WL, galaxy clustering, CMB lensing, and all associated cross-correlations (including ISW). The parameter constraint values shown above the contour plots are obtained from the full $12 \times 2$pt combination (in dark orange).\label{fig:neutrino_mass_constraints_fid}}
\end{figure}

\begin{figure}[h]
    \centering
    \includegraphics{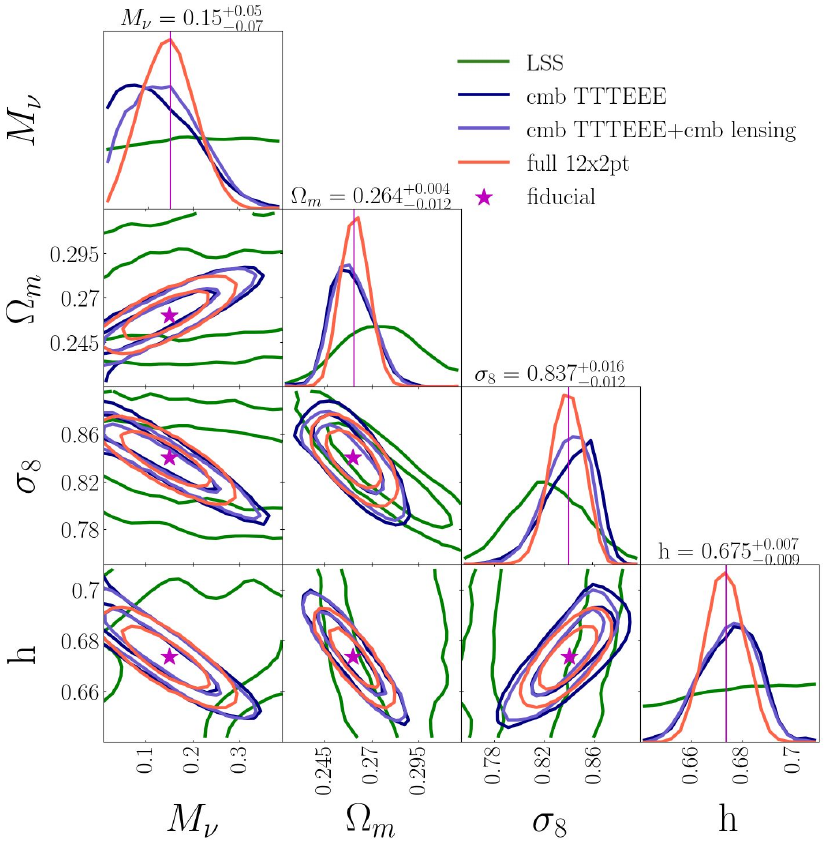}
    \caption{Same as Fig.~\ref{fig:neutrino_mass_constraints_fid} but with the high neutrino mass ($M_\nu=0.15\mathrm{eV}$) mock datavector.\label{fig:neutrino_mass_constraints_highmnu}}
\end{figure}

In this section, we present the constraining power on $M_\nu$ for various combinations of datasets using a mock data vector in the baseline and \highmnu{}s in Fig.~\ref{fig:neutrino_mass_constraints_fid} \& Fig.~\ref{fig:neutrino_mass_constraints_highmnu}. A change in the neutrino energy density generally affects both the background evolution of the scale factor and the evolution of the metric potentials at linear-order in perturbation theory. We can therefore measure the neutrino mass by finding the signatures of these effects on the different cosmological probes. One problem commonly encountered with single-probe analyses of neutrino masses is degeneracies in the way that $M_\nu$ and the other parameters in the $\Lambda$CDM model affect the measured spectra. For example, in the CMB alone case, there are clear degeneracies between $M_\nu$ and $\sigma_8$, $h$ and $\Omega_m$ (dark blue contours in Fig.~\ref{fig:neutrino_mass_constraints_fid} \& Fig.~\ref{fig:neutrino_mass_constraints_highmnu}). On the LSS side, there is a degeneracy in CMB data between $M_\nu$ and $\tau$ (indirectly through the constraint on the amplitude of the initial power spectrum $A_s$) related to the CMB lensing amplitude, as well as a degeneracy with galaxy bias~\cite{Boyle:2020rxq}. Whilst the full combination of LSS data (WL, galaxy clustering, CMB lensing, and associated cross-correlations including ISW) is not able to directly constrain $M_\nu$ (see green contours in Fig.~\ref{fig:neutrino_mass_constraints_fid} \& Fig.\ref{fig:neutrino_mass_constraints_highmnu}), the addition of this data enables breaking degeneracies in the constraints from the CMB alone (see also~\cite{Vagnozzi:2017ovm, Tanseri:2022zfe, Doux:2017tsv,DiValentino:2021hoh}). In particular, the addition of LSS data provides a CMB-independent measure of both $\sigma_8$ and $\Omega_m$ and hence enhances the $M_\nu$ constraint by breaking the degeneracy with these parameters. In the baseline cosmology with $M_\nu=0.06 \mathrm{eV}$, we are able to place an upper limit $M_\nu < 0.16\mathrm{eV}$ at the 95\% CL when combining CMB+LSS compared to $M_\nu < 0.25\mathrm{eV}$ using CMB TTTEEE data alone representing an improvement of $\sim 35\%$. The combined limit is slightly less constraining than the \textit{Planck}+BAO result of $M_\nu < 0.12\mathrm{eV}$ reported in~\cite{Planck:2018vyg}. We hypothesize this is because in our case the LSS data helps to exclude the very low $M_\nu < 0.06\mathrm{eV}$ region (see the left-hand side of the dark orange 1D-marginalized $M_\nu$ constraint in Fig.~\ref{fig:neutrino_mass_constraints_fid}). This biases the 95\% CL constraint to larger values meaning it is not a fair measure of the constraining ability of this data combination with respect to $M_\nu$. Another reason may be due to the internal $A_L$ tension in \textit{Planck} PR3 data biasing this constraint towards small $M_\nu$~\cite{DiValentino:2021imh}. In the \highmnu{} we are able to detect a neutrino mass at the $\sim 3\sigma$-level ($M_\nu=0.15 \pm 0.06 \mathrm{eV}$). This is similar to the constraining power found in Ref.~\cite{Chudaykin:2022rnl} using a combination of CMB data from \textit{Planck} and SPT-3G, LSS and supernova data. 

\subsubsection{Impact of $A_L$}\label{subsubsec:a_l_mnu-results}

Recently, several studies using data independent from \textit{Planck} have been able to place weak limits on the neutrino mass at regions of the parameter space ($M_\nu \sim 0.2 \mathrm{eV}$) that are at face-value strongly ruled out by the \textit{Planck}-derived upper limits (see for example ~\cite{DiValentino:2021imh, Chudaykin:2022rnl}). One postulated reason for this discrepancy is due to the internal lensing tension in the \textit{Planck} 2018 dataset which may bias the neutrino mass constraints found using this data towards $M_\nu=0\mathrm{eV}$. As described in $\S$\ref{subsec:fid_model}, the internal lensing consistency of a CMB dataset can be parameterized via $A_L$. The expected value of this parameter is $A_L = 1$ in the standard $\Lambda$CDM model. However, the \textit{Planck} collaboration PR3 analysis finds $A_L$ = 1.18 ± 0.065 from primary anisotropy CMB data (detection of $A_L$ > 1 at 99\% CL). The effect of lensing on the CMB is anti-correlated with $M_\nu$ and therefore such an anomaly can bias neutrino mass constraints towards lower values (see discussion in e.g.~\cite{DiValentino:2021imh,ACT:2023kun}). It remains unclear whether this is due to unaccounted physics (see for example ~\cite{Domenech:2019cyh}) or due to a systematic error in the \textit{Planck} data. It should also be noted that a recent re-analyses with the \texttt{NPIPE} pipeline presented in Ref.~\cite{Rosenberg:2022sdy} and Ref.~\cite{Tristram:2023haj} find $A_L=1.095\pm0.056$ and $A_L=1.039\pm0.052$ respectively representing significantly milder tensions with the expected result in $\Lambda CDM$. In any case, it is interesting to consider the impact of $A_L$ on neutrino mass constraints in the context of this novel combined probes pipeline.

\begin{figure}
  \centering
  \begin{subfigure}[b]{0.47\textwidth}
    \centering
    \includegraphics[width=\textwidth]{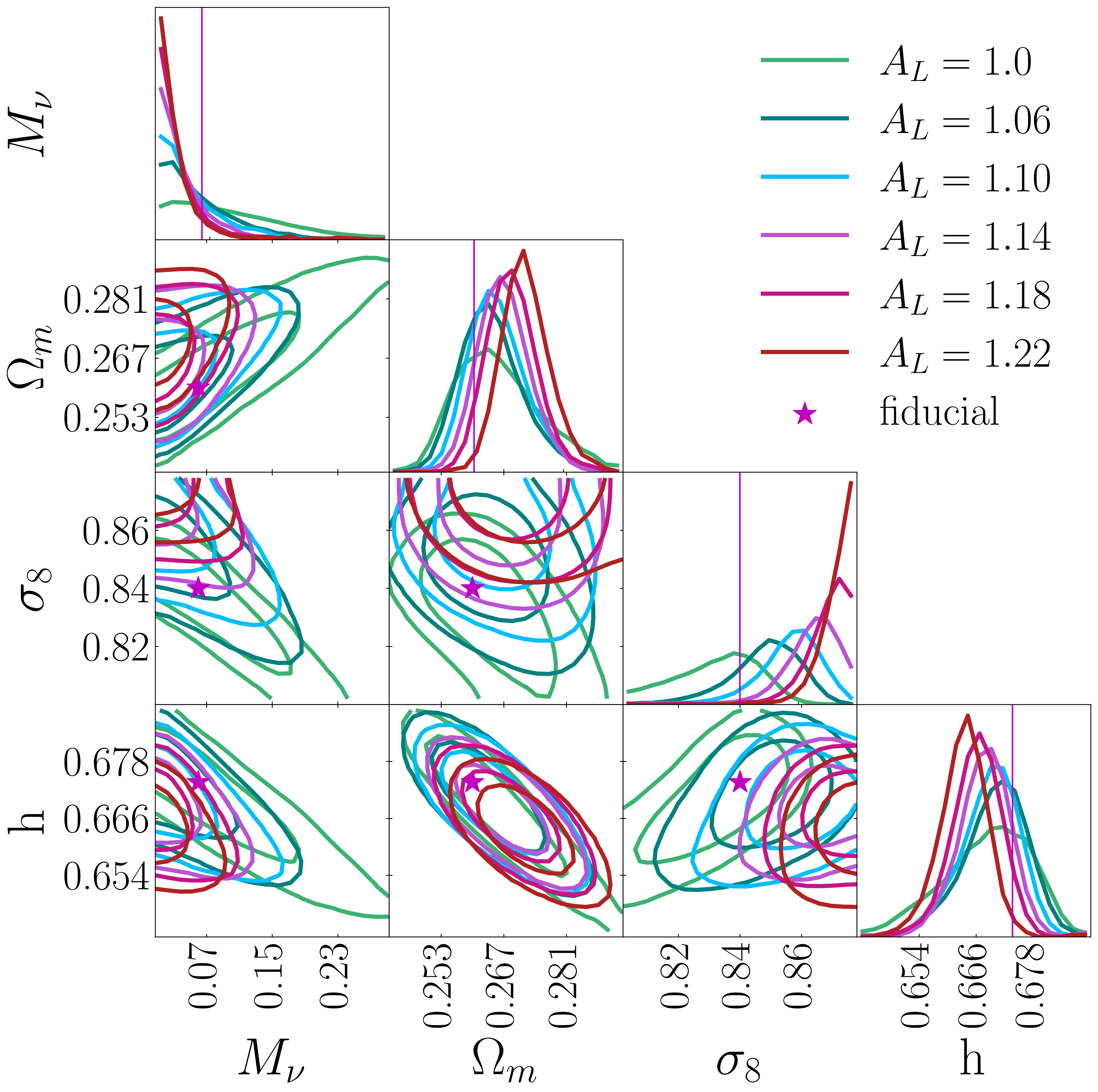}
    \caption{CMB TTTEEE}
  \end{subfigure}
  \hfill
  \begin{subfigure}[b]{0.47\textwidth}
    \centering
    \includegraphics[width=\textwidth]{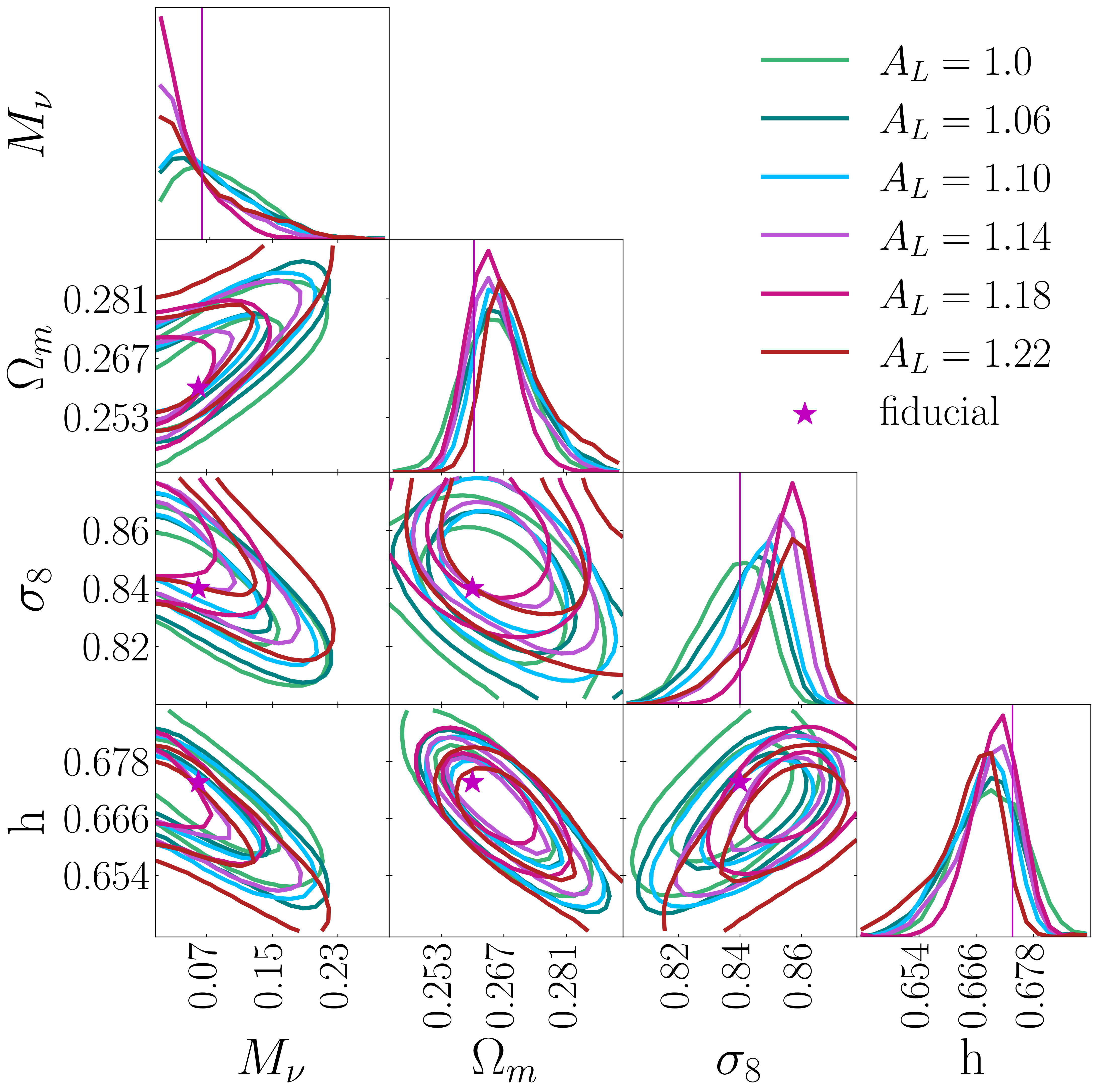}
    \caption{Full combined probes}
  \end{subfigure}

 \caption{Exploration of the impact of an $A_L$-like systematic for a data vector at the fiducial cosmology ($M_\nu=0.06\mathrm{eV}$): \textbf{left} CMB alone \textbf{right} full $12 \times 2$pt combination.\label{fig:alens_explore_fid}}
\end{figure}

\begin{figure}[!htb]
  \centering

  \begin{subfigure}[b]{0.47\textwidth}
    \includegraphics[width=\textwidth]{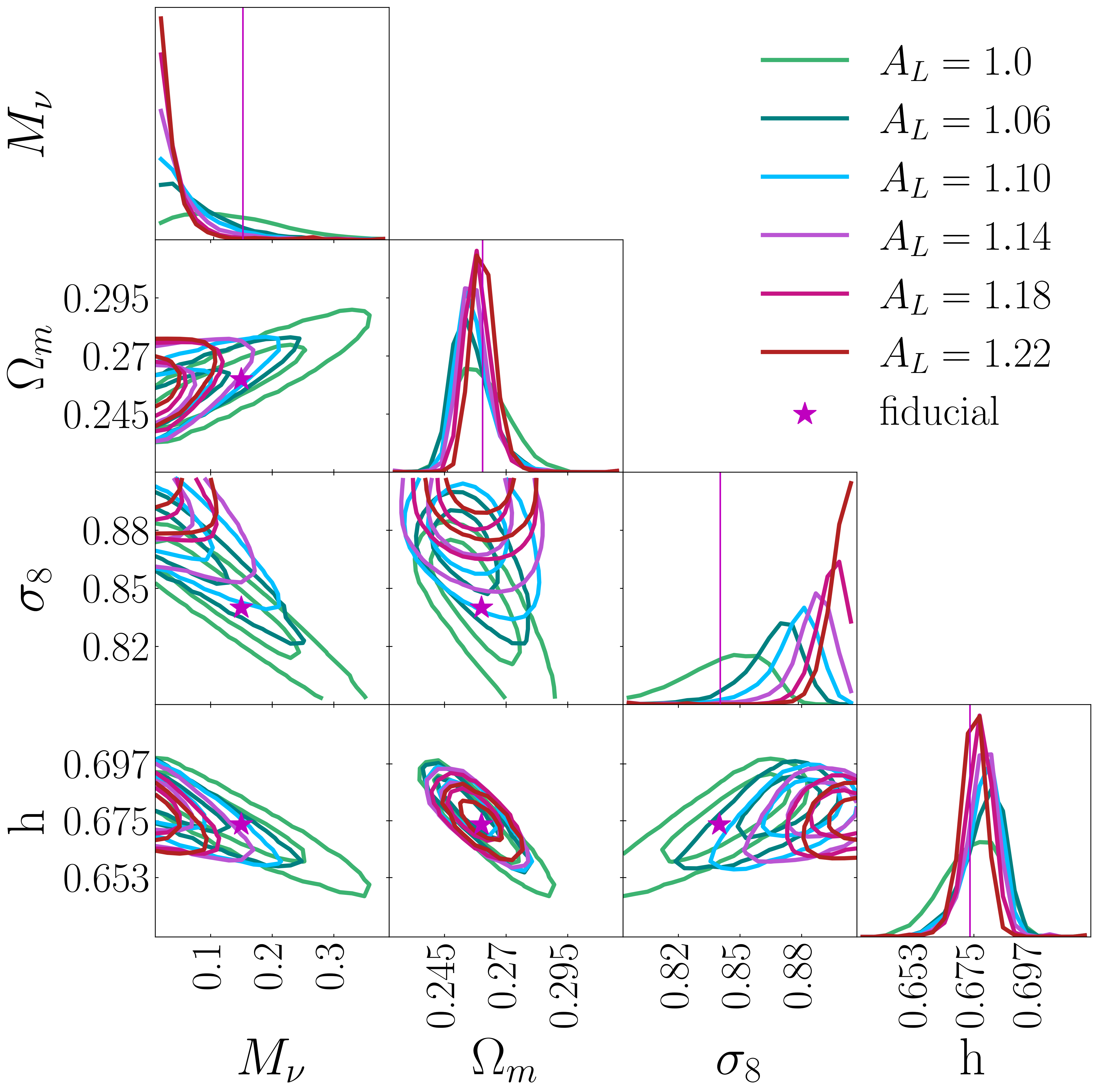}
    \caption{CMB TTTEEE}
  \end{subfigure}
  \hfill
  \begin{subfigure}[b]{0.47\textwidth}
    \centering
    \includegraphics[width=\textwidth]{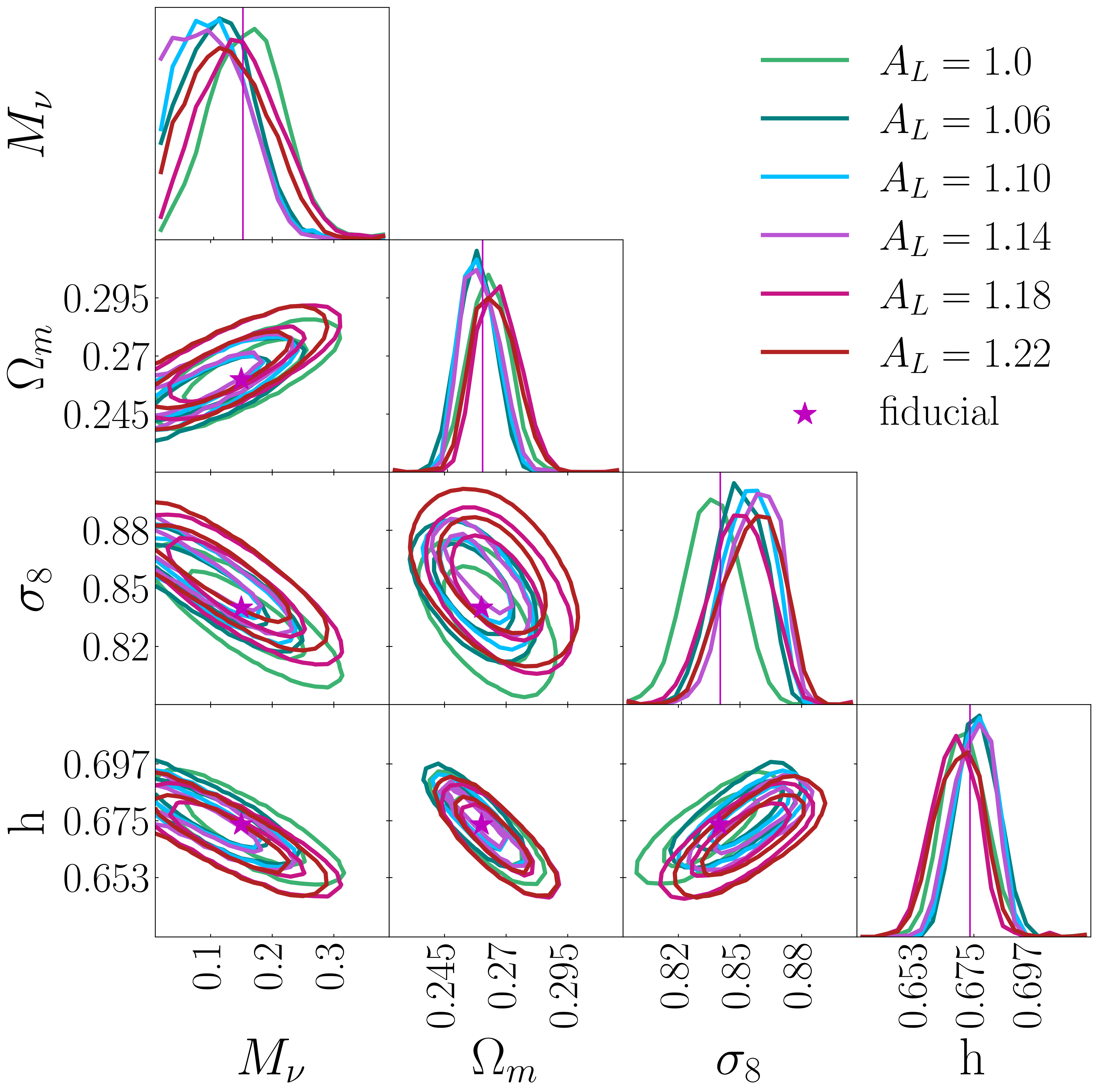}
    \caption{Full combined probes}
  \end{subfigure}

 \caption{Same as Fig.~\ref{fig:alens_explore_fid} but in the \highmnu{}.\label{fig:alens_explore_high_mnu}}
\end{figure}

In Figures~\ref{fig:alens_explore_fid} \& \ref{fig:alens_explore_high_mnu}, we explore the impact of artificially including an $A_L$ systematic on the neutrino mass constraint. To achieve this, we create mock CMB primary TTTEEE data vectors with varying $A_L$ by changing this parameter in the \texttt{CLASS} code. However, we do not vary this parameter in the subsequent analysis, thereby treating this modification as an unaccounted-for systematic in the data. Given the anti-correlation between $A_L$ and $M_\nu$, this leads to a bias in our $M_\nu$ constraints which we quantify in this section. The left-hand plots in Figures~\ref{fig:alens_explore_fid} \& \ref{fig:alens_explore_high_mnu} show that when considering CMB data alone there is a strong induced bias towards low $M_\nu$. For example, when using the baseline cosmology, the CMB alone constraint with $A_L=1$ is $M_\nu < 0.25$ (95 \% CL). As the underlying $A_L$ in the signal is increased, the upper limit constraint on $M_\nu$ monotonically decreases. In particular, for the central $A_L$ value found with \textit{Planck} PR3 data ($A_L$ = 1.180~\cite{Planck:2018lbu}), we would find an upper limit of $M_\nu < 0.13\mathrm{eV}$. Considering the right-hand plot of Fig.~\ref{fig:alens_explore_fid}, we find that the inclusion of LSS data in the full-combined case produces neutrino mass constraints that are more robust to systematic issues of this type. In particular, for $A_L=1.18$ the upper limit is $M_\nu < 0.15\mathrm{eV}$ compared to the no-systematic ($A_L=1.0$) baseline of $M_\nu < 0.16\mathrm{eV}$. The addition of external data also lessens the impact of the $A_L$ internal systematic on the $\sigma_8$ and $h$ constraints. The corresponding plot in the \highmnu{} (Fig.~\ref{fig:alens_explore_high_mnu}) also demonstrates that the neutrino mass constraint is also stabilized by the addition of LSS data. In particular, the constraint found for $A_L=1.18$, is $M_\nu=0.14 \pm 0.07$ compared to $M_\nu=0.15 \pm 0.06$ for the no-systematic baseline. These results suggest that robust neutrino mass constraints can be derived by combining CMB and LSS data even in the presence of an $A_L$-like systematic in the CMB data.

\section{Conclusions}
\label{sec:conclusion}

In this work, we introduce a new pipeline to rapidly infer cosmological parameters from a multi-probe analysis including a combination of LSS and CMB data. The pipeline is built upon correlated, map-level realizations of the LSS probes (including CMB lensing), generated by applying the \ufalcon{} package to the \cosmogrid{} suite of simulations. We use these realizations to estimate a non-Gaussian covariance matrix and a mock data vector for the LSS probes. To overcome the challenge of large data vectors for combined probe analyses, we integrate both the MOPED and PCA data compression schemes into the pipeline. We additionally make use of a neural network emulator for accelerated theoretical predictions. We validate our approach by comparing simulations and theoretical predictions, and we also compare our derived constraints to earlier analyses (see Appendix~\ref{sec:valid}). We apply our framework to a simulated 12 $\times$ 2pt tomographic analysis of KiDS, BOSS, and \textit{Planck}, and forecast constraints for both a standard $\Lambda$CDM model and $\nu \Lambda$CDM. In the standard scenario, we are able to determine $h$, and $\Omega_m$ to within 0.5\% and 1.5\% respectively from the full $12 \times 2$pt combination, representing a $\sim 30\%$ improvement compared to the official \textit{Planck} CMB TTTEEE + CMB lensing analysis~\cite{Planck:2018vyg}. 

As part of this analysis, we demonstrate the power of differentiable emulators both in terms of rapid parameter space exploration and in terms of accurately predicting derivatives, $d C_\ell/ d \vec{\theta}$, which we find to be crucial for robust MOPED-based data compression. Compared to PCA compression, MOPED provides a more extreme compression in all cases, and using this we demonstrate a lossless compression of the full combined-probes data vector from 856 to just 24 data points. We also release a \textit{Planck} CMB TTTEEE+lensing likelihood compressed to just 16 data points and show that this produces constraints in excellent agreement with the official results. 

We additionally explore neutrino mass constraints in the multiprobe setting. We find that, whilst the majority of the constraining power comes from primary CMB data, the combination of CMB lensing, WL, and galaxy clustering data helps to exclude regions of parameter space by breaking degeneracies. In this optimistic setting with all simulated data sharing the same underlying cosmology, we forecast a constraint of $M_\nu < 0.16\mathrm{eV}$ for the fiducial case where $M_\nu=0.06\mathrm{eV}$ and determine a $\sim 3\sigma$ detection in the \highmnu{} ($M_\nu=0.15 \pm0.06$). We reconfirm that an $A_L$-like internal tension in CMB data can bias determinations of $M_\nu$~\cite{Couchot:2017pvz, Motloch:2019gux, DiValentino:2021imh}, for example when analyzing CMB data alone we find constraints of $M_\nu < 0.25\mathrm{eV}$ for $A_L=1.0$ and $M_\nu < 0.13\mathrm{eV}$ for $A_L=1.18$ (the central value in \textit{Planck} PR3 data~\cite{Planck:2018lbu}). However, the impact of $A_L$-like systematics on determinations of $M_\nu$ is significantly mitigated by the addition of LSS data, which helps to anchor the constraints towards the correct cosmological parameters in both the fiducial and \highmnu{}s.

This analysis relies on several simplifying assumptions. We ignore the covariance between CMB primary fluctuations and the LSS probes including CMB lensing/ISW which, whilst justified for current data, would likely need to be accounted for when analyzing future CMB experiments (especially the CMB primary-lensing covariance see~\cite{Trendafilova:2023oni, Schmittfull:2013uea}). Further, we do not model several systematic effects concerning the cross-correlations between CMB lensing and LSS, for example so-called ``N 3/2'' biases~\cite{Fabbian:2019tik} as well as biases in the CMB lensing reconstruction due to cosmic infrared background (CIB), radio sources and the tSZ signal (all of which can be in general correlated with the LSS probes)~\cite{vanEngelen:2013rla}. These effects are small for the surveys simulated in this work though may have important consequences for future CMB lensing cross-correlation analyses. We do not model the cross-correlations between intrinsic alignment and other LSS tracers which may enable better constraints to be placed on the $A_IA$ parameter~\cite{Heymans:2020gsg}. We neglect baryonic feedback effects on WL and restrict to linear bias models for galaxy clustering. In the future, it will be interesting to relax these assumptions and extend our analysis to smaller scales, which we expect to boost the constraining power of the LSS data. Another natural extension of the pipeline is to add additional datasets: On the CMB side, we envisage including ACT CMB primary or CMB lensing data~\cite{ACT:2023kun}, as well as the recently released likelihoods from \textit{Planck} PR4 maps~\cite{Rosenberg:2022sdy, Tristram:2023haj}. On the LSS side, it would be interesting to include DES data~\cite{DES:2005dhi, DES:2017myr, DES:2021wwk, Halder:2023kfy} and the data from the Dark Energy Spectroscopic Instrument (DESI)~\cite{DESI:2016fyo, DESI:2023ytc}. The addition of tSZ data would also be highly beneficial, especially for constraining Baryonic feedback when extending the LSS analysis to small scales ~\cite{Troster:2021gsz, Fang:2023efj}.

This analysis presents a further demonstration of the power of combined probes analyses for deriving strong constraints, robust to systematics. The results and extensive validation tests presented as part of this work strongly motivate applying the combined probes pipeline to current and future survey data, thus deriving robust constraints on $\Lambda$CDM and its extensions.

\appendix
\section{Validating likelihood implementation} \label{sec:valid}

\subsection{\textit{Planck} 2018 primary} \label{subsec:valid_planck_alone}
In order to validate our approach for the individual surveys, we compare our results with official analyses where possible. For the \textit{Planck} primary $C_\ell^{TT}, C_\ell^{TE}$ and $C_\ell^{EE}$ case, we compare our emulator implementation with the official chains downloaded from the \textit{Planck} legacy archive~\footnote{\url{https://pla.esac.esa.int/home}}. Note these chains are for the full likelihood that is not marginalized over the foreground parameters whereas in the pipeline we implement the \texttt{Plik\_lite} likelihood. In addition to this, we also validate and release, a MOPED compressed high-$\ell$ TTTEEE likelihood consisting of just 6 data points. Combining with the 5 bins of the \texttt{planck-low-py} log-normal Gaussian binned likelihood from~\cite{Prince:2021fdv} the full \textit{Planck} likelihood can be compressed to just 11 data points that faithfully reproduce the full-likelihood within $\Lambda$CDM. Our constraints (both in the uncompressed and compressed case) match with the official analysis to within 1\% in both the $\sigma$ and mean for each 1D marginalized parameter constraint.  
\begin{figure}[htbp!]\label{fig:cmb_alone_compare}
\centering
\includegraphics[scale=0.3]{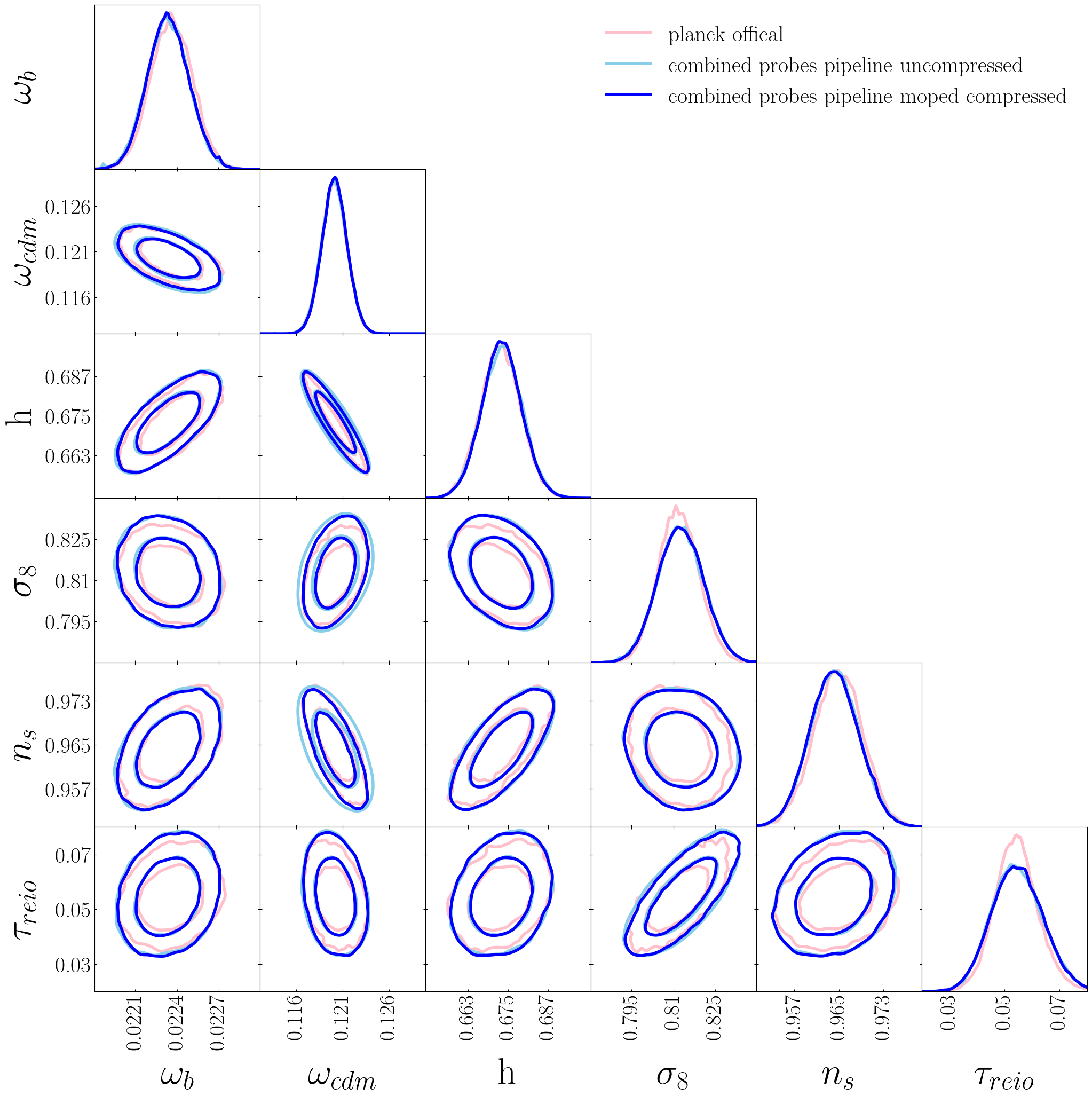}
\caption{Comparison of the \textit{Planck} official high-$\ell$ TTTEEE + low-$\ell$ T/E chains~\cite{Planck:2018vyg} with the chains derived from the implementation in the combined probes pipeline and the MOPED compressed version.}
\end{figure}

\subsection{\textit{Planck} 2018 lensing}\label{subsec:valid_planck_lensing}
\begin{figure}
  \centering

  \begin{subfigure}[b]{0.49\textwidth}
    \includegraphics[width=\textwidth]{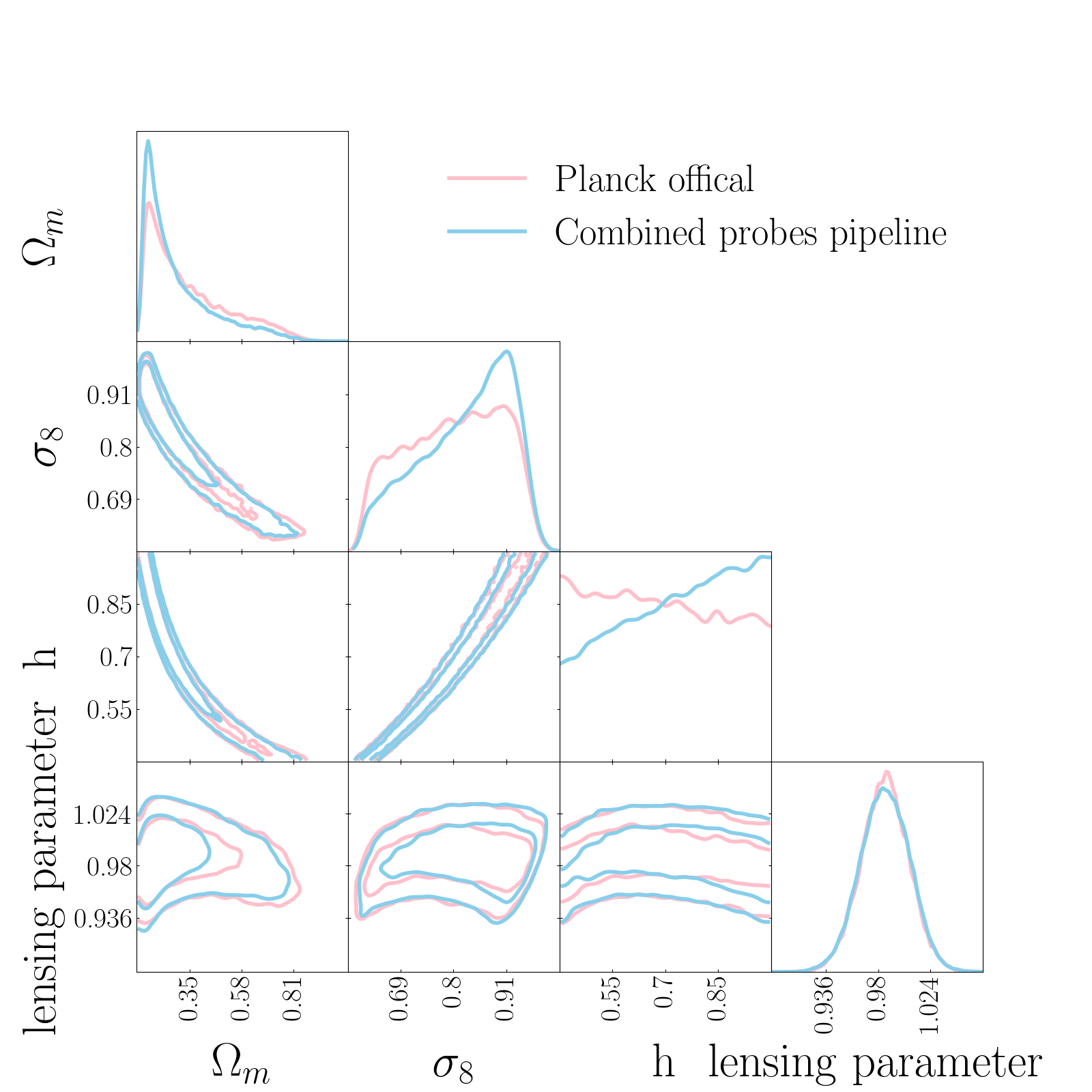}
    \caption{Lensing alone}
  \end{subfigure}
  \hfill
  \begin{subfigure}[b]{0.49\textwidth}
    \centering
    \includegraphics[width=\textwidth]{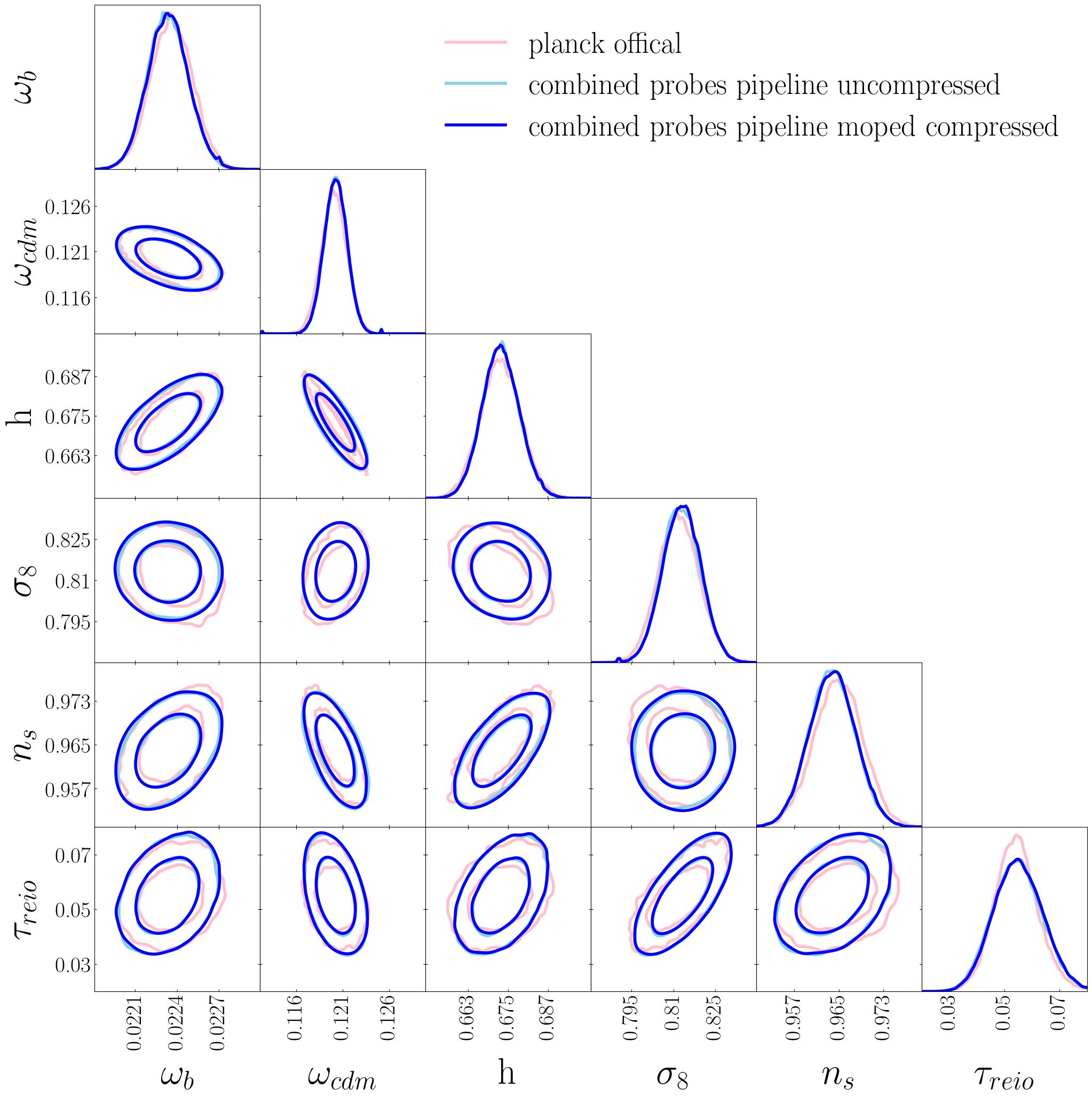}
    \caption{CMB primary + lensing}
  \end{subfigure}

 \caption{Validating the \textit{Planck} CMB lensing implementation: \textbf{left} \textit{Planck} CMB lensing alone comparison with official chains~\cite{Planck:2018lbu} \textbf{right} CMB TTTEEE+lensing comparison between uncompressed, MOPED compressed and official chains.\label{fig:lensing_cmb_comparison}}
\end{figure}

For \textit{Planck} CMB lensing, we compare our emulator implementation with the official chains in two scenarios: firstly on the left-hand plot of Fig.~\ref{fig:lensing_cmb_comparison} we derive ``lensing only'' contours. For these, we place the same priors on the varied cosmological parameters as in Ref.~\cite{Planck:2018lbu} (gaussian priors of $\omega_b=0.0222 \pm 0.0005$, $n_s=0.96 \pm 0.02$, a uniform prior of $0.4 < h < 1$ and fixing $\tau_{\text{reio}}=0.055$)\footnote{Note that $\Omega_m$ is also varied in this analysis with a flat uninformative prior: $0.1<\Omega_m<0.9$} and compare to the official chains. In addition to the cosmological parameters we also derive the posterior distribution of the ``lensing parameter'' defined by~\cite{Planck:2018lbu}: 
\begin{equation}
    \mathrm{lensing \ parameter} = \frac{\sigma_8}{0.8}(\frac{\Omega_m}{0.3})^{0.23}(\frac{\Omega_m h^2}{0.13})^{-0.32},
\end{equation}
this parameter describes the degeneracy direction probed by CMB-lensing data~\cite{Planck:2015mym}. We find excellent agreement in each cosmological parameter and, especially in the lensing parameter validating that the likelihood implementation is accurate in the ``lensing only'' regime. 

On the right-hand plot of Fig.~\ref{fig:lensing_cmb_comparison}, we compare our implementation of the \textit{Planck} CMB lensing likelihood in the case that the primary CMB spectra ($C_\ell^{TT}, C_\ell^{TE}$ and $C_\ell^{EE}$) are also varied. As stated in $\S$\ref{subsubsec:CMB_like}, we are using the ``CMB marginalized'' likelihood for $C_\ell^{\kappa \kappa}$ which has only been validated in the case of analyzing CMB lensing alone. We, therefore, check that we are in good agreement (sub percent level match in the mean and $\sigma$ of the marginalized 1D parameter distributions) with the official CMB + lensing results from \textit{Planck} which are unmarginalized and so account for the effect of the CMB primary spectra on the lensing normalization and $N1$ bias at each step of the inference~\cite{Planck:2018vyg}. In addition to this, we also validate and release, a MOPED compressed ``CMB marginalized'' lensing likelihood consisting of just 5 data points (one for each parameter of the $\Lambda$CDM model excluding $\tau_{\text{reio}}$ which is not varied). Combining with the \texttt{planck-low-py} log-normal Gaussian binned likelihood from~\cite{Prince:2021fdv} and the MOPED compressed \textit{Planck} high-$\ell$ TTTEEE likelihood, the full \textit{Planck} CMB + lensing likelihood can be compressed to just 16 data points that reproduce the full-likelihood within $\Lambda$CDM (see blue contours on right-hand plot of Fig.~\ref{fig:lensing_cmb_comparison}). The code to make the plot on the right-hand side of Fig.~\ref{fig:lensing_cmb_comparison} can be found in the \texttt{planck\_compressed} repository.

\subsection{KiDS-1000}\label{subsec:valid_kids_1000}
\begin{figure}
\centering
\includegraphics[scale=0.19]{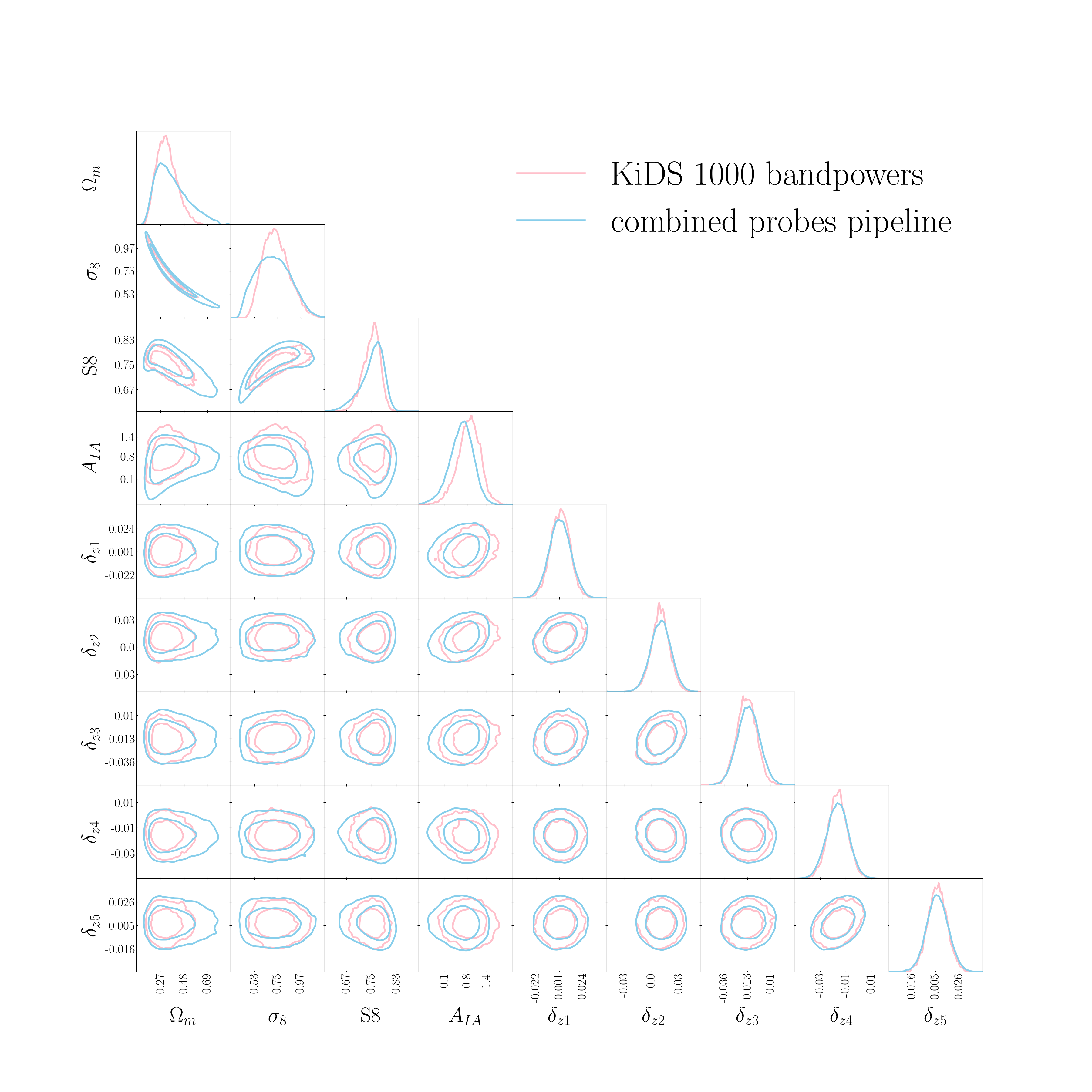}
\caption{Comparison of the official KiDS-1000 bandpower results~\cite{KiDS:2020suj} with the pseudo-$C_\ell$s implementation in the combined probes pipeline. \label{fig:kids1000_compare}}
\end{figure}

\begin{table}[h]
    \centering
    \begin{tabular}{cccc}
        parameter this work & prior this work & parameter KiDS-1000 & prior KiDS-1000\\
        \hline
        \hline 
        $\omega_b$ & [0.019, 0.026] & $\omega_b$ & [0.019, 0.026]\\
        $n_s$ & [0.84, 1.1] & $n_s$ & [0.84, 1.1]\\
        $h$ & [0.64, 0.82] & $h$ & [0.64, 0.82]\\
        $\Omega_m$ & [0.1,0.6] & $\omega_{\text{cdm}}$ & [0.051,0.255]\\
        $\sigma_8$ & [0.2, 1.2] & $S8$ & [0.1,1.3]\\
        $A_{\mathrm{IA}}$ & [-6,6] & $A_{\mathrm{IA}}$ & [-6,6]\\
        $\Delta_{z1}-\Delta_{z5}$ & $\mathcal{N}(\mu_z, \textbf{C}_z)$ & $\Delta_{z1}-\Delta_{z5}$ & $\mathcal{N}(\mu_z, \textbf{C}_z)$\\
        --- & --- & $A_{\text{bary}}$ & [2, 3.13] \\
        \hline
    \end{tabular}
    \caption{Comparison of sampling parameters and priors used for KiDS-1000 WL comparison. Our constraints are derived when varying the parameters within the priors from the left-hand side of the table, whilst the `official results' are deriving sampling over the parameters within the priors on the right-hand side of the table~\cite{KiDS:2020suj}. Note the official KiDS-1000 analysis also varies the ``$A_{\text{bary}}$'' parameter (see Ref.~\cite{KiDS:2020suj} for more details) to incorporate the effect of baryonic feedback, whilst we do not vary this parameter in our set-up.\label{tab:priors_for_kids}}
    \end{table}
    
We make several different analysis choices compared to the official KiDS-1000 analysis~\cite{KiDS:2020suj}. In particular, we estimate the covariance matrix directly from simulations rather than using an analytic method. We also make a simplification in that we do not vary the baryonic physics parameters of \texttt{HMCode}~\cite{Mead:2016zqy}\footnote{The KiDS team tested the impact of this on real data analysis and found sub $1 \sigma$ shifts in the value of $S8$.}. Finally, our emulators are trained over a set of fixed parameters that are not identical to those used in the official KiDS-1000 bandpower analysis. We match the priors where possible to make this comparison (see Table~\ref{tab:priors_for_kids}). The comparison between our implementation and the band power results is shown in Fig.~\ref{fig:kids1000_compare}. The official results give a slightly lower $S8$ compared to our implementation which can be due to the different covariance matrix, the lack of a baryonic feedback implementation and the sampled parameters/priors not being exactly matched. We note that our results match very closely to the ``power spectra'' results of~\cite{Fluri:2022rvb} which used the same set of simulations in computing the covariance matrix.

\section{Emulator set-up and accuracy}\label{sec:emu}

\subsection{Distribution of spectra over networks and prior ranges}\label{subsec:emu_dist}

\begin{table}[!h]
    \centering
    \begin{tabular}{ccc}
        parameter & prior min & prior max \\
        \hline
        \hline 
        $\tau_{\text{reio}}$ & 0.02 & 0.12\\
        $\omega_b$ & 0.019 & 0.026 \\
        $\omega_{\text{cdm}}$ & 0.08 & 0.2 \\
        $n_s$ & 0.8 & 1.2 \\
        $h$ & 0.4 & 1.0\\
        $\sigma_8$ & 0.5 & 1.1 \\
        $M_\nu$ & 0.001 & 0.4 \\
        $A_L$ & 0 & 5\\
        \hline
    \end{tabular}
    \caption{Priors used for the three dedicated CMB primary networks which output $C_\ell^{TT}$, $C_\ell^{TE}$, and $C_\ell^{EE}$ respectively. \label{tab:cmb_priors}}
    \end{table}
    
\begin{table}[!h]
    \centering
    \begin{tabular}{ccc}
        parameter & prior min & prior max \\
        \hline
        \hline 
        $\omega_b$ & 0.018 & 0.026 \\
        $\Omega_{m}$ & 0.1 & 0.9 \\
        $n_s$ & 0.84 & 1.1\\
        $h$ & 0.6 & 0.9 \\
        $\sigma_8$ & 0.3 & 1.3 \\
        $M_\nu$ & 0.001 & 0.4 \\
        $A_\mathrm{IA}$ & -6 & 6 \\
        $\Delta_{z1}-\Delta_{z5}$ & -0.05 & 0.05 \\
        bias lowz & 0.1 & 5 \\
        bias cmass & 0.1 & 5 \\
        \hline
    \end{tabular}
    \caption{Priors used for the LSS emulators in this pipeline. Identical priors are used for the CMB lensing auto correlation $C_\ell^{\kappa \kappa}$ except that the WL/clustering nuisance parameters are not varied.\label{tab:lss_priors}}
\end{table}

To evaluate the suitability of a given emulator for use in cosmological inference it is important to consider the impact on the eventual parameter constraints. We follow a procedure similar to Ref.~\cite{SpurioMancini:2021ppk} in assessing the emulator accuracy in terms of the relative difference between the true $C_\ell$ derived from a theory code and the emulated spectrum compared with the error associated with the data used in the inference step.
This pragmatic approach to emulator accuracy informs our decision to train just six emulators to cover the $C_\ell$s in the pipeline. The data associated with the CMB probes ($C_\ell^{TT}$, $C_\ell^{EE}$,$C_\ell^{TE}$,$C_\ell^{\kappa \kappa}$) has a small error budget and hence we require a high emulator fidelity. We, therefore, train a dedicated emulator for each of these $C_\ell$s. We also train the models to learn the mapping from cosmological parameters to the full unbinned spectra (from $2<\ell<2508$) so that the model does not additionally have to learn a survey-specific binning scheme (we hence apply the required binning at the inference step). 
The data associated with the LSS probes, on the other hand, has a larger error margin and hence we find that we do not need dedicated models: just two networks are able to predict all of the LSS auto- and cross-correlation $C_\ell$s to sufficient accuracy. In this case, each emulator outputs a concatenation of multiple binned $C_\ell$s which already include the effect of convolution and deconvolution with the respective survey mixing matrices (computed using \texttt{NaMaster}). 
The networks are trained over different parameter sets given we only include the nuisance parameters associated with the particular spectra that the network is trained on. The parameters and associated prior ranges for the emulators are shown in Tables ~\ref{tab:cmb_priors} \& ~\ref{tab:lss_priors}.
    
\subsection{Emulator accuracy}\label{subsec:emu_acc}

\begin{figure*}[h]
\centering
\includegraphics[scale=0.14]{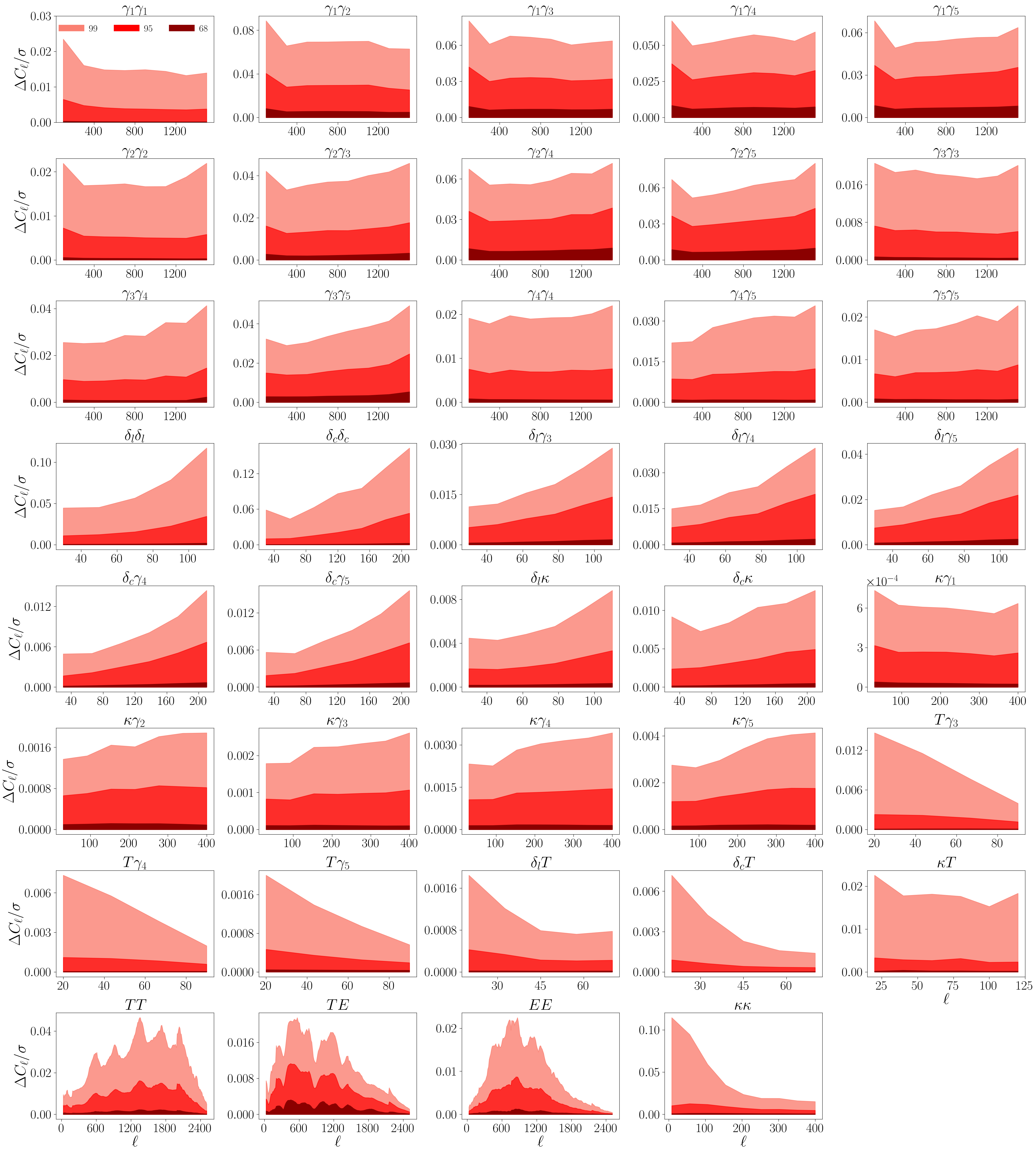}
\caption{The distribution over 60000 testing points of the relative difference between the emulator predictions and the theory code predictions, scaled by the $1 \sigma$ error bars of the modeled surveys. The different colors represent the $68/95/99\%$ limits of the distributions of each spectrum.\label{fig:emu_accuracy}}
\end{figure*}

We follow the approach taken in~\cite{SpurioMancini:2021ppk} in testing the emulator accuracy by considering the fractional error in the emulation of each of a set of 60000 test spectra divided by the error given approximately by:
\begin{equation}
    \begin{split}
        \sigma(C_\ell) = \sqrt{\mathrm{Cov}(\ell, \ell)}.
    \end{split}
\end{equation}
Here ``$\mathrm{Cov}$'' refers to the full covariance matrix from our fiducial set-up i.e. the official \textit{Planck} covariance matrix for the CMB primary $C_\ell$s and our simulation-based covariance for the LSS $C_\ell$s. We follow this procedure for each of the emulators in the pipeline and plot the distribution of the errors. We checked that 99\% of the testing samples have a relative error $< 0.15\sigma$ and hence that the emulator accuracy is unlikely to bias our derived constraints. In Fig.~\ref{fig:emu_accuracy} we show the distributions for all of the spectra in our pipeline.

\section{Covariance matrix validation}

\subsection{Covariance matrix convergence} \label{subsec:covmat_conv}

\begin{figure}[h]
\centering
\includegraphics[scale=0.25]{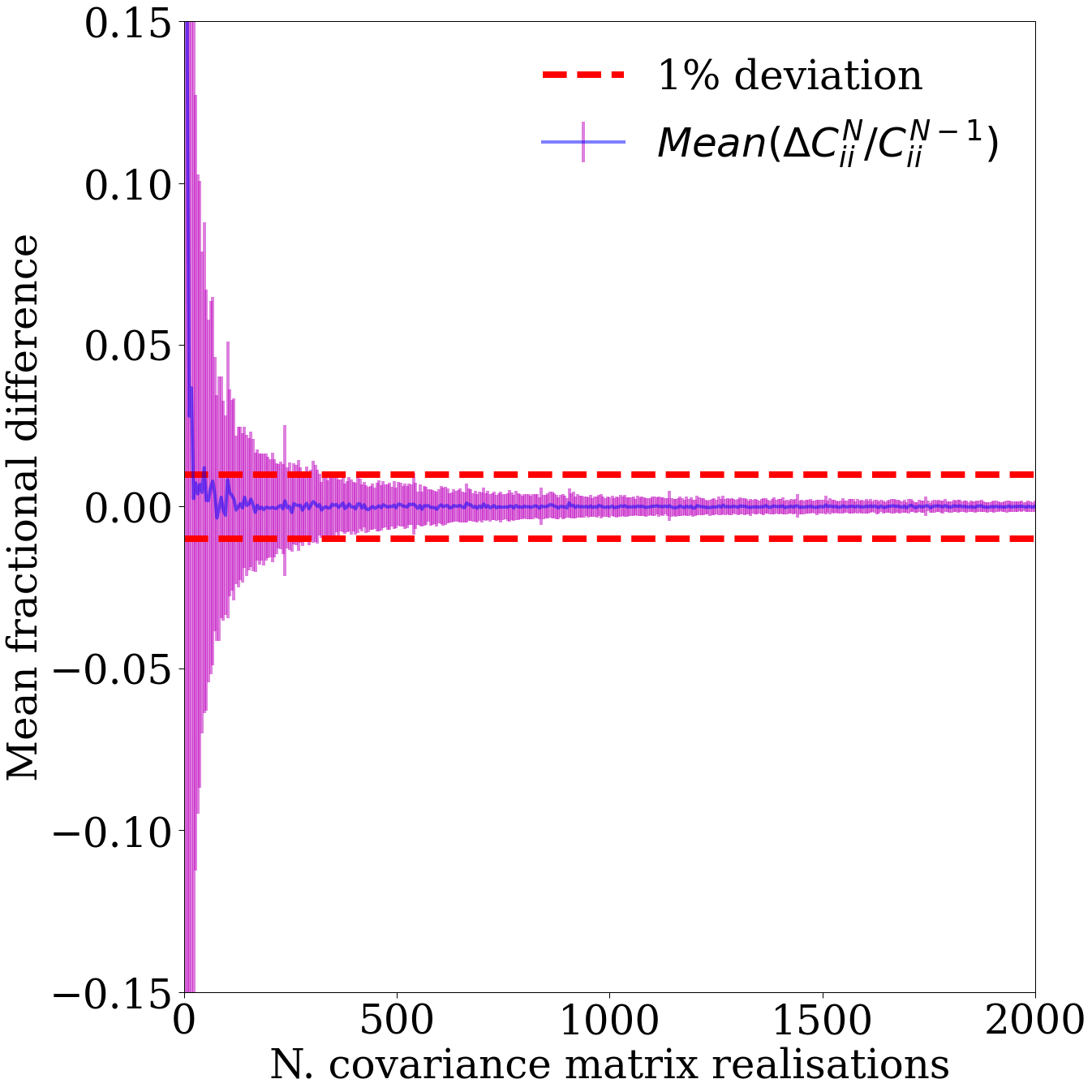}
\caption{The mean fractional change in the diagonals of the covariance matrix ($Mean(\Delta C^N_{ii}/C^{N-1}_{ii})$) as a function of the number of realizations used to generate the matrix. The error bars in purple represent the standard deviation of this fractional change across the different diagonal covariance matrix elements.\label{fig:convariance_conv}}
\end{figure}

To validate that our covariance matrix is converged, we measure the fractional change in the diagonal elements of our covariance matrix as a function of the number of realizations used to estimate it (see Fig.~\ref{fig:convariance_conv}). We find that the 2000 realizations we use (200 underlying simulations each with 10 random noise realizations) are sufficient to produce a sub-percent mean fractional change with a negligible spread across the different diagonal components.

\subsection{Comparison with Gaussian analytic covariance}\label{subsec:compare_to_gauss} 
\begin{figure}[htbp!]
\centering
\includegraphics[scale=0.18]{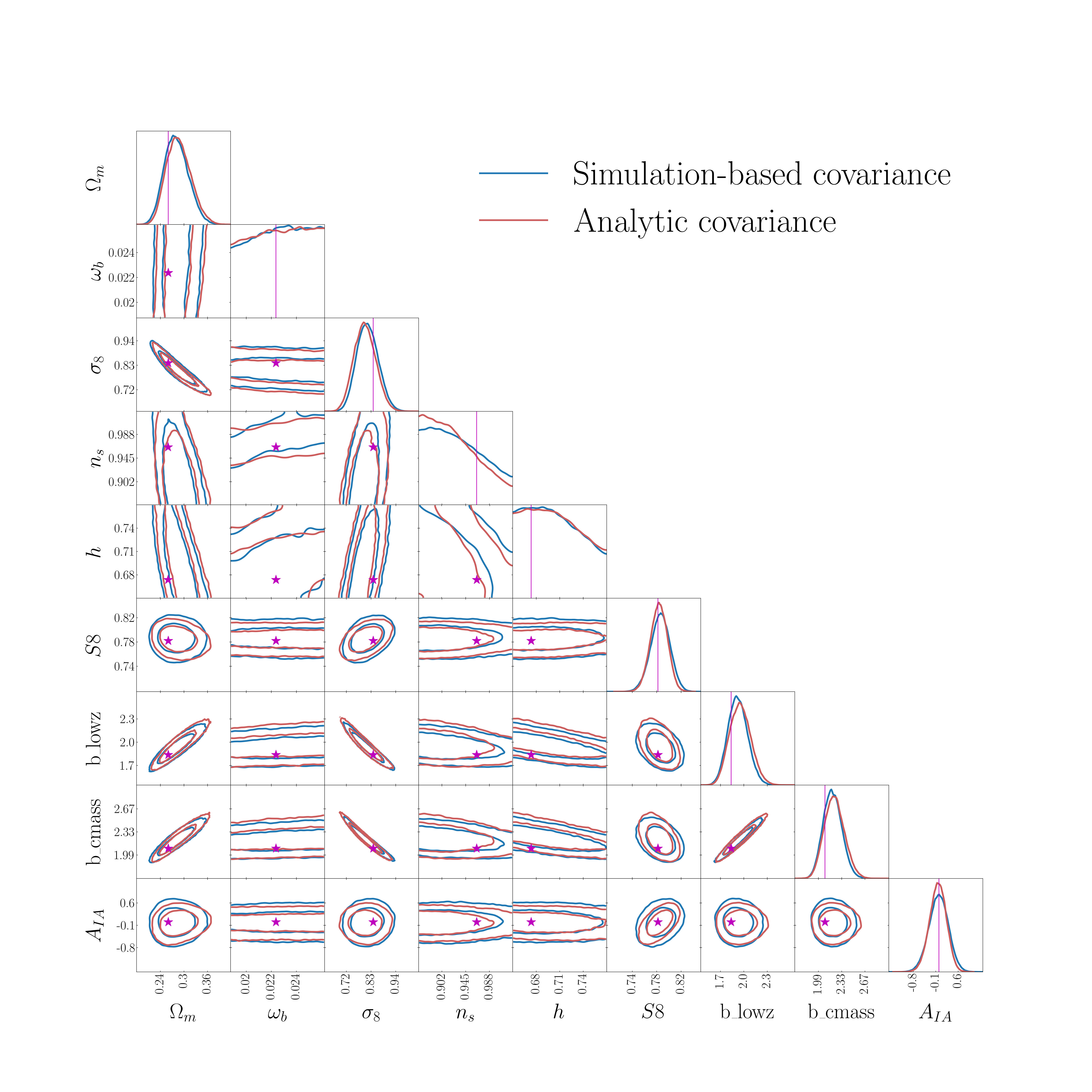}
\caption{Comparison of constraints using an analytic covariance vs simulation-based covariance for a combination of WL, galaxy clustering, CMB lensing, and associated cross-correlations including ISW. \label{fig:gauss_vs_nongauss}}
\end{figure}

A common approach to estimating covariances in cosmology is to assume a negligible non-Gaussian component to total covariance. The covariance matrix can then be computed analytically as a Gaussian piece plus some corrections (e.g. the super-sample covariance). It is possible to write an analytic expression for the Gaussian contribution to the covariance matrix given a) a mock data vector at some fiducial cosmology, b) an estimated noise vector c) the survey mask~\cite{Alonso:2018jzx}. To compare our simulation-based approach to the analytic method, we first generate an analytic covariance using the \texttt{NaMaster} framework which handles solving for the Gaussian covariance given the inputs described above. We checked that there are deviations of no more than 10\% over the diagonal elements of these matrices. Then, in Fig.~\ref{fig:gauss_vs_nongauss}, we compare the results of our full LSS analysis (combining WL, galaxy clustering, CMB lensing, and all associated cross-correlations including ISW). We observe excellent agreement between the constraints, which is a non-trivial result given the completely independent methods used to estimate the two covariance matrices. This agreement provides further evidence that our simulation-based estimation of the covariance matrix works as expected.

\acknowledgments
Some of the results in this paper have been derived using the healpy and HEALPix packages. AR thanks Jiachuan Xu for help in implementing the \textit{Planck} CMB lensing auto-$C_\ell$ likelihood. AR would also like to thank in no particular order Krishna Naidoo, Simone Ferraro, Alex Krolewski, Heather Prince, Carlos Garcia-Garcia, Tilman Tröster, Arthur Loureiro, Arne Thomsen, Silvan Fischbacher, Joël Mayor, Beatrice Moser,  Dominik Zürcher, Raphaël Sgier, Azucena Garvía Bosshard and Janis Fluri for helpful discussions. We also thank Uwe Schmitt for continuous software support throughout this project especially in releasing the \ufalcon{} package. We acknowledge the support of Euler Cluster by High Performance Computing Group from ETHZ Scientific IT Services that we used for most of our computations.

We used functionalities provided by \texttt{numpy}~\cite{harris2020array}, \texttt{scipy}~\cite{2020SciPy-NMeth}, and
\texttt{matplotlib}~\cite{Hunter:2007} for this work. Job arrays were submitted using esub-epipe~\cite{Zurcher:2022clh, Zurcher:2020dvu}. Corner plots were created with \texttt{trianglechain}~\cite{Fischbacher:2022gua, Kacprzak:2022oit}. We used \texttt{emcee}~\cite{Foreman-Mackey:2012any} to run our MCMC chains.

% query inspire, trying to verify the data (this will probably spot

\newpage{}
\bibliographystyle{jcap}
\bibliography{xcorr.bib}

\providecommand{\href}[2]{#2}\begingroup\raggedright\begin{thebibliography}{100}

\bibitem{EUCLID:2011zbd}
{\scshape EUCLID} collaboration, \emph{{Euclid Definition Study Report}},
  \href{https://arxiv.org/abs/1110.3193}{{\ttfamily 1110.3193}}.

\bibitem{LSSTScience:2009jmu}
{\scshape LSST Science, LSST Project} collaboration, \emph{{LSST Science Book,
  Version 2.0}},  \href{https://arxiv.org/abs/0912.0201}{{\ttfamily
  0912.0201}}.

\bibitem{SimonsObservatory:2018koc}
{\scshape Simons Observatory} collaboration, \emph{{The Simons Observatory:
  Science goals and forecasts}},
  \href{https://doi.org/10.1088/1475-7516/2019/02/056}{\emph{JCAP} {\bfseries
  02} (2019) 056} [\href{https://arxiv.org/abs/1808.07445}{{\ttfamily
  1808.07445}}].

\bibitem{Akeson:2019biv}
R.~Akeson et~al., \emph{{The Wide Field Infrared Survey Telescope: 100 Hubbles
  for the 2020s}},  \href{https://arxiv.org/abs/1902.05569}{{\ttfamily
  1902.05569}}.

\bibitem{CMB-S4:2016ple}
{\scshape CMB-S4} collaboration, \emph{{CMB-S4 Science Book, First Edition}},
  \href{https://arxiv.org/abs/1610.02743}{{\ttfamily 1610.02743}}.

\bibitem{Ruiz-Zapatero:2021rzl}
J.~Ruiz-Zapatero et~al., \emph{{Geometry versus growth - Internal consistency
  of the flat $\Lambda$CDM model with KiDS-1000}},
  \href{https://doi.org/10.1051/0004-6361/202141350}{\emph{Astron. Astrophys.}
  {\bfseries 655} (2021) A11}
  [\href{https://arxiv.org/abs/2105.09545}{{\ttfamily 2105.09545}}].

\bibitem{Nicola:2017ryw}
A.~Nicola, A.~Amara and A.~Refregier, \emph{{Integrated cosmological probes:
  Concordance quantified}},
  \href{https://doi.org/10.1088/1475-7516/2017/10/045}{\emph{JCAP} {\bfseries
  10} (2017) 045} [\href{https://arxiv.org/abs/1706.06593}{{\ttfamily
  1706.06593}}].

\bibitem{Kacprzak:2022oit}
T.~Kacprzak and J.~Fluri, \emph{{DeepLSS: Breaking Parameter Degeneracies in
  Large-Scale Structure with Deep-Learning Analysis of Combined Probes}},
  \href{https://doi.org/10.1103/PhysRevX.12.031029}{\emph{Phys. Rev. X}
  {\bfseries 12} (2022) 031029}
  [\href{https://arxiv.org/abs/2203.09616}{{\ttfamily 2203.09616}}].

\bibitem{Rhodes:2013fyq}
J.~Rhodes et~al., \emph{{Exploiting Cross Correlations and Joint Analyses}},
  \href{https://doi.org/10.1016/j.astropartphys.2014.02.009}{\emph{Astropart.
  Phys.} {\bfseries 63} (2015) 42}
  [\href{https://arxiv.org/abs/1309.5388}{{\ttfamily 1309.5388}}].

\bibitem{Weinberg:2013agg}
D.H.~Weinberg, M.J.~Mortonson, D.J.~Eisenstein, C.~Hirata, A.G.~Riess and
  E.~Rozo, \emph{{Observational Probes of Cosmic Acceleration}},
  \href{https://doi.org/10.1016/j.physrep.2013.05.001}{\emph{Phys. Rept.}
  {\bfseries 530} (2013) 87} [\href{https://arxiv.org/abs/1201.2434}{{\ttfamily
  1201.2434}}].

\bibitem{Baxter:2022enq}
E.J.~Baxter et~al., \emph{{Snowmass2021: Opportunities from Cross-survey
  Analyses of Static Probes}},
  \href{https://arxiv.org/abs/2203.06795}{{\ttfamily 2203.06795}}.

\bibitem{DES:2017myr}
{\scshape DES} collaboration, \emph{{Dark Energy Survey year 1 results:
  Cosmological constraints from galaxy clustering and weak lensing}},
  \href{https://doi.org/10.1103/PhysRevD.98.043526}{\emph{Phys. Rev. D}
  {\bfseries 98} (2018) 043526}
  [\href{https://arxiv.org/abs/1708.01530}{{\ttfamily 1708.01530}}].

\bibitem{HSC:2018mrq}
{\scshape HSC} collaboration, \emph{{Cosmology from cosmic shear power spectra
  with Subaru Hyper Suprime-Cam first-year data}},
  \href{https://doi.org/10.1093/pasj/psz010}{\emph{Publ. Astron. Soc. Jap.}
  {\bfseries 71} (2019) 43} [\href{https://arxiv.org/abs/1809.09148}{{\ttfamily
  1809.09148}}].

\bibitem{Heymans:2020gsg}
C.~Heymans et~al., \emph{{KiDS-1000 Cosmology: Multi-probe weak gravitational
  lensing and spectroscopic galaxy clustering constraints}},
  \href{https://doi.org/10.1051/0004-6361/202039063}{\emph{Astron. Astrophys.}
  {\bfseries 646} (2021) A140}
  [\href{https://arxiv.org/abs/2007.15632}{{\ttfamily 2007.15632}}].

\bibitem{Hildebrandt:2018yau}
H.~Hildebrandt et~al., \emph{{KiDS+VIKING-450: Cosmic shear tomography with
  optical and infrared data}},
  \href{https://doi.org/10.1051/0004-6361/201834878}{\emph{Astron. Astrophys.}
  {\bfseries 633} (2020) A69}
  [\href{https://arxiv.org/abs/1812.06076}{{\ttfamily 1812.06076}}].

\bibitem{Piccirilli:2022myi}
G.~Piccirilli, M.~Migliaccio, E.~Branchini and A.~Dolfi, \emph{{A
  cross-correlation analysis of CMB lensing and radio galaxy maps}},
  \href{https://doi.org/10.1051/0004-6361/202244799}{\emph{Astron. Astrophys.}
  {\bfseries 671} (2023) A42}
  [\href{https://arxiv.org/abs/2208.07774}{{\ttfamily 2208.07774}}].

\bibitem{Singh:2016xey}
S.~Singh, R.~Mandelbaum and J.R.~Brownstein, \emph{{Cross-correlating Planck
  CMB lensing with SDSS: Lensing-lensing and galaxy-lensing
  cross-correlations}}, \href{https://doi.org/10.1093/mnras/stw2482}{\emph{Mon.
  Not. Roy. Astron. Soc.} {\bfseries 464} (2017) 2120}
  [\href{https://arxiv.org/abs/1606.08841}{{\ttfamily 1606.08841}}].

\bibitem{White:2021yvw}
M.~White et~al., \emph{{Cosmological constraints from the tomographic
  cross-correlation of DESI Luminous Red Galaxies and Planck CMB lensing}},
  \href{https://doi.org/10.1088/1475-7516/2022/02/007}{\emph{JCAP} {\bfseries
  02} (2022) 007} [\href{https://arxiv.org/abs/2111.09898}{{\ttfamily
  2111.09898}}].

\bibitem{Doux:2017tsv}
C.~Doux, M.~Penna-Lima, S.D.P.~Vitenti, J.~Tr\'eguer, E.~Aubourg and K.~Ganga,
  \emph{{Cosmological constraints from a joint analysis of cosmic microwave
  background and spectroscopic tracers of the large-scale structure}},
  \href{https://doi.org/10.1093/mnras/sty2160}{\emph{Mon. Not. Roy. Astron.
  Soc.} {\bfseries 480} (2018) 5386}
  [\href{https://arxiv.org/abs/1706.04583}{{\ttfamily 1706.04583}}].

\bibitem{Krolewski:2019yrv}
A.~Krolewski, S.~Ferraro, E.F.~Schlafly and M.~White, \emph{{unWISE tomography
  of Planck CMB lensing}},
  \href{https://doi.org/10.1088/1475-7516/2020/05/047}{\emph{JCAP} {\bfseries
  05} (2020) 047} [\href{https://arxiv.org/abs/1909.07412}{{\ttfamily
  1909.07412}}].

\bibitem{HerschelATLAS:2014txv}
{\scshape Herschel ATLAS} collaboration, \emph{{Cross-correlation between the
  CMB lensing potential measured by Planck and high-z sub-mm galaxies detected
  by the Herschel-ATLAS survey}},
  \href{https://doi.org/10.1088/0004-637X/802/1/64}{\emph{Astrophys. J.}
  {\bfseries 802} (2015) 64} [\href{https://arxiv.org/abs/1410.4502}{{\ttfamily
  1410.4502}}].

\bibitem{Planck:2015fcm}
{\scshape Planck} collaboration, \emph{{Planck 2015 results. XXI. The
  integrated Sachs-Wolfe effect}},
  \href{https://doi.org/10.1051/0004-6361/201525831}{\emph{Astron. Astrophys.}
  {\bfseries 594} (2016) A21}
  [\href{https://arxiv.org/abs/1502.01595}{{\ttfamily 1502.01595}}].

\bibitem{Krolewski:2021znk}
A.~Krolewski and S.~Ferraro, \emph{{The Integrated Sachs Wolfe effect: unWISE
  and Planck constraints on dynamical dark energy}},
  \href{https://doi.org/10.1088/1475-7516/2022/04/033}{\emph{JCAP} {\bfseries
  04} (2022) 033} [\href{https://arxiv.org/abs/2110.13959}{{\ttfamily
  2110.13959}}].

\bibitem{Robertson:2020xom}
N.C.~Robertson et~al., \emph{{Strong detection of the CMB lensing and galaxy
  weak lensing cross-correlation from ACT-DR4, Planck Legacy, and KiDS-1000}},
  \href{https://doi.org/10.1051/0004-6361/202039975}{\emph{Astron. Astrophys.}
  {\bfseries 649} (2021) A146}
  [\href{https://arxiv.org/abs/2011.11613}{{\ttfamily 2011.11613}}].

\bibitem{Fang:2023efj}
X.~Fang, E.~Krause, T.~Eifler, S.~Ferraro, K.~Benabed, P.R.~S. et~al.,
  \emph{{Cosmology from weak lensing, galaxy clustering, CMB lensing and tSZ:
  I. 10x2pt Modelling Methodology}},
  \href{https://arxiv.org/abs/2308.01856}{{\ttfamily 2308.01856}}.

\bibitem{Sgier:2021bzf}
R.~Sgier, C.~Lorenz, A.~Refregier, J.~Fluri, D.~Z\"urcher and F.~Tarsitano,
  \emph{{Combined $13\times2$-point analysis of the Cosmic Microwave Background
  and Large-Scale Structure: implications for the $S_8$-tension and neutrino
  mass constraints}},  \href{https://arxiv.org/abs/2110.03815}{{\ttfamily
  2110.03815}}.

\bibitem{Garcia-Garcia:2021unp}
C.~Garc\'\i{}a-Garc\'\i{}a, J.R.~Zapatero, D.~Alonso, E.~Bellini,
  P.G.~Ferreira, E.-M.~Mueller et~al., \emph{{The growth of density
  perturbations in the last \ensuremath{\sim}10 billion years from tomographic
  large-scale structure data}},
  \href{https://doi.org/10.1088/1475-7516/2021/10/030}{\emph{JCAP} {\bfseries
  10} (2021) 030} [\href{https://arxiv.org/abs/2105.12108}{{\ttfamily
  2105.12108}}].

\bibitem{Nicola:2016qrc}
A.~Nicola, A.~Refregier and A.~Amara, \emph{{Integrated Cosmological Probes:
  Extended Analysis}},
  \href{https://doi.org/10.1103/PhysRevD.95.083523}{\emph{Phys. Rev. D}
  {\bfseries 95} (2017) 083523}
  [\href{https://arxiv.org/abs/1612.03121}{{\ttfamily 1612.03121}}].

\bibitem{Nicola:2016eua}
A.~Nicola, A.~Refregier and A.~Amara, \emph{{Integrated approach to cosmology:
  Combining CMB, large-scale structure and weak lensing}},
  \href{https://doi.org/10.1103/PhysRevD.94.083517}{\emph{Phys. Rev. D}
  {\bfseries 94} (2016) 083517}
  [\href{https://arxiv.org/abs/1607.01014}{{\ttfamily 1607.01014}}].

\bibitem{DES:2022xxr}
{\scshape DES, SPT} collaboration, \emph{{Joint analysis of Dark Energy Survey
  Year 3 data and CMB lensing from SPT and Planck. II. Cross-correlation
  measurements and cosmological constraints}},
  \href{https://doi.org/10.1103/PhysRevD.107.023530}{\emph{Phys. Rev. D}
  {\bfseries 107} (2023) 023530}
  [\href{https://arxiv.org/abs/2203.12440}{{\ttfamily 2203.12440}}].

\bibitem{Wenzl:2021rrq}
L.~Wenzl, C.~Doux, C.~Heinrich, R.~Bean, B.~Jain, O.~Dor\'e et~al.,
  \emph{{Cosmology with the Roman Space Telescope \textendash{} Synergies with
  CMB lensing}}, \href{https://doi.org/10.1093/mnras/stac790}{\emph{Mon. Not.
  Roy. Astron. Soc.} {\bfseries 512} (2022) 5311}
  [\href{https://arxiv.org/abs/2112.07681}{{\ttfamily 2112.07681}}].

\bibitem{Krause:2016jvl}
E.~Krause and T.~Eifler, \emph{{cosmolike \textendash{} cosmological likelihood
  analyses for photometric galaxy surveys}},
  \href{https://doi.org/10.1093/mnras/stx1261}{\emph{Mon. Not. Roy. Astron.
  Soc.} {\bfseries 470} (2017) 2100}
  [\href{https://arxiv.org/abs/1601.05779}{{\ttfamily 1601.05779}}].

\bibitem{Eifler:2013fit}
T.~Eifler, E.~Krause, P.~Schneider and K.~Honscheid, \emph{{Combining Probes of
  Large-Scale Structure with CosmoLike}},
  \href{https://doi.org/10.1093/mnras/stu251}{\emph{Mon. Not. Roy. Astron.
  Soc.} {\bfseries 440} (2014) 1379}
  [\href{https://arxiv.org/abs/1302.2401}{{\ttfamily 1302.2401}}].

\bibitem{BOSS:2016wmc}
{\scshape BOSS} collaboration, \emph{{The clustering of galaxies in the
  completed SDSS-III Baryon Oscillation Spectroscopic Survey: cosmological
  analysis of the DR12 galaxy sample}},
  \href{https://doi.org/10.1093/mnras/stx721}{\emph{Mon. Not. Roy. Astron.
  Soc.} {\bfseries 470} (2017) 2617}
  [\href{https://arxiv.org/abs/1607.03155}{{\ttfamily 1607.03155}}].

\bibitem{Planck:2018vyg}
{\scshape Planck} collaboration, \emph{{Planck 2018 results. VI. Cosmological
  parameters}},
  \href{https://doi.org/10.1051/0004-6361/201833910}{\emph{Astron. Astrophys.}
  {\bfseries 641} (2020) A6}
  [\href{https://arxiv.org/abs/1807.06209}{{\ttfamily 1807.06209}}].

\bibitem{Calabrese:2008rt}
E.~Calabrese, A.~Slosar, A.~Melchiorri, G.F.~Smoot and O.~Zahn, \emph{{Cosmic
  Microwave Weak lensing data as a test for the dark universe}},
  \href{https://doi.org/10.1103/PhysRevD.77.123531}{\emph{Phys. Rev. D}
  {\bfseries 77} (2008) 123531}
  [\href{https://arxiv.org/abs/0803.2309}{{\ttfamily 0803.2309}}].

\bibitem{DiValentino:2021imh}
E.~Di~Valentino and A.~Melchiorri, \emph{{Neutrino Mass Bounds in the Era of
  Tension Cosmology}},
  \href{https://doi.org/10.3847/2041-8213/ac6ef5}{\emph{Astrophys. J. Lett.}
  {\bfseries 931} (2022) L18}
  [\href{https://arxiv.org/abs/2112.02993}{{\ttfamily 2112.02993}}].

\bibitem{Rosenberg:2022sdy}
E.~Rosenberg, S.~Gratton and G.~Efstathiou, \emph{{CMB power spectra and
  cosmological parameters from Planck PR4 with CamSpec}},
  \href{https://doi.org/10.1093/mnras/stac2744}{\emph{Mon. Not. Roy. Astron.
  Soc.} {\bfseries 517} (2022) 4620}
  [\href{https://arxiv.org/abs/2205.10869}{{\ttfamily 2205.10869}}].

\bibitem{Tristram:2023haj}
M.~Tristram et~al., \emph{{Cosmological parameters derived from the final (PR4)
  Planck data release}},  \href{https://arxiv.org/abs/2309.10034}{{\ttfamily
  2309.10034}}.

\bibitem{Heavens:1999am}
A.~Heavens, R.~Jimenez and O.~Lahav, \emph{{Massive lossless data compression
  and multiple parameter estimation from galaxy spectra}},
  \href{https://doi.org/10.1046/j.1365-8711.2000.03692.x}{\emph{Mon. Not. Roy.
  Astron. Soc.} {\bfseries 317} (2000) 965}
  [\href{https://arxiv.org/abs/astro-ph/9911102}{{\ttfamily
  astro-ph/9911102}}].

\bibitem{pearson:1901op}
K.P.~F.R.S., \emph{Liii. on lines and planes of closest fit to systems of
  points in space}, \href{https://doi.org/10.1080/14786440109462720}{\emph{The
  London, Edinburgh, and Dublin Philosophical Magazine and Journal of Science}
  {\bfseries 2} (1901) 559}
  [\href{https://arxiv.org/abs/https://doi.org/10.1080/14786440109462720}{{\ttfamily
  https://doi.org/10.1080/14786440109462720}}].

\bibitem{Prince:2019hse}
H.~Prince and J.~Dunkley, \emph{{Data compression in cosmology: A compressed
  likelihood for Planck data}},
  \href{https://doi.org/10.1103/PhysRevD.100.083502}{\emph{Phys. Rev. D}
  {\bfseries 100} (2019) 083502}
  [\href{https://arxiv.org/abs/1909.05869}{{\ttfamily 1909.05869}}].

\bibitem{Kacprzak:2022pww}
T.~Kacprzak, J.~Fluri, A.~Schneider, A.~Refregier and J.~Stadel,
  \emph{{CosmoGridV1: a simulated $w$CDM theory prediction for map-level
  cosmological inference}},  \href{https://arxiv.org/abs/2209.04662}{{\ttfamily
  2209.04662}}.

\bibitem{Potter:2016ttn}
D.~Potter, J.~Stadel and R.~Teyssier, \emph{{PKDGRAV3: Beyond Trillion Particle
  Cosmological Simulations for the Next Era of Galaxy Surveys}},
  \href{https://arxiv.org/abs/1609.08621}{{\ttfamily 1609.08621}}.

\bibitem{Tram:2018znz}
T.~Tram, J.~Brandbyge, J.~Dakin and S.~Hannestad, \emph{{Fully relativistic
  treatment of light neutrinos in $N$-body simulations}},
  \href{https://doi.org/10.1088/1475-7516/2019/03/022}{\emph{JCAP} {\bfseries
  03} (2019) 022} [\href{https://arxiv.org/abs/1811.00904}{{\ttfamily
  1811.00904}}].

\bibitem{Sgier:2018soj}
R.~Sgier, A.~R\'efr\'egier, A.~Amara and A.~Nicola, \emph{{Fast generation of
  covariance matrices for weak lensing}},
  \href{https://doi.org/10.1088/1475-7516/2019/01/044}{\emph{JCAP} {\bfseries
  01} (2019) 044} [\href{https://arxiv.org/abs/1801.05745}{{\ttfamily
  1801.05745}}].

\bibitem{Sgier:2020das}
R.~Sgier, J.~Fluri, J.~Herbel, A.~R\'efr\'egier, A.~Amara, T.~Kacprzak et~al.,
  \emph{{Fast Lightcones for Combined Cosmological Probes}},
  \href{https://doi.org/10.1088/1475-7516/2021/02/047}{\emph{JCAP} {\bfseries
  02} (2021) 047} [\href{https://arxiv.org/abs/2007.05735}{{\ttfamily
  2007.05735}}].

\bibitem{Gorski:2004by}
K.M.~G\'orski, E.~Hivon, A.J.~Banday, B.D.~Wandelt, F.K.~Hansen, M.~Reinecke
  et~al., \emph{{HEALPix - A Framework for high resolution discretization, and
  fast analysis of data distributed on the sphere}},
  \href{https://doi.org/10.1086/427976}{\emph{Astrophys. J.} {\bfseries 622}
  (2005) 759} [\href{https://arxiv.org/abs/astro-ph/0409513}{{\ttfamily
  astro-ph/0409513}}].

\bibitem{Refregier:2003ct}
A.~Refregier, \emph{{Weak gravitational lensing by large scale structure}},
  \href{https://doi.org/10.1146/annurev.astro.41.111302.102207}{\emph{Ann. Rev.
  Astron. Astrophys.} {\bfseries 41} (2003) 645}
  [\href{https://arxiv.org/abs/astro-ph/0307212}{{\ttfamily
  astro-ph/0307212}}].

\bibitem{Kilbinger:2014cea}
M.~Kilbinger, \emph{{Cosmology with cosmic shear observations: a review}},
  \href{https://doi.org/10.1088/0034-4885/78/8/086901}{\emph{Rept. Prog. Phys.}
  {\bfseries 78} (2015) 086901}
  [\href{https://arxiv.org/abs/1411.0115}{{\ttfamily 1411.0115}}].

\bibitem{Teyssier:2008zd}
R.~Teyssier, S.~Pires, S.~Prunet, D.~Aubert, C.~Pichon, A.~Amara et~al.,
  \emph{{Full-Sky Weak Lensing Simulation with 70 Billion Particles}},
  \href{https://doi.org/10.1051/0004-6361/200810657}{\emph{Astron. Astrophys.}
  {\bfseries 497} (2009) 335}
  [\href{https://arxiv.org/abs/0807.3651}{{\ttfamily 0807.3651}}].

\bibitem{Wallis:2017lwt}
C.G.R.~Wallis, M.A.~Price, J.D.~McEwen, T.D.~Kitching, B.~Leistedt and
  A.~Plouviez, \emph{{Mapping dark matter on the celestial sphere with weak
  gravitational lensing}},
  \href{https://doi.org/10.1093/mnras/stab3235}{\emph{Mon. Not. Roy. Astron.
  Soc.} {\bfseries 509} (2021) 4480}
  [\href{https://arxiv.org/abs/1703.09233}{{\ttfamily 1703.09233}}].

\bibitem{Desjacques:2016bnm}
V.~Desjacques, D.~Jeong and F.~Schmidt, \emph{{Large-Scale Galaxy Bias}},
  \href{https://doi.org/10.1016/j.physrep.2017.12.002}{\emph{Phys. Rept.}
  {\bfseries 733} (2018) 1} [\href{https://arxiv.org/abs/1611.09787}{{\ttfamily
  1611.09787}}].

\bibitem{Lewis:2006fu}
A.~Lewis and A.~Challinor, \emph{{Weak gravitational lensing of the CMB}},
  \href{https://doi.org/10.1016/j.physrep.2006.03.002}{\emph{Phys. Rept.}
  {\bfseries 429} (2006) 1}
  [\href{https://arxiv.org/abs/astro-ph/0601594}{{\ttfamily
  astro-ph/0601594}}].

\bibitem{Zonca:2019vzt}
A.~Zonca, L.~Singer, D.~Lenz, M.~Reinecke, C.~Rosset, E.~Hivon et~al.,
  \emph{{healpy: equal area pixelization and spherical harmonics transforms for
  data on the sphere in Python}},
  \href{https://doi.org/10.21105/joss.01298}{\emph{Journal of Open Source
  Software} {\bfseries 4} (2019) 1298}.

\bibitem{SachsWolfe1967}
R.~{Sachs} and A.~{Wolfe}, \emph{{Perturbations of a Cosmological Model and
  Angular Variations of the Microwave Background}},
  \href{https://doi.org/10.1086/148982}{\emph{ApJ} {\bfseries 147} (1967) 73}.

\bibitem{ReesSciama1968}
M.J.~{Rees} and D.W.~{Sciama}, \emph{{Large-scale Density Inhomogeneities in
  the Universe}}, \href{https://doi.org/10.1038/217511a0}{\emph{Nature}
  {\bfseries 217} (1968) 511}.

\bibitem{Naidoo:2021ylw}
K.~Naidoo, P.~Fosalba, L.~Whiteway and O.~Lahav, \emph{{Full-sky integrated
  Sachs\textendash{}Wolfe maps for the MICE grand challenge lightcone
  simulation}}, \href{https://doi.org/10.1093/mnras/stab1962}{\emph{Mon. Not.
  Roy. Astron. Soc.} {\bfseries 506} (2021) 4344}
  [\href{https://arxiv.org/abs/2103.14654}{{\ttfamily 2103.14654}}].

\bibitem{Seljak:1995eu}
U.~Seljak, \emph{{Rees-Sciama effect in a CDM universe}},
  \href{https://doi.org/10.1086/176991}{\emph{Astrophys. J.} {\bfseries 460}
  (1996) 549} [\href{https://arxiv.org/abs/astro-ph/9506048}{{\ttfamily
  astro-ph/9506048}}].

\bibitem{Shapiro_2012}
C.~Shapiro, R.G.~Crittenden and W.J.~Percival, \emph{The complementarity of
  redshift-space distortions and the integrated sachs-wolfe effect: a 3d
  spherical analysis},
  \href{https://doi.org/10.1111/j.1365-2966.2012.20785.x}{\emph{Monthly Notices
  of the Royal Astronomical Society} {\bfseries 422} (2012) 2341}.

\bibitem{Grasshorn_Gebhardt_2021}
H.S.G.~Gebhardt and O.~Dor{\'{e} }, \emph{Fabulous code for spherical
  fourier-bessel decomposition},
  \href{https://doi.org/10.1103/physrevd.104.123548}{\emph{Physical Review D}
  {\bfseries 104} (2021) }.

\bibitem{SFB_IEEE}
Q.~Wang, O.~Ronneberger and H.~Burkhardt, \emph{Rotational invariance based on
  fourier analysis in polar and spherical coordinates},
  \href{https://doi.org/10.1109/TPAMI.2009.29}{\emph{IEEE Transactions on
  Pattern Analysis and Machine Intelligence} {\bfseries 31} (2009) 1715}.

\bibitem{KiDS:2020suj}
M.~Asgari et~al., \emph{{KiDS-1000 Cosmology: Cosmic shear constraints and
  comparison between two point statistics}},
  \href{https://doi.org/10.1051/0004-6361/202039070}{\emph{Astron. Astrophys.}
  {\bfseries 645} (2021) A104}
  [\href{https://arxiv.org/abs/2007.15633}{{\ttfamily 2007.15633}}].

\bibitem{Limber:1954zz}
D.N.~Limber, \emph{{The Analysis of Counts of the Extragalactic Nebulae in
  Terms of a Fluctuating Density Field. II}},
  \href{https://doi.org/10.1086/145870}{\emph{Astrophys. J.} {\bfseries 119}
  (1954) 655}.

\bibitem{Simon:2006gm}
P.~Simon, \emph{{How accurate is Limber's equation?}},
  \href{https://doi.org/10.1051/0004-6361:20066352}{\emph{Astron. Astrophys.}
  {\bfseries 473} (2007) 711}
  [\href{https://arxiv.org/abs/astro-ph/0609165}{{\ttfamily
  astro-ph/0609165}}].

\bibitem{Kaiser:1991qi}
N.~Kaiser, \emph{{Weak gravitational lensing of distant galaxies}},
  \href{https://doi.org/10.1086/171151}{\emph{Astrophys. J.} {\bfseries 388}
  (1992) 272}.

\bibitem{Kaiser:1996tp}
N.~Kaiser, \emph{{Weak lensing and cosmology}},
  \href{https://doi.org/10.1086/305515}{\emph{Astrophys. J.} {\bfseries 498}
  (1998) 26} [\href{https://arxiv.org/abs/astro-ph/9610120}{{\ttfamily
  astro-ph/9610120}}].

\bibitem{LoVerde:2008re}
M.~LoVerde and N.~Afshordi, \emph{{Extended Limber Approximation}},
  \href{https://doi.org/10.1103/PhysRevD.78.123506}{\emph{Phys. Rev. D}
  {\bfseries 78} (2008) 123506}
  [\href{https://arxiv.org/abs/0809.5112}{{\ttfamily 0809.5112}}].

\bibitem{Tarsitano:2020ddh}
F.~Tarsitano, U.~Schmitt, A.~Refregier, J.~Fluri, R.~Sgier, A.~Nicola et~al.,
  \emph{{Predicting cosmological observables with PyCosmo}},
  \href{https://doi.org/10.1016/j.ascom.2021.100484}{\emph{Astron. Comput.}
  {\bfseries 36} (2021) 100484}
  [\href{https://arxiv.org/abs/2005.00543}{{\ttfamily 2005.00543}}].

\bibitem{Kiessling:2015sma}
A.~Kiessling et~al., \emph{{Galaxy Alignments: Theory, Modelling
  \textbackslash{}\& Simulations}},
  \href{https://doi.org/10.1007/s11214-015-0203-6}{\emph{Space Sci. Rev.}
  {\bfseries 193} (2015) 67}
  [\href{https://arxiv.org/abs/1504.05546}{{\ttfamily 1504.05546}}].

\bibitem{Kirk:2015nma}
D.~Kirk et~al., \emph{{Galaxy alignments: Observations and impact on
  cosmology}}, \href{https://doi.org/10.1007/s11214-015-0213-4}{\emph{Space
  Sci. Rev.} {\bfseries 193} (2015) 139}
  [\href{https://arxiv.org/abs/1504.05465}{{\ttfamily 1504.05465}}].

\bibitem{Joachimi:2015mma}
B.~Joachimi et~al., \emph{{Galaxy alignments: An overview}},
  \href{https://doi.org/10.1007/s11214-015-0177-4}{\emph{Space Sci. Rev.}
  {\bfseries 193} (2015) 1} [\href{https://arxiv.org/abs/1504.05456}{{\ttfamily
  1504.05456}}].

\bibitem{Bridle:2007ft}
S.~Bridle and L.~King, \emph{{Dark energy constraints from cosmic shear power
  spectra: impact of intrinsic alignments on photometric redshift
  requirements}}, \href{https://doi.org/10.1088/1367-2630/9/12/444}{\emph{New
  J. Phys.} {\bfseries 9} (2007) 444}
  [\href{https://arxiv.org/abs/0705.0166}{{\ttfamily 0705.0166}}].

\bibitem{Bardeen:1985tr}
J.M.~Bardeen, J.R.~Bond, N.~Kaiser and A.S.~Szalay, \emph{{The Statistics of
  Peaks of Gaussian Random Fields}},
  \href{https://doi.org/10.1086/164143}{\emph{Astrophys. J.} {\bfseries 304}
  (1986) 15}.

\bibitem{Kaiser:1984sw}
N.~Kaiser, \emph{{On the Spatial correlations of Abell clusters}},
  \href{https://doi.org/10.1086/184341}{\emph{Astrophys. J. Lett.} {\bfseries
  284} (1984) L9}.

\bibitem{Blas:2011rf}
D.~Blas, J.~Lesgourgues and T.~Tram, \emph{{The Cosmic Linear Anisotropy
  Solving System (CLASS) II: Approximation schemes}},
  \href{https://doi.org/10.1088/1475-7516/2011/07/034}{\emph{JCAP} {\bfseries
  07} (2011) 034} [\href{https://arxiv.org/abs/1104.2933}{{\ttfamily
  1104.2933}}].

\bibitem{Refregier:2017seh}
A.~Refregier, L.~Gamper, A.~Amara and L.~Heisenberg, \emph{{PyCosmo: An
  Integrated Cosmological Boltzmann Solver}},
  \href{https://doi.org/10.1016/j.ascom.2018.08.001}{\emph{Astron. Comput.}
  {\bfseries 25} (2018) 38} [\href{https://arxiv.org/abs/1708.05177}{{\ttfamily
  1708.05177}}].

\bibitem{Moser:2021rej}
B.~Moser, C.S.~Lorenz, U.~Schmitt, A.~Refregier, J.~Fluri, R.~Sgier et~al.,
  \emph{{Symbolic implementation of extensions of the PyCosmo~Boltzmann
  solver}}, \href{https://doi.org/10.1016/j.ascom.2022.100603}{\emph{Astron.
  Comput.} {\bfseries 40} (2022) 100603}
  [\href{https://arxiv.org/abs/2112.08395}{{\ttfamily 2112.08395}}].

\bibitem{Mead:2016zqy}
A.~Mead, C.~Heymans, L.~Lombriser, J.~Peacock, O.~Steele and H.~Winther,
  \emph{{Accurate halo-model matter power spectra with dark energy, massive
  neutrinos and modified gravitational forces}},
  \href{https://doi.org/10.1093/mnras/stw681}{\emph{Mon. Not. Roy. Astron.
  Soc.} {\bfseries 459} (2016) 1468}
  [\href{https://arxiv.org/abs/1602.02154}{{\ttfamily 1602.02154}}].

\bibitem{Petri:2015ura}
A.~Petri, J.~Liu, Z.~Haiman, M.~May, L.~Hui and J.M.~Kratochvil,
  \emph{{Emulating the CFHTLenS Weak Lensing data: Cosmological Constraints
  from moments and Minkowski functionals}},
  \href{https://doi.org/10.1103/PhysRevD.91.103511}{\emph{Phys. Rev. D}
  {\bfseries 91} (2015) 103511}
  [\href{https://arxiv.org/abs/1503.06214}{{\ttfamily 1503.06214}}].

\bibitem{Bolliet:2023sst}
B.~Bolliet, A.~Spurio~Mancini, J.C.~Hill, M.~Madhavacheril, H.T.~Jense,
  E.~Calabrese et~al., \emph{{High-accuracy emulators for observables in
  $\Lambda$CDM, $N_\mathrm{eff}$, $\Sigma m_\nu$, and $w$ cosmologies}},
  \href{https://arxiv.org/abs/2303.01591}{{\ttfamily 2303.01591}}.

\bibitem{Gong:2023nzy}
Z.~Gong, A.~Halder, A.~Barreira, S.~Seitz and O.~Friedrich, \emph{{Cosmology
  from the integrated shear 3-point correlation function: simulated likelihood
  analyses with machine-learning emulators}},
  \href{https://arxiv.org/abs/2304.01187}{{\ttfamily 2304.01187}}.

\bibitem{Fischbacher:2022gua}
S.~Fischbacher, T.~Kacprzak, J.~Blazek and A.~Refregier, \emph{{Redshift
  requirements for cosmic shear with intrinsic alignment}},
  \href{https://arxiv.org/abs/2207.01627}{{\ttfamily 2207.01627}}.

\bibitem{Piras:2023aub}
D.~Piras and A.~Spurio~Mancini, \emph{{CosmoPower-JAX: high-dimensional
  Bayesian inference with differentiable cosmological emulators}},
  \href{https://arxiv.org/abs/2305.06347}{{\ttfamily 2305.06347}}.

\bibitem{SpurioMancini:2021ppk}
A.~Spurio~Mancini, D.~Piras, J.~Alsing, B.~Joachimi and M.P.~Hobson,
  \emph{{CosmoPower: emulating cosmological power spectra for accelerated
  Bayesian inference from next-generation surveys}},
  \href{https://doi.org/10.1093/mnras/stac064}{\emph{Mon. Not. Roy. Astron.
  Soc.} {\bfseries 511} (2022) 1771}
  [\href{https://arxiv.org/abs/2106.03846}{{\ttfamily 2106.03846}}].

\bibitem{Euclid:2018mlb}
{\scshape Euclid} collaboration, \emph{{Euclid preparation: II. The
  EuclidEmulator -- A tool to compute the cosmology dependence of the nonlinear
  matter power spectrum}},
  \href{https://doi.org/10.1093/mnras/stz197}{\emph{Mon. Not. Roy. Astron.
  Soc.} {\bfseries 484} (2019) 5509}
  [\href{https://arxiv.org/abs/1809.04695}{{\ttfamily 1809.04695}}].

\bibitem{Angulo:2020vky}
R.E.~Angulo, M.~Zennaro, S.~Contreras, G.~Aric\`o, M.~Pellejero-Iba\~nez and
  J.~St\"ucker, \emph{{The BACCO simulation project: exploiting the full power
  of large-scale structure for cosmology}},
  \href{https://doi.org/10.1093/mnras/stab2018}{\emph{Mon. Not. Roy. Astron.
  Soc.} {\bfseries 507} (2021) 5869}
  [\href{https://arxiv.org/abs/2004.06245}{{\ttfamily 2004.06245}}].

\bibitem{tensorflow2015-whitepaper}
M.~Abadi, A.~Agarwal, P.~Barham, E.~Brevdo, Z.~Chen, C.~Citro et~al.,
  \emph{{TensorFlow}: Large-scale machine learning on heterogeneous systems},
  2015.

\bibitem{adamxyz}
D.P.~{Kingma} and J.~{Ba}, \emph{{Adam: A Method for Stochastic Optimization}},
  \href{https://doi.org/10.48550/arXiv.1412.6980}{\emph{arXiv e-prints} (2014)
  arXiv:1412.6980} [\href{https://arxiv.org/abs/1412.6980}{{\ttfamily
  1412.6980}}].

\bibitem{Hildebrandt:2020rno}
H.~Hildebrandt et~al., \emph{{KiDS-1000 catalogue: Redshift distributions and
  their calibration}},
  \href{https://doi.org/10.1051/0004-6361/202039018}{\emph{Astron. Astrophys.}
  {\bfseries 647} (2021) A124}
  [\href{https://arxiv.org/abs/2007.15635}{{\ttfamily 2007.15635}}].

\bibitem{Planck:2019evm}
{\scshape Planck} collaboration, \emph{{Planck 2018 results. VII. Isotropy and
  Statistics of the CMB}},
  \href{https://doi.org/10.1051/0004-6361/201935201}{\emph{Astron. Astrophys.}
  {\bfseries 641} (2020) A7}
  [\href{https://arxiv.org/abs/1906.02552}{{\ttfamily 1906.02552}}].

\bibitem{Planck:2018lbu}
{\scshape Planck} collaboration, \emph{{Planck 2018 results. VIII.
  Gravitational lensing}},
  \href{https://doi.org/10.1051/0004-6361/201833886}{\emph{Astron. Astrophys.}
  {\bfseries 641} (2020) A8}
  [\href{https://arxiv.org/abs/1807.06210}{{\ttfamily 1807.06210}}].

\bibitem{Kuijken:2019gsa}
K.~Kuijken et~al., \emph{{The fourth data release of the Kilo-Degree Survey:
  ugri imaging and nine-band optical-IR photometry over 1000 square degrees}},
  \href{https://doi.org/10.1051/0004-6361/201834918}{\emph{Astron. Astrophys.}
  {\bfseries 625} (2019) A2}
  [\href{https://arxiv.org/abs/1902.11265}{{\ttfamily 1902.11265}}].

\bibitem{edgexyz}
A.~{Edge}, W.~{Sutherland}, K.~{Kuijken}, S.~{Driver}, R.~{McMahon}, S.~{Eales}
  et~al., \emph{{The VISTA Kilo-degree Infrared Galaxy (VIKING) Survey:
  Bridging the Gap between Low and High Redshift}}, {\emph{The Messenger}
  {\bfseries 154} (2013) 32}.

\bibitem{Giblin:2020quj}
B.~Giblin et~al., \emph{{KiDS-1000 catalogue: Weak gravitational lensing shear
  measurements}},
  \href{https://doi.org/10.1051/0004-6361/202038850}{\emph{Astron. Astrophys.}
  {\bfseries 645} (2021) A105}
  [\href{https://arxiv.org/abs/2007.01845}{{\ttfamily 2007.01845}}].

\bibitem{Fluri:2022rvb}
J.~Fluri, T.~Kacprzak, A.~Lucchi, A.~Schneider, A.~Refregier and T.~Hofmann,
  \emph{{Full wCDM analysis of KiDS-1000 weak lensing maps using deep
  learning}}, \href{https://doi.org/10.1103/PhysRevD.105.083518}{\emph{Phys.
  Rev. D} {\bfseries 105} (2022) 083518}
  [\href{https://arxiv.org/abs/2201.07771}{{\ttfamily 2201.07771}}].

\bibitem{Troster:2021gsz}
T.~Tr\"oster et~al., \emph{{Joint constraints on cosmology and the impact of
  baryon feedback: Combining KiDS-1000 lensing with the thermal
  Sunyaev\textendash{}Zeldovich effect from Planck and ACT}},
  \href{https://doi.org/10.1051/0004-6361/202142197}{\emph{Astron. Astrophys.}
  {\bfseries 660} (2022) A27}
  [\href{https://arxiv.org/abs/2109.04458}{{\ttfamily 2109.04458}}].

\bibitem{Reid:2015gra}
B.~Reid et~al., \emph{{SDSS-III Baryon Oscillation Spectroscopic Survey Data
  Release 12: galaxy target selection and large scale structure catalogues}},
  \href{https://doi.org/10.1093/mnras/stv2382}{\emph{Mon. Not. Roy. Astron.
  Soc.} {\bfseries 455} (2016) 1553}
  [\href{https://arxiv.org/abs/1509.06529}{{\ttfamily 1509.06529}}].

\bibitem{SDSS-III:2015hof}
{\scshape SDSS-III} collaboration, \emph{{The Eleventh and Twelfth Data
  Releases of the Sloan Digital Sky Survey: Final Data from SDSS-III}},
  \href{https://doi.org/10.1088/0067-0049/219/1/12}{\emph{Astrophys. J. Suppl.}
  {\bfseries 219} (2015) 12}
  [\href{https://arxiv.org/abs/1501.00963}{{\ttfamily 1501.00963}}].

\bibitem{Swanson:2007aj}
M.E.C.~Swanson, M.~Tegmark, A.J.S.~Hamilton and J.C.~Hill, \emph{{Methods for
  Rapidly Processing Angular Masks of Next-Generation Galaxy Surveys}},
  \href{https://doi.org/10.1111/j.1365-2966.2008.13296.x}{\emph{Mon. Not. Roy.
  Astron. Soc.} {\bfseries 387} (2008) 1391}
  [\href{https://arxiv.org/abs/0711.4352}{{\ttfamily 0711.4352}}].

\bibitem{Hamilton:2003ea}
A.J.S.~Hamilton and M.~Tegmark, \emph{{A Scheme to deal accurately and
  efficiently with complex angular masks in galaxy surveys}},
  \href{https://doi.org/10.1111/j.1365-2966.2004.07490.x}{\emph{Mon. Not. Roy.
  Astron. Soc.} {\bfseries 349} (2004) 115}
  [\href{https://arxiv.org/abs/astro-ph/0306324}{{\ttfamily
  astro-ph/0306324}}].

\bibitem{Loureiro:2018qva}
A.~Loureiro et~al., \emph{{Cosmological measurements from angular power spectra
  analysis of BOSS DR12 tomography}},
  \href{https://doi.org/10.1093/mnras/stz191}{\emph{Mon. Not. Roy. Astron.
  Soc.} {\bfseries 485} (2019) 326}
  [\href{https://arxiv.org/abs/1809.07204}{{\ttfamily 1809.07204}}].

\bibitem{Planck:2018nkj}
{\scshape Planck} collaboration, \emph{{Planck 2018 results. I. Overview and
  the cosmological legacy of Planck}},
  \href{https://doi.org/10.1051/0004-6361/201833880}{\emph{Astron. Astrophys.}
  {\bfseries 641} (2020) A1}
  [\href{https://arxiv.org/abs/1807.06205}{{\ttfamily 1807.06205}}].

\bibitem{Krolewski:2021yqy}
A.~Krolewski, S.~Ferraro and M.~White, \emph{{Cosmological constraints from
  unWISE and Planck CMB lensing tomography}},
  \href{https://doi.org/10.1088/1475-7516/2021/12/028}{\emph{JCAP} {\bfseries
  12} (2021) 028} [\href{https://arxiv.org/abs/2105.03421}{{\ttfamily
  2105.03421}}].

\bibitem{Hivon:2001jp}
E.~Hivon, K.M.~Gorski, C.B.~Netterfield, B.P.~Crill, S.~Prunet and F.~Hansen,
  \emph{{Master of the cosmic microwave background anisotropy power spectrum: a
  fast method for statistical analysis of large and complex cosmic microwave
  background data sets}},
  \href{https://doi.org/10.1086/338126}{\emph{Astrophys. J.} {\bfseries 567}
  (2002) 2} [\href{https://arxiv.org/abs/astro-ph/0105302}{{\ttfamily
  astro-ph/0105302}}].

\bibitem{Alonso:2018jzx}
{\scshape LSST Dark Energy Science} collaboration, \emph{{A unified
  pseudo-$C_\ell$ framework}},
  \href{https://doi.org/10.1093/mnras/stz093}{\emph{Mon. Not. Roy. Astron.
  Soc.} {\bfseries 484} (2019) 4127}
  [\href{https://arxiv.org/abs/1809.09603}{{\ttfamily 1809.09603}}].

\bibitem{SDSS:2006egz}
{\scshape SDSS} collaboration, \emph{{The Clustering of Luminous Red Galaxies
  in the Sloan Digital Sky Survey Imaging Data}},
  \href{https://doi.org/10.1111/j.1365-2966.2007.11593.x}{\emph{Mon. Not. Roy.
  Astron. Soc.} {\bfseries 378} (2007) 852}
  [\href{https://arxiv.org/abs/astro-ph/0605302}{{\ttfamily
  astro-ph/0605302}}].

\bibitem{Fang:2019xat}
X.~Fang, E.~Krause, T.~Eifler and N.~MacCrann, \emph{{Beyond Limber: Efficient
  computation of angular power spectra for galaxy clustering and weak
  lensing}}, \href{https://doi.org/10.1088/1475-7516/2020/05/010}{\emph{JCAP}
  {\bfseries 05} (2020) 010}
  [\href{https://arxiv.org/abs/1911.11947}{{\ttfamily 1911.11947}}].

\bibitem{LSSTDarkEnergyScience:2022lno}
{\scshape LSST Dark Energy Science} collaboration, \emph{{The N5K Challenge:
  Non-Limber Integration for LSST Cosmology}},
  \href{https://doi.org/10.21105/astro.2212.04291}{\emph{Open Journal of
  Astrophysics} {\bfseries 6} (2023) 1}
  [\href{https://arxiv.org/abs/2212.04291}{{\ttfamily 2212.04291}}].

\bibitem{Planck:2019nip}
{\scshape Planck} collaboration, \emph{{Planck 2018 results. V. CMB power
  spectra and likelihoods}},
  \href{https://doi.org/10.1051/0004-6361/201936386}{\emph{Astron. Astrophys.}
  {\bfseries 641} (2020) A5}
  [\href{https://arxiv.org/abs/1907.12875}{{\ttfamily 1907.12875}}].

\bibitem{Dunkley:2013vu}
J.~Dunkley et~al., \emph{{The Atacama Cosmology Telescope: likelihood for
  small-scale CMB data}},
  \href{https://doi.org/10.1088/1475-7516/2013/07/025}{\emph{JCAP} {\bfseries
  07} (2013) 025} [\href{https://arxiv.org/abs/1301.0776}{{\ttfamily
  1301.0776}}].

\bibitem{Prince:2021fdv}
H.~Prince and J.~Dunkley, \emph{{Compressed Python likelihood for large scale
  temperature and polarization from Planck}},
  \href{https://doi.org/10.1103/PhysRevD.105.023518}{\emph{Phys. Rev. D}
  {\bfseries 105} (2022) 023518}
  [\href{https://arxiv.org/abs/2104.05715}{{\ttfamily 2104.05715}}].

\bibitem{Hu:2001kj}
W.~Hu and T.~Okamoto, \emph{{Mass reconstruction with cmb polarization}},
  \href{https://doi.org/10.1086/341110}{\emph{Astrophys. J.} {\bfseries 574}
  (2002) 566} [\href{https://arxiv.org/abs/astro-ph/0111606}{{\ttfamily
  astro-ph/0111606}}].

\bibitem{Planck:2015mym}
{\scshape Planck} collaboration, \emph{{Planck 2015 results. XV. Gravitational
  lensing}}, \href{https://doi.org/10.1051/0004-6361/201525941}{\emph{Astron.
  Astrophys.} {\bfseries 594} (2016) A15}
  [\href{https://arxiv.org/abs/1502.01591}{{\ttfamily 1502.01591}}].

\bibitem{Schmittfull:2013uea}
M.M.~Schmittfull, A.~Challinor, D.~Hanson and A.~Lewis, \emph{{Joint analysis
  of CMB temperature and lensing-reconstruction power spectra}},
  \href{https://doi.org/10.1103/PhysRevD.88.063012}{\emph{Phys. Rev. D}
  {\bfseries 88} (2013) 063012}
  [\href{https://arxiv.org/abs/1308.0286}{{\ttfamily 1308.0286}}].

\bibitem{Peloton:2016kbw}
J.~Peloton, M.~Schmittfull, A.~Lewis, J.~Carron and O.~Zahn, \emph{{Full
  covariance of CMB and lensing reconstruction power spectra}},
  \href{https://doi.org/10.1103/PhysRevD.95.043508}{\emph{Phys. Rev. D}
  {\bfseries 95} (2017) 043508}
  [\href{https://arxiv.org/abs/1611.01446}{{\ttfamily 1611.01446}}].

\bibitem{Hartlap:2006kj}
J.~Hartlap, P.~Simon and P.~Schneider, \emph{{Why your model parameter
  confidences might be too optimistic: Unbiased estimation of the inverse
  covariance matrix}},
  \href{https://doi.org/10.1051/0004-6361:20066170}{\emph{Astron. Astrophys.}
  {\bfseries 464} (2007) 399}
  [\href{https://arxiv.org/abs/astro-ph/0608064}{{\ttfamily
  astro-ph/0608064}}].

\bibitem{Tegmark:1996bz}
M.~Tegmark, A.~Taylor and A.~Heavens, \emph{{Karhunen-Loeve eigenvalue problems
  in cosmology: How should we tackle large data sets?}},
  \href{https://doi.org/10.1086/303939}{\emph{Astrophys. J.} {\bfseries 480}
  (1997) 22} [\href{https://arxiv.org/abs/astro-ph/9603021}{{\ttfamily
  astro-ph/9603021}}].

\bibitem{Heavens:2017efz}
A.~Heavens, E.~Sellentin, D.~de~Mijolla and A.~Vianello, \emph{{Massive data
  compression for parameter-dependent covariance matrices}},
  \href{https://doi.org/10.1093/mnras/stx2326}{\emph{Mon. Not. Roy. Astron.
  Soc.} {\bfseries 472} (2017) 4244}
  [\href{https://arxiv.org/abs/1707.06529}{{\ttfamily 1707.06529}}].

\bibitem{Heavens:2020spq}
A.~Heavens, E.~Sellentin and A.~Jaffe, \emph{{Extreme data compression while
  searching for new physics}},
  \href{https://doi.org/10.1093/mnras/staa2589}{\emph{Mon. Not. Roy. Astron.
  Soc.} {\bfseries 498} (2020) 3440}
  [\href{https://arxiv.org/abs/2006.06706}{{\ttfamily 2006.06706}}].

\bibitem{Campagne:2023ter}
J.-E.~Campagne, F.~Lanusse, J.~Zuntz, A.~Boucaud, S.~Casas, M.~Karamanis
  et~al., \emph{{JAX-COSMO: An End-to-End Differentiable and GPU Accelerated
  Cosmology Library}},
  \href{https://doi.org/10.21105/astro.2302.05163}{\emph{Open J. Astrophys.}
  {\bfseries 6} (2023) 1} [\href{https://arxiv.org/abs/2302.05163}{{\ttfamily
  2302.05163}}].

\bibitem{Zurcher:2022clh}
D.~Z\"urcher, J.~Fluri, V.~Ajani, S.~Fischbacher, A.~Refregier and T.~Kacprzak,
  \emph{{Towards a full $w$CDM map-based analysis for weak lensing surveys}},
  \href{https://arxiv.org/abs/2206.01450}{{\ttfamily 2206.01450}}.

\bibitem{Vagnozzi:2017ovm}
S.~Vagnozzi, E.~Giusarma, O.~Mena, K.~Freese, M.~Gerbino, S.~Ho et~al.,
  \emph{{Unveiling $\nu$ secrets with cosmological data: neutrino masses and
  mass hierarchy}},
  \href{https://doi.org/10.1103/PhysRevD.96.123503}{\emph{Phys. Rev. D}
  {\bfseries 96} (2017) 123503}
  [\href{https://arxiv.org/abs/1701.08172}{{\ttfamily 1701.08172}}].

\bibitem{Giusarma:2016phn}
E.~Giusarma, M.~Gerbino, O.~Mena, S.~Vagnozzi, S.~Ho and K.~Freese,
  \emph{{Improvement of cosmological neutrino mass bounds}},
  \href{https://doi.org/10.1103/PhysRevD.94.083522}{\emph{Phys. Rev. D}
  {\bfseries 94} (2016) 083522}
  [\href{https://arxiv.org/abs/1605.04320}{{\ttfamily 1605.04320}}].

\bibitem{Archidiacono:2016lnv}
M.~Archidiacono, T.~Brinckmann, J.~Lesgourgues and V.~Poulin, \emph{{Physical
  effects involved in the measurements of neutrino masses with future
  cosmological data}},
  \href{https://doi.org/10.1088/1475-7516/2017/02/052}{\emph{JCAP} {\bfseries
  02} (2017) 052} [\href{https://arxiv.org/abs/1610.09852}{{\ttfamily
  1610.09852}}].

\bibitem{Joachimi:2020abi}
B.~Joachimi et~al., \emph{{KiDS-1000 methodology: Modelling and inference for
  joint weak gravitational lensing and spectroscopic galaxy clustering
  analysis}}, \href{https://doi.org/10.1051/0004-6361/202038831}{\emph{Astron.
  Astrophys.} {\bfseries 646} (2021) A129}
  [\href{https://arxiv.org/abs/2007.01844}{{\ttfamily 2007.01844}}].

\bibitem{BaleatoLizancos:2023zpl}
A.~Baleato~Lizancos and M.~White, \emph{{The impact of anisotropic redshift
  distributions on angular clustering}},
  \href{https://doi.org/10.1088/1475-7516/2023/07/044}{\emph{JCAP} {\bfseries
  07} (2023) 044} [\href{https://arxiv.org/abs/2305.15406}{{\ttfamily
  2305.15406}}].

\bibitem{Foreman-Mackey:2012any}
D.~Foreman-Mackey, D.W.~Hogg, D.~Lang and J.~Goodman, \emph{{emcee: The MCMC
  Hammer}}, \href{https://doi.org/10.1086/670067}{\emph{Publ. Astron. Soc.
  Pac.} {\bfseries 125} (2013) 306}
  [\href{https://arxiv.org/abs/1202.3665}{{\ttfamily 1202.3665}}].

\bibitem{Handley:2019wlz}
W.~Handley and P.~Lemos, \emph{{Quantifying tensions in cosmological
  parameters: Interpreting the DES evidence ratio}},
  \href{https://doi.org/10.1103/PhysRevD.100.043504}{\emph{Phys. Rev. D}
  {\bfseries 100} (2019) 043504}
  [\href{https://arxiv.org/abs/1902.04029}{{\ttfamily 1902.04029}}].

\bibitem{Boyle:2020rxq}
A.~Boyle and F.~Schmidt, \emph{{Neutrino mass constraints beyond linear order:
  cosmology dependence and systematic biases}},
  \href{https://doi.org/10.1088/1475-7516/2021/04/022}{\emph{JCAP} {\bfseries
  04} (2021) 022} [\href{https://arxiv.org/abs/2011.10594}{{\ttfamily
  2011.10594}}].

\bibitem{Tanseri:2022zfe}
I.~Tanseri, S.~Hagstotz, S.~Vagnozzi, E.~Giusarma and K.~Freese, \emph{{Updated
  neutrino mass constraints from galaxy clustering and CMB lensing-galaxy
  cross-correlation measurements}},
  \href{https://doi.org/10.1016/j.jheap.2022.07.002}{\emph{JHEAp} {\bfseries
  36} (2022) 1} [\href{https://arxiv.org/abs/2207.01913}{{\ttfamily
  2207.01913}}].

\bibitem{DiValentino:2021hoh}
E.~Di~Valentino, S.~Gariazzo and O.~Mena, \emph{{Most constraining cosmological
  neutrino mass bounds}},
  \href{https://doi.org/10.1103/PhysRevD.104.083504}{\emph{Phys. Rev. D}
  {\bfseries 104} (2021) 083504}
  [\href{https://arxiv.org/abs/2106.15267}{{\ttfamily 2106.15267}}].

\bibitem{Chudaykin:2022rnl}
A.~Chudaykin, D.~Gorbunov and N.~Nedelko, \emph{{Exploring ${\Lambda}$CDM
  extensions with SPT-3G and Planck data: 4$\sigma$ evidence for neutrino
  masses and implications of extended dark energy models for cosmological
  tensions}},  \href{https://arxiv.org/abs/2203.03666}{{\ttfamily 2203.03666}}.

\bibitem{ACT:2023kun}
{\scshape ACT} collaboration, \emph{{The Atacama Cosmology Telescope: DR6
  Gravitational Lensing Map and Cosmological Parameters}},
  \href{https://arxiv.org/abs/2304.05203}{{\ttfamily 2304.05203}}.

\bibitem{Domenech:2019cyh}
G.~Dom\`enech and M.~Kamionkowski, \emph{{Lensing anomaly and oscillations in
  the primordial power spectrum}},
  \href{https://doi.org/10.1088/1475-7516/2019/11/040}{\emph{JCAP} {\bfseries
  11} (2019) 040} [\href{https://arxiv.org/abs/1905.04323}{{\ttfamily
  1905.04323}}].

\bibitem{Couchot:2017pvz}
F.~Couchot, S.~Henrot-Versill\'e, O.~Perdereau, S.~Plaszczynski,
  B.~Rouill\'e~d'Orfeuil, M.~Spinelli et~al., \emph{{Cosmological constraints
  on the neutrino mass including systematic uncertainties}},
  \href{https://doi.org/10.1051/0004-6361/201730927}{\emph{Astron. Astrophys.}
  {\bfseries 606} (2017) A104}
  [\href{https://arxiv.org/abs/1703.10829}{{\ttfamily 1703.10829}}].

\bibitem{Motloch:2019gux}
P.~Motloch and W.~Hu, \emph{{Lensinglike tensions in the $Planck$ legacy
  release}}, \href{https://doi.org/10.1103/PhysRevD.101.083515}{\emph{Phys.
  Rev. D} {\bfseries 101} (2020) 083515}
  [\href{https://arxiv.org/abs/1912.06601}{{\ttfamily 1912.06601}}].

\bibitem{Trendafilova:2023oni}
C.~Trendafilova, \emph{{The Impact of Cross-Covariances Between the CMB and
  Reconstructed Lensing Power}},
  \href{https://arxiv.org/abs/2308.11588}{{\ttfamily 2308.11588}}.

\bibitem{Fabbian:2019tik}
G.~Fabbian, A.~Lewis and D.~Beck, \emph{{CMB lensing reconstruction biases in
  cross-correlation with large-scale structure probes}},
  \href{https://doi.org/10.1088/1475-7516/2019/10/057}{\emph{JCAP} {\bfseries
  10} (2019) 057} [\href{https://arxiv.org/abs/1906.08760}{{\ttfamily
  1906.08760}}].

\bibitem{vanEngelen:2013rla}
A.~van Engelen, S.~Bhattacharya, N.~Sehgal, G.P.~Holder, O.~Zahn and D.~Nagai,
  \emph{{CMB Lensing Power Spectrum Biases from Galaxies and Clusters using
  High-angular Resolution Temperature Maps}},
  \href{https://doi.org/10.1088/0004-637X/786/1/13}{\emph{Astrophys. J.}
  {\bfseries 786} (2014) 13} [\href{https://arxiv.org/abs/1310.7023}{{\ttfamily
  1310.7023}}].

\bibitem{DES:2005dhi}
{\scshape DES} collaboration, \emph{{The Dark Energy Survey}},
  \href{https://arxiv.org/abs/astro-ph/0510346}{{\ttfamily astro-ph/0510346}}.

\bibitem{DES:2021wwk}
{\scshape DES} collaboration, \emph{{Dark Energy Survey Year 3 results:
  Cosmological constraints from galaxy clustering and weak lensing}},
  \href{https://doi.org/10.1103/PhysRevD.105.023520}{\emph{Phys. Rev. D}
  {\bfseries 105} (2022) 023520}
  [\href{https://arxiv.org/abs/2105.13549}{{\ttfamily 2105.13549}}].

\bibitem{Halder:2023kfy}
A.~Halder, Z.~Gong, A.~Barreira, O.~Friedrich, S.~Seitz and D.~Gruen,
  \emph{{Beyond 3\texttimes{}2-point cosmology: the integrated shear and galaxy
  3-point correlation functions}},
  \href{https://doi.org/10.1088/1475-7516/2023/10/028}{\emph{JCAP} {\bfseries
  10} (2023) 028} [\href{https://arxiv.org/abs/2305.17132}{{\ttfamily
  2305.17132}}].

\bibitem{DESI:2016fyo}
{\scshape DESI} collaboration, \emph{{The DESI Experiment Part I:
  Science,Targeting, and Survey Design}},
  \href{https://arxiv.org/abs/1611.00036}{{\ttfamily 1611.00036}}.

\bibitem{DESI:2023ytc}
{\scshape DESI} collaboration, \emph{{The Early Data Release of the Dark Energy
  Spectroscopic Instrument}},
  \href{https://arxiv.org/abs/2306.06308}{{\ttfamily 2306.06308}}.

\bibitem{harris2020array}
C.R.~Harris, K.J.~Millman, S.J.~van~der Walt, R.~Gommers, P.~Virtanen,
  D.~Cournapeau et~al., \emph{Array programming with {NumPy}},
  \href{https://doi.org/10.1038/s41586-020-2649-2}{\emph{Nature} {\bfseries
  585} (2020) 357}.

\bibitem{2020SciPy-NMeth}
P.~Virtanen, R.~Gommers, T.E.~Oliphant, M.~Haberland, T.~Reddy, D.~Cournapeau
  et~al., \emph{{{SciPy} 1.0: Fundamental Algorithms for Scientific Computing
  in Python}}, \href{https://doi.org/10.1038/s41592-019-0686-2}{\emph{Nature
  Methods} {\bfseries 17} (2020) 261}.

\bibitem{Hunter:2007}
J.D.~Hunter, \emph{Matplotlib: A 2d graphics environment},
  \href{https://doi.org/10.1109/MCSE.2007.55}{\emph{Computing in Science \&
  Engineering} {\bfseries 9} (2007) 90}.

\bibitem{Zurcher:2020dvu}
D.~Z\"urcher, J.~Fluri, R.~Sgier, T.~Kacprzak and A.~Refregier,
  \emph{{Cosmological Forecast for non-Gaussian Statistics in large-scale weak
  Lensing Surveys}},
  \href{https://doi.org/10.1088/1475-7516/2021/01/028}{\emph{JCAP} {\bfseries
  01} (2021) 028} [\href{https://arxiv.org/abs/2006.12506}{{\ttfamily
  2006.12506}}].

\end{thebibliography}\endgroup
\end{document}